\documentclass[twocolumn]{aastex63}
\usepackage{array,multirow}
\usepackage{amsmath}
\usepackage{wrapfig}
\usepackage{acronym}
\usepackage{enumitem}

\usepackage{lineno}

\newcommand{\changed}[1]{\textcolor{black}{#1}}
\newcommand{\changedtwo}[1]{\textcolor{black}{#1}}

\acrodef{HMXB}{high-mass X-ray binary}
\acrodefplural{HMXB}[HMXBs]{high-mass X-ray binaries}
\acrodef{LMXB}{low-mass X-ray binary}
\acrodefplural{LMXB}[LMXBs]{low-mass X-ray binaries}
\acrodef{BH}{black hole}
\acrodef{CO}{compact object}
\acrodef{BBH}{binary black hole}
\acrodef{NS}{neutron star}
\acrodef{WD}{white dwarf}
\acrodef{GW}{gravitational-wave}
\acrodef{DCO}{double compact object}
\acrodef{LIGO}{Laser Interferometer Gravitational-Wave Observatory}
\acrodef{HST}{\textit{Hubble Space Telescope}}
\acrodef{SMC}{Small Magellanic Cloud}
\acrodef{LMC}{Large Magellanic Cloud}
\acrodef{XRB}{X-ray binary}
\acrodef{ULX}{ultraluminous X-ray source}
\acrodef{IMF}{initial mass function}
\acrodef{ZAMS}{zero-age main-sequence}
\acrodef{SN}{supernova}
\acrodefplural{SN}[SNe]{supernovae}

\submitjournal{ApJ}
\shorttitle{The Missing Link Between HMXBs and GW Sources}

\graphicspath{{./}{figures/}}

\begin{document}

\title{The Missing Link Between Black Holes in High-Mass X-ray Binaries and Gravitational-Wave Sources: \\Observational Selection Effects}

\author[0000-0002-8883-3351]{Camille Liotine}
\affiliation{Center for Interdisciplinary Exploration and Research in Astrophysics (CIERA), Northwestern University, \\ 1800 Sherman, Evanston, IL, 60201, USA}
\affiliation{Department of Physics and Astronomy, Northwestern University, 2145 Sheridan Road, Evanston, IL 60208, USA}

\author[0000-0002-0147-0835]{Michael Zevin}
\affiliation{Kavli Institute for Cosmological Physics, The University of Chicago, 5640 South Ellis Avenue, Chicago, Illinois 60637, USA}
\affiliation{Enrico Fermi Institute, The University of Chicago, 933 East 56th Street, Chicago, Illinois 60637, USA}

\author[0000-0003-3870-7215]{Christopher Berry}
\affiliation{Institute for Gravitational Research, University of Glasgow, Kelvin Building, University Avenue, Glasgow, G12 8QQ, Scotland}
\affiliation{Center for Interdisciplinary Exploration and Research in Astrophysics (CIERA), Northwestern University, \\ 1800 Sherman, Evanston, IL, 60201, USA}

\author[0000-0002-2077-4914]{Zoheyr Doctor}
\affiliation{Center for Interdisciplinary Exploration and Research in Astrophysics (CIERA), Northwestern University, \\ 1800 Sherman, Evanston, IL, 60201, USA}

\author[0000-0001-9236-5469]{Vicky Kalogera}
\affiliation{Center for Interdisciplinary Exploration and Research in Astrophysics (CIERA), Northwestern University, \\ 1800 Sherman, Evanston, IL, 60201, USA}
\affiliation{Department of Physics and Astronomy, Northwestern University, 2145 Sheridan Road, Evanston, IL 60208, USA}

\begin{abstract}

 There are few observed high-mass X-ray binaries (HMXBs) that harbor massive black holes, and none are likely to result in a binary black hole (BBH) that merges within a Hubble time; however, we know that massive merging BBHs exist from gravitational-wave observations. 
 We investigate the role that X-ray and gravitational-wave observational selection effects play in determining the properties of their respective detected binary populations. 
 We \changed{find} that, as a result of selection effects, \changed{detectable} HMXBs and \changed{detectable} BBHs form at different redshifts and metallicities, with \changed{detectable} HMXBs forming at much lower redshifts and higher metallicities than \changed{detectable} BBHs. 
 We also find disparities in the mass distributions of these populations, with \changed{detectable} merging BBH progenitors pulling to higher component masses relative to the full \changed{detectable} HMXB population. 
Fewer than $3\%$ of \changed{detectable} HMXBs host black holes $> 35M_{\odot}$ in our simulated populations.
 Furthermore, we find the probability that a detectable HMXB will merge as a BBH system within a Hubble time is $\simeq 0.6\%$. 
 Thus, it is unsurprising that no currently observed HMXB is predicted to form a merging BBH with high probability.
\end{abstract}


\section{Introduction} \label{sec:intro}

Current \ac{GW} sources consist of merging \changed{\acp{DCO} with} \ac{NS} or \ac{BH} components~\citep{abbott_gwtc-1_2019, abbott_gwtc-2_2021, abbott_gwtc-3_2021, abbott_gwtc-21_2022}. 
Their progenitors are important markers of binary evolution, as they provide information on the origin of \acp{DCO} and inform which evolutionary scenarios dominate their formation~\citep{podsiadlowski_formation_2003, tauris_formation_2003, kratter_formation_2011, miller_masses_2015, heuvel_high-mass_2018, abbott_astrophysical_2016}. 
Such progenitors include \acp{HMXB}, which are binary systems that contain a massive OB-type donor star with an accreting \ac{CO} that can be either a \ac{NS} or a \ac{BH}~\citep[e.g.,][]{verbunt_origin_1993, remillard_x-ray_2006}. 
The donor star in \ac{HMXB} systems is massive enough to itself form a \ac{CO}, making \acp{HMXB} prime candidates for the progenitors of \ac{GW} sources \changed{detectable} by current ground-based detectors.

\acp{HMXB} are predominantly wind fed, meaning the accretion process is fueled by the stellar wind of the donor star~\citep{blondin_wind_1997}. 
At its simplest level, accretion can be \changed{modeled using the classical Bondi--Hoyle method}~\citep{1944MNRAS.104..273B} and drives X-ray emission that can be observed with contemporary X-ray surveys~\citep[e.g.,][]{oh_105-month_2018, krivonos_15_2021, pavlinsky_art-xc_2021, predehl_erosita_2021}. 
\changed{When} accretion is Eddington-limited, the X-ray accretion luminosity is also limited, \changed{e.g., emission is limited} to $\sim10^{39}~\mathrm{erg\,s}^{-1}$ for a $10~M_{\odot}$ \ac{BH} accretor. 
All observed wind-fed \acp{HMXB} with confirmed \ac{BH} accretors have high Roche lobe filling factors~\citep{orosz_1565-solar-mass_2007, orosz_new_2009, miller-jones_cygnus_2021}, which means that these systems could soon begin Roche lobe overflow mass transfer. 
\changed{High Roche lobe filling factors may also be important in determining which \acp{HMXB} are detectable, as they are critical in forming focused accretion streams that drive \ac{BH} disk formation~\citep{hirai_conditions_2021}}.

Approximately $80$ Galactic \acp{HMXB} have been observed, but not all systems have well-constrained binary properties~\citep{krivonos_integralibis_2012, clavel_nustar_2019, kretschmar_advances_2019}. 
Most of these \acp{HMXB} are thought to have \ac{BH} accretors, but the majority have yet to be dynamically confirmed with spectroscopic observations \citep{motta_integral_2021}. 
Cyg X-1, a Galactic \ac{HMXB} containing the first-ever dynamically confirmed stellar-mass \ac{BH}~\citep{bolton_identification_1972, webster_cygnus_1972}, is estimated to contain a $21.2^{+2.2}_{-2.2}~M_{\odot}$ \ac{BH} accretor and a ${40.6^{+7.7}_{-7.1}~M_{\odot}}$ main-sequence donor~\citep{miller-jones_cygnus_2021}. 
Another well-studied Galactic \ac{HMXB} is Cyg X-3, which is estimated to contain a \ac{CO} accretor of $2.4^{+2.1}_{-1.1}~M_{\odot}$ and a Wolf--Rayet donor of $10.3^{+3.9}_{-2.8}~M_{\odot}$~\citep{zdziarski_cyg_2013}. 
While the mass of the accretor falls near the maximum mass allowed by the uncertain \ac{NS} equation of state~\citep[][]{lattimer_neutron_2021}, radio, infrared and X-ray properties of the system suggest that it is a low-mass \ac{BH}~\citep{zdziarski_cyg_2013}. 
Thus, of the few Galactic \acp{HMXB} that have well-constrained binary properties, only Cyg X-1 is thought to contain a \ac{BH} $\gtrsim 20 M_{\odot}$.

Observational campaigns throughout the past two decades have uncovered a large number of extragalactic \ac{HMXB} candidates~\citep[e.g.,][]{fabbiano_populations_2006, haberl_high-mass_2016, lazzarini_young_2018, rice_x-ray_2021}. 
Of these sources, one of the few that has well-resolved binary properties is LMC X-1, the brightest X-ray source in the Large Magellanic Cloud~\citep{mark_detection_1969}. 
\citet{orosz_new_2009} estimate it to have a donor mass of $(31.8\pm3.5)~M_{\odot}$ and a \ac{BH} accretor mass of $(10.9\pm1.4)~M_{\odot}$. 
Another resolved extragalactic \changed{\ac{HMXB}} is M33 X-7, an \ac{HMXB} in the spiral galaxy M33 estimated to have a $38^{+22}_{-10} M_{\odot}$ donor star and a $11.4^{+3.3}_{-1.7}~M_{\odot}$ \ac{BH}~\citep{ramachandran_phase-resolved_2022}. 
Similar to Galactic \changed{\acp{HMXB}}, the population of well-constrained extragalactic sources is small, and in this case none are found to harbor \ac{BH} accretors $>20~M_{\odot}$. 

In addition to dynamically-observed \acp{BH} in \changed{\acp{HMXB}}, many stellar-mass \acp{BH} have been discovered with the \acl{LIGO}~\changed{\citep[\acs{LIGO};\acused{LIGO}][]{aasi_advanced_2015}} and Virgo~\changed{\citep{acernese_advanced_2015}} \ac{GW} detectors as \ac{DCO} mergers~\changed{\citep{abbott_observation_2016, abbott_gwtc-3_2021}}. 
The global \ac{GW} detector network made the first observation of \acp{GW} from a merging \ac{BBH} with component masses of $34.6^{+4.4}_{-2.6}~M_{\odot}$ and $30.0^{+2.9}_{-4.6}~M_{\odot}$ in 2015~\citep{abbott_observation_2016, abbott_gwtc-21_2022}. 
The \ac{LIGO} Scientific, Virgo and KAGRA Collaboration has identified $90$ probable \ac{DCO} merger candidates during their first three observing runs~\citep{abbott_gwtc-1_2019, abbott_gwtc-2_2021, abbott_gwtc-21_2022, abbott_gwtc-3_2021}, including two sources that have component masses consistent with a \ac{NS}--\ac{NS} merger~\citep{abbott_gw170817_2017, abbott_gw190425_2020} and a few sources with component masses consistent with a \ac{NS}--\ac{BH} merger~\citep{abbott_gw190814_2020, abbott_gwtc-3_2021, abbott_observation_2021, abbott_gwtc-21_2022}. 

Using \ac{GW} observations, population analyses are able to constrain aspects of the underlying \ac{BBH} mass distribution, with the most recent analyses identifying substructure in the mass distribution beyond the simplest phenomenological models~\citep{tiwari_emergence_2021, abbott_population_2022, edelman_aint_2022}. 
The global peak of the mass distribution is at $\simeq10M_{\odot}$, and there is strong evidence of a secondary peak at $\simeq35M_{\odot}$. 
Though there is clear support for masses above what would be allowed for a simple power law with a cutoff, the $99$th percentile of the underlying mass distribution is $44^{+9.2}_{-5.1}M_{\odot}$~\citep{abbott_population_2022}.
Distributions of \ac{BH} source properties, such as their mass, can be used to probe the astrophysics of \ac{BBH} formation and evolution~\citep[][]{belczynski_first_2016, barrett_accuracy_2018, fishbach_most_2020, zevin_one_2021,mandel_merging_2018}.

Since \acp{HMXB} will potentially form \changed{\acp{DCO}}, they are prime candidates for the progenitors of compact binary mergers detected via \acp{GW}. 
However, there is uncertainty as to whether \ac{HMXB} progenitors significantly contribute to the merging compact binary population.
For example, the \changed{\ac{HMXB}} donor star and \ac{CO} may merge later in their evolution before becoming a compact binary, get disrupted during the supernova that forms the second \ac{CO}, or have too wide of an orbital separation at \ac{DCO} formation to merge within a Hubble time. 

Several studies have been conducted to predict the fate of observed \acp{HMXB}. 
\citet{neijssel_wind_2021} predict the fate of Cyg X-1 with its updated \ac{BH} and donor star mass estimates from \citet{miller-jones_cygnus_2021} by evolving it forward with the \texttt{COMPAS} population synthesis code~\citep{riley_rapid_2022}. 
They find that the system will most likely form a \ac{BH}--\ac{NS} binary that has a $\simeq 7\%$ chance of remaining bound after the \ac{NS} natal kick. 
With revised models of mass transfer and natal kicks that produce heavier donor remnant masses, \citet{neijssel_wind_2021} predict that Cyg X-1 could potentially form a \ac{BBH} albeit at a low probability; in these scenarios, they find the probability Cyg X-1 will merge as a \ac{BBH} within a Hubble time is $\simeq$ 4--5\%. 
Similar studies have been done for LMC X-1~\citep{belczynski_high_2012} and Cyg X-3~\citep{belczynski_cyg_2013}. 
Of these two binaries, only Cyg X-3 is predicted to potentially merge as a \ac{BBH} within a Hubble time, although this outcome heavily relies on where the component masses fall within the measured observational uncertainties as well as the assumed models of binary evolution.  
Thus, it is unlikely that any currently observed \acp{HMXB} will form merging \acp{BBH} within a Hubble time.

In addition to the lack of currently observed \acp{HMXB} that are thought to be merging \ac{BBH} progenitors, there are differences in the \ac{BH} masses of the \changed{observed \ac{HMXB}} and merging \ac{BBH} populations.
While the primary \ac{BH} mass distribution predicted from \ac{GW} observations extends to masses higher than the peak near $35M_{\odot}$~\citep{abbott_population_2022}, there are no observed \acp{HMXB} with \ac{BH} accretor masses that fall near or beyond this limit. 
Rather than arising from fundamental astrophysical differences, the disparities in the observed \ac{HMXB} and \ac{GW} \ac{BH} population masses may actually be a product of detector selection effects alone. 
\citet{fishbach_apples_2022} compare the observed mass distributions of \acp{BH} in \changed{\acp{HMXB}} and \acp{BH} observed with \acp{GW}, and find that when \ac{GW} detector selection effects are accounted for, there are currently no statistically significant mass differences between the \changed{\ac{HMXB}} and \ac{GW} \ac{BH} populations.
However, it is critical that X-ray and \ac{GW} observational selection effects are \textit{jointly} examined, as they may produce differing outcomes in the observed populations.

The spin distributions of X-ray and \ac{GW} \ac{BH} populations can also provide insight on their evolutionary history and the role of selection effects in determining their properties.
\citet{fishbach_apples_2022} examine the \ac{BH} spin distributions of \ac{BH}--\changed{\acp{HMXB}} and \ac{GW}--\acp{BBH} and find them to be inconsistent with one another. 
It is possible, however, that differing binary evolutionary channels between the two populations may contribute to this discrepancy.
For example, \citet{gallegos-garcia_high-spin_2022} find that high-spin \changed{\acp{HMXB}} formed through Case-A mass transfer can only form merging \acp{BBH} within a small parameter space, and thus it is not surprising that the observed spin distributions are observed to be different.

In this paper, we show that observational selection effects play an important role in determining the observed \ac{BH} \changed{population masses} and the evolutionary predictions of detected \acp{HMXB}.
We compare detectable \ac{HMXB} populations with detectable \ac{GW} populations using simulated astrophysical samples of binaries, and we quantify the probability that \changed{detectable} \acp{HMXB} will form \ac{BH}--\ac{BH} binaries that will merge in a Hubble time, accounting for X-ray selection effects. 
We do not attempt to reproduce \changed{\ac{HMXB}} or \ac{GW} observations in detail; rather, we seek to understand how observational selection effects may produce the aforementioned differences in mass ranges and evolutionary predictions of \acp{HMXB}. 

In Section~\ref{sec:method}, we discuss our methods for sampling populations of binaries distributed throughout the universe, as well as how we model their X-ray and \ac{GW} emission properties. 
We develop a formalism for quantifying the probability of obtaining a given source in our sampled population, taking into account observational selection effects. 
In Section~\ref{sec:results}, we discuss parameter distribution properties for detectable \acp{HMXB} and detectable merging \acp{BBH} in our sample, as well as evolutionary probabilities for these sources. 
Finally, in Section~\ref{sec:discussion}, we discuss our main results, caveats in our analysis, and the implications of our findings in the context of X-ray and \ac{GW} observations. 
Throughout this work, we assume \textit{Planck 2018} cosmological parameters~\citep{aghanim_planck_2020}, including values for the Hubble constant ($H_0 = 67.7 \, \mathrm{km} \, \mathrm{s}^{-1} \mathrm{Mpc}^{-1}$) and the mass and dark energy density parameters ($\Omega_\mathrm{m}=0.31$ and $\Omega_{\Lambda}=0.69$, respectively). 
\changed{All data and code files supporting the findings reported in this paper are provided as supplementary information on Zenodo\footnote{\href{https://doi.org/10.5281/zenodo.7216270}{doi.org/10.5281/zenodo.7216270}} and Github\footnote{\href{https://github.com/celiotine/hmxb.bbh.selection.effects}{github.com/celiotine/hmxb.bbh.selection.effects}}, respectively.}

\section{Methods} \label{sec:method}

We examine populations of \acp{HMXB} with the rapid binary population synthesis code \texttt{COSMIC}~\citep[version 3.4;][]{breivik_cosmic_2020}. 
\texttt{COSMIC} is based on the single-stellar evolution formulae from \citet{hurley_comprehensive_2000} and the binary evolution prescriptions from \citet{hurley_evolution_2002}. 
\texttt{COSMIC} includes many updates to these prescriptions, such as those for OB stellar winds~\citep{vink_mass-loss_2001}, Wolf--Rayet star winds~\citep{vink_metallicity_2005}, the initiation of unstable mass transfer~\citep{belczynski_compact_2008, claeys_theoretical_2014, neijssel_effect_2019}, remnant formation \citep{fryer_compact_2012}, and pair instability supernovae~\citep{woosley_pulsational_2017, marchant_pulsational_2019, spera_merging_2019}. 
We generate binary populations with \texttt{COSMIC} across a grid of metallicities, and then sample systems from these populations and distribute them across redshifts.

\subsection{Binary Population Models}

We use \texttt{COSMIC} to generate populations of \acp{HMXB}. 
To do this, we allow the formation of \changed{\acp{DCO}} where the first-born object is a \ac{BH} or \ac{NS} and the second-born object is a \ac{WD}, \ac{NS}, or \ac{BH}. 
We only keep binaries that remain bound after the first \ac{SN} event and have stellar companion masses $>5M_{\odot}$ at the formation time of the first-born \ac{CO}.
We make these cuts because we define \acp{HMXB} to be systems that are concurrently bound and contain one accreting \ac{CO} and one stellar object $\geq$ 5~$M_{\odot}$ at some point in their evolution. 
These cuts ensure that our initial sample space allows for \ac{HMXB} formation but is not restricted to only merging \ac{BBH} progenitors. 

We simulate 16 \ac{DCO} populations across a discrete log-spaced metallicity grid that spans $(1/200) Z_{\odot}$ to $(7/4) Z_{\odot}$, as this metallicity range is approximately that of the grids used for stellar evolution within \texttt{COSMIC}~\citep{hurley_comprehensive_2000}.
Our population models include a set of key assumptions. 
\changed{We draw primary masses using the initial mass function from \citet{kroupa_variation_2001}, and initial orbital periods and eccentricities following the prescriptions from~\citet{sana_binary_2012}. 
We fix the binary fraction to be 0.7.
We limit the minimum mass ratio to be set such that the pre-main sequence lifetime of the secondary is not longer than the full lifetime of the primary if it were to evolve as a single star.}
We assume a solar metallicity of $Z_{\odot} = 0.017$~\citep{grevesse_standard_1998} for all calculations.
To calculate \ac{CO} remnant properties from the pre-supernova properties of a star, we employ the Delayed remnant mass prescription from \citet{fryer_compact_2012} that allows for \ac{CO} formation in the lower mass gap~\citep[e.g.,][]{zevin_exploring_2020}. 
We assume a conservative upper-limit for the maximum \ac{NS} mass of $3~M_\odot$ \citep{rhoades_maximum_1974, kalogera_maximum_1996}. 
To determine respective remnant masses for pulsational pair-instability supernovae and pair-instability supernovae, we use fits to Table 1 of \citet{marchant_pulsational_2019}. 
For common-envelope evolution, we use the prescription from \cite{belczynski_compact_2008} to determine the critical mass ratio for the onset of unstable mass transfer.
We set $\alpha = 1$ for the common-envelope efficiency parameter~\citep{livio_common_1988, ivanova_common_2013}, and use a variable prescription for the envelope binding energy factor $\lambda$~\citep{claeys_theoretical_2014}. 
Last, we set the wind-accretion efficiency factor to $0.5$, and enforce Eddington-limited accretion in all binaries. 

These assumptions represent a single fiducial point in a high-dimensional parameter space of binary evolution uncertainties that can have a significant impact on the population properties of compact binaries and their progenitors~\citep[e.g.,][]{barrett_accuracy_2018, belczynski_uncertain_2022, broekgaarden_impact_2022}. 
Because we do not attempt to reproduce observations in detail and rather seek to understand how selection effects impact \ac{HMXB} and merging \ac{BBH} populations, a fiducial model is satisfactory for our purposes.
We reserve a more systematic exploration of this parameter space for future work, and we comment on possible impacts to our results in Section \ref{sec:discussion}.

\subsection{Population Sampling} \label{sec:pop_samp}

From our set of \texttt{COSMIC} models run at discrete metallicities, we sample two populations of binaries: the first is a local population sampled out to redshift 0.05 (\changed{denoted} $z_{<0.05}$), and the second is a population sampled out to redshift 20 (\changed{denoted} $z_{<20}$). 
Because our \texttt{COSMIC} populations assume a single burst of star formation and evolve all systems for a Hubble time, we must assign binary redshifts and metallicities in postprocessing. 
We sample a $z_{<0.05}$ population to acquire robust \changed{detectable} \ac{HMXB} population statistics that cannot be acquired in a $z_{<20}$ sample, as the effective sample size of detectable \acp{HMXB} in the $z_{<20}$ population is small (there are $\simeq 100$ detectable \acp{HMXB} in the $z_{<20}$ sample compared to $\simeq 7 \times10^3$ in the $z_{<0.05}$ sample). 
Our choice to use a locally-sampled population of \acp{HMXB} in our analysis is discussed in more detail in Section \ref{sec:xrb_results}.

We draw redshifts and metallicites jointly from a two-dimensional redshift--metallicity grid using the weights 
\begin{equation}\label{eq:redshift_metallicity_weights}
    \mathcal{W}(z,Z) = \zeta(Z) P(Z|z) \psi[z(t)]\ \frac{\mathrm{d}t}{\mathrm{d}z}.
\end{equation}
Here, $\zeta(Z)$ is the \changed{number of \acp{HMXB} formed per unit stellar mass formed} at metallicity $Z$; $P(Z|z)$ is the probability of drawing metallicity $Z$ at redshift $z$; $\psi[z(t)]$ is the star formation rate at redshift $z$ attained by marginalizing over metallicity in a grid of stellar mass formed per unit lookback time and metallicity, and the term $\mathrm{d}t/\mathrm{d}z$ is to change of variables from time to redshift space for $\psi[z(t)]$,
\begin{equation}
    \frac{\mathrm{d}t}{\mathrm{d}z} = \frac{1}{(1+z)E(z)},
\end{equation}
where $E(z)$ is the cosmological factor for a flat universe~\citep{ryden_introduction_2002},
\begin{equation} \label{eqn:cosmo_factor}
    E(z) = \sqrt{\Omega_\mathrm{m}(1+z)^3 + \Omega_{\nu}(1+z)^4 + \Omega_{\Lambda}}.
\end{equation} 

We determine $P(Z|z)$ using publicly-available data from the \texttt{Illustris-TNG} simulation~\citep{nelson_illustristng_2019}.
This simulation provides stellar mass formed per unit lookback time and metallicity in a $100~\mathrm{Mpc}^3$ comoving box \citep[][Section 2.3]{zevin_suspicious_2022}.
The bounds of the distribution for $P(Z|z)$ are $Z/Z_{\odot}=5\times10^{-5}$ and $Z/Z_{\odot}=0.04$, which approximately correspond to the lower and upper bounds of our \texttt{COSMIC} population metallicity grid. 
Though the redshift--metallicity evolution $P(Z|z)$ is highly uncertain, especially at high redshifts, the star formation rate density evolution predicted from \texttt{Illustris-TNG} falls within the range of uncertainty \citep[e.g.,][]{chruslinska_chemical_2022}, and the evolution of the metallicity distribution is more astrophysically motivated than analytic approaches, such as truncated log-normal distributions. 

We jointly draw $N_\mathrm{total}=10^6$ redshifts and metallicities using the relative weights from Eq.~\eqref{eq:redshift_metallicity_weights}, and randomly sample binaries from the discrete \texttt{COSMIC} metallicity models based on which discrete metallicity model is closest to a drawn metallicity in log-space. 
This provides a \ac{DCO} population distributed across metallicity and redshift that is representative of a population of binaries in the universe.

\subsection{X-Ray Binary Detection Flux} \label{sec:xrb_calc}

As we are interested in binaries that can potentially lead to \ac{BBH} mergers, we consider \ac{BH}--\acp{HMXB} in our sample to be systems that are concurrently bound and contain one \ac{BH} object and one stellar object $\geq$ 5~$M_{\odot}$ at some point in their evolution. 
The \changed{\ac{XRB}} phase begins at first \ac{BH} formation and ends at second \ac{CO} formation or when the donor mass falls below 5~$M_{\odot}$. 

Since the time series resolution needed to calculate the X-ray emission of \acp{HMXB} is much finer than what \texttt{COSMIC} can handle for large populations, we re-evolve \acp{HMXB} in our sample with a small timestep resolution of $10^3$ years. 
We only apply this fine resolution during the HMXB phase to ensure computational efficiency. 
This is possible due to the ability of \texttt{COSMIC} to restart binaries in the middle of their evolution and evolve them forward~\citep{breivik_cosmic_2020}.\footnote{Updated \texttt{COSMIC} functionality is described at  \href{https://cosmic-popsynth.github.io/docs/stable/examples/index.html}{cosmic-popsynth.github.io/docs/stable/examples/index.html}.} 

We adopt the method from \cite{podsiadlowski_formation_2003} to calculate the \ac{HMXB} accretion luminosity,
\begin{equation} \label{eq:lum}
L_\mathrm{acc} = \eta \dot{M}_\mathrm{acc}c^2,
\end{equation}
where $\dot{M}_\mathrm{acc}$ is the Eddington-limited accretion rate of matter onto the \ac{BH} and $\eta$ is an efficiency factor for the \ac{BH} conversion of rest mass into radiative energy. 
For simplicity, we calculate $\eta$ assuming zero initial \ac{BH} spin and that it is determined entirely by the last stable particle orbit. For $M_\mathrm{BH} < \sqrt{6}M_{\mathrm{BH},0}$, the efficiency $\eta$ is 
\begin{equation}\label{eq:eta}
    \eta = 1 - \sqrt{1 - \bigg(\frac{M_{\mathrm{BH}}}{3M_{\mathrm{BH},0}}\bigg)^2},
\end{equation}
where $M_{\mathrm{BH}}$ is the \ac{BH} mass at a given time and $M_{\mathrm{BH},0}$ is the initial \ac{BH} gravitating mass--energy at formation according to \cite{bardeen_kerr_1970}. 
If $M_{\mathrm{BH}}$ exceeds $\sqrt{6}M_{\mathrm{BH},0}$, then $\eta$ is taken to be 0.42. 
The accretion rate is related to the change in \ac{BH} mass by
\begin{equation}\label{eq:mdot_BH}
    \dot{M}_{\mathrm{BH}} = (1-\eta)\dot{M}_{\mathrm{acc}},
\end{equation}
as energy released as radiation will not contribute to the BH mass. 
\texttt{COSMIC} includes wind prescriptions for mass loss in its evolution, so we do not need to explicitly calculate wind accretion. 
All changes in \ac{BH} mass from both Roche lobe overflow and wind mass transfer are included in $\dot{M}_{\mathrm{BH}}$.

To determine if a given \ac{HMXB} is detectable, we impose a flux limit for observations. 
The X-ray flux is
\begin{equation}
    F = \frac{L_{\mathrm{acc}}}{4\pi D_\mathrm{L}(z)^2},
\end{equation}
where $D_\mathrm{L}(z)$ is the luminosity distance of the binary at the start of the \ac{HMXB} phase. 
An \ac{HMXB} is considered detectable when its flux exceeds $5 \times 10^{-15}$~erg s$^{-1}$ cm$^{-2}$, which is an observational threshold commonly used in X-ray surveys \citep[e.g.,][]{chandra_x-ray_center_chandra_2021}.

\subsection{Gravitational-Wave Detection Probability} \label{sec:gw_pdet}

We calculate the \ac{GW} detection probabilities $p_{\mathrm{det}}$ of \ac{DCO} binaries in our sample that have merged by today ($z=0$). 
We evaluate the source signal-to-noise ratio (S/N) $\rho$ as
\begin{equation}
    \rho = \rho_0 \omega,
\end{equation}
where $\rho_0$ is the \changed{maximal} S/N of a \changed{face-on, overhead} source with component masses $m_1$ and $m_2$ at redshift $z$, and $\omega$ is a  projection  factor  that  depends  on  the relative  angular  orientation  of  the  source  and  the detector~\citep{dominik_double_2015}. 

We approximate $\rho_0$ following \cite{fishbach_does_2018} as
\begin{equation}
     \rho_0 = 8\left[\frac{\mathcal{M}(1+z_\mathrm{m})}{\mathcal{M}_8}\right]^{5/6}\frac{D_{\mathrm{L},8}}{D_\mathrm{L}},
\end{equation}
where $\mathcal{M} =  (m_1 m_2)^{3/5} / (m_1 + m_2)^{1/5}$ is the source chirp mass, $D_\mathrm{L}$ is the luminosity distance of the source, $z_\mathrm{m}$ is the merger redshift, and the constants $\mathcal{M}_8$ and $D_{\mathrm{L},8}$ are set to $10 \;M_{\odot}$ and $1~\mathrm{Gpc}$, respectively, such that they represent the typical distances~\citep{chen_distance_2021} at which sources are detectable by Advanced LIGO at design sensitivity~\citep{aasi_advanced_2015}. 
This scaling approximates the amplitude of a \ac{GW} coalescence signal to first order. 
For the projection factor $\omega$, we use the analytical expression for a single-detector network from \citet{dominik_double_2015}, which is equivalent to $\Theta/4$ in \citet{finn_observing_1993}. 
We calculate the detection probability $p_{\mathrm{det}}$ by Monte Carlo sampling $\omega$ and taking the fraction of the resulting source S/N that exceed a detection threshold of $\rho = 8$~\citep{thorne_gravitational_1997, chen_distance_2021}.

\subsection{Population Probabilities} \label{sec:probs}

We are interested in calculating the probability of obtaining sources in our sampled populations. 
In the most general form, we define the probability of finding a given source in state $X$ in our sample as
\begin{equation} \label{eqn:probs_general}
    P({X}) = \frac{N_{X}}{N_{\mathrm{total}}},
\end{equation}
where $N_X$ is the number of sources in state $X$ and $N_{\mathrm{total}}$ is the number of sources in the full sample of interest.
Because we need to count sources with different lifetimes distributed across space and time relative to an observer, a more involved framework is necessary to obtain values for $N_X$ and $N_{\mathrm{total}}$. 

We first need to count the number of binaries of interest $N_X$ in a given spatial comoving volume $V_\mathrm{c}$ with spatial points $\vec{x}$ surveyed from comoving times $t_1=t_1(\vec{x})$ to $t_2=t_2(\vec{x})$ in the source frame. 
Each source in a small spatial density of sources $n(\vec{x},t)$ has its own worldline along which it may be in some state $X$ for some time, which we track with an indicator function $I(\vec{x},t)$ that is nonzero when in state $X$ and zero otherwise, and normalized such that for a single source $\int \int I(\vec{x},t) \,\mathrm{d}t \mathrm{d}\vec{x} = 1$. 
Thus, we count the number of systems in state $X$ as
\begin{equation} \label{eqn:definition}
    N_X = \int_{V_\mathrm{c}} \int_{t_1(\vec{x})}^{t_2(\vec{x})} {n(\vec{x}, t) I(\vec{x},t)}\,\mathrm{d}t \mathrm{d}\vec{x}.
\end{equation}

\subsubsection{Long-lived Sources}\label{sec:longlived}

For any type of source whose lifetime is significantly longer than the length of the observing window ${T = (t_2-t_1)}$, the number of counted sources $N_{\mathrm{long}}$ does not change during the observing period. 
This implies that the spatial density is only a function of position $n(\vec{x})$. 
The indicator function takes on a form $I(x)/T$, and all observing-time dependence disappears after integrating over time. 
One instance where this is the case is with detectable \acp{HMXB}. 
As detailed in Section~\ref{sec:xrb_results}, even \acp{HMXB} that emit \changed{above the X-ray detection threshold} for a comparatively short amount of time emit on the order of thousands of years, making it valid to count them as long-lived sources compared to the timescale of astronomical surveys. 
This long-lived limit also holds for binaries that exist in some evolutionary state for a significant length of time, e.g., bound \acp{BBH} that have not merged.

With this simplification considered, we now count the number of long-lived sources as
\begin{equation}
    N_{\mathrm{long}} = \int_{V_\mathrm{c}} n(\vec{x})\,\mathrm{d}\vec{x}.
\end{equation}
This equation can be rewritten in terms of redshift,
\begin{equation} \label{eqn:convert_to_z}
    N_{\mathrm{long}} = \int_{z_1}^{z_2} n(z)\frac{\mathrm{d}V_\mathrm{c}}{\mathrm{d}z}\mathrm{d}z,
\end{equation}
where the redshifts $z_1, z_2$ are the bounds of the volume $V_\mathrm{c}$, $n(z)$ is the number density of sources in terms of redshift, and the factor $\mathrm{d}V_\mathrm{c}/\mathrm{d}z$ is the comoving volume element corresponding to redshift $z$, which is given by~\citep{ryden_introduction_2002},
\begin{equation}
    \frac{\mathrm{d}V_\mathrm{c}}{\mathrm{d}z} = \frac{4\pi c}{H_0} \frac{D_\mathrm{c}(z)^2}{E(z)},
\end{equation}
where $D_\mathrm{c}(z)$ is the comoving distance at redshift $z$ and $E(z)$ is defined in Eq.~\eqref{eqn:cosmo_factor}.
Changing variables from spatial coordinates to redshift significantly simplifies the evaluation of the integral for $N_{\mathrm{long}}$.

Since we use a discrete simulation for our calculations, Eq.~\eqref{eqn:convert_to_z} must be approximated as a discrete sum over individual sources in our \texttt{COSMIC} populations and a Riemann sum over redshift,
\begin{equation} \label{eqn:long_sum_final}
      N_{\mathrm{long}} = \sum_i \sum_j \frac{I_i^X(z_j)}{V_{\mathrm{box}}} \frac{\mathrm{d}V_\mathrm{c}}{\mathrm{d}z_j}\Delta z_j ,
\end{equation}
where we have re-written the spatial density of sources as $n(z_j)=\sum_iI_i^X(z_j)/V_{\mathrm{box}}$.
Here, $I_i^{X}(z_j)$ indicates if the $i$-th system in the sample is in state $X$ at redshift $z_j$, and $V_{\mathrm{box}}$ is the volume of the comoving box that we consider for our resampled population. 
Eq.~\eqref{eqn:long_sum_final} allows for a quick, simplified computation of $N_{\mathrm{long}}$.

\subsubsection{Short-lived Sources}\label{sec:shortlived}

We can also apply Eq.~\eqref{eqn:definition} to calculate the number of sources $N_{\mathrm{short}}$ in the case that their lifetimes are significantly shorter than the length of the observing window. 
This includes \ac{BBH} mergers in the frequency band of ground-based \ac{GW} detectors which only exist as detectable sources for at most a few seconds before coalescence. 
In this case, the length of the source lifetime approaches zero, which causes the indicator function to act as a delta function,
\begin{equation}
     N_{\mathrm{short}} = \int_{V_\mathrm{c}} \int_{t_1(\vec{x})}^{t_2(\vec{x})} n(\vec{x},t) \delta(t-t'(\vec{x}))\,\mathrm{d}t \mathrm{d}\vec{x} ,
\end{equation}
where $t'(\vec{x})$ marks the time at which state $X$ occurs at spatial position $\vec{x}$. 
In the continuum limit, this expression is an integral over the rate density of sources $r(\vec{x}, t)$,
\begin{equation} \label{eqn:rate_density_int}
    N_{\mathrm{short}} = \int_{V_\mathrm{c}} \int_{t_1(\vec{x})}^{t_2(\vec{x})}  r(\vec{x}, t)\,\mathrm{d}t \mathrm{d}\vec{x}.
\end{equation}
When we observe for some time window $T$, the time dependence of the rate density in Eq.~\eqref{eqn:rate_density_int} integrates to $T$ with a factor of $1/(1+z)$ to account for transitioning between the source frame and observer frame time. 
As with Eq.~\eqref{eqn:convert_to_z}, we use a change of variables to write the volume integral as an integral over redshift,
\begin{equation}
    N_{\mathrm{short}} = T \int_{z_1}^{z_2} r(z) \frac{\mathrm{d}V_\mathrm{c}}{\mathrm{d}z} \frac{1}{1+z}\mathrm{d}z.
\end{equation}
\changed{where $r(z)$ is the rate density of sources as a function of redshift.}
Rewriting this equation as a sum over individual sources, we obtain
\begin{equation} \label{eqn:short_sum_final}
    N_{\mathrm{short}} = T \sum_i \sum_j \frac{I_i^X(t_j, t_j + \tau)}{\tau V_{\mathrm{box}}} \frac{\mathrm{d}V_\mathrm{c}}{\mathrm{d}z_j} \frac{1}{1+z_j}\Delta z_j,
\end{equation}
where we have re-written the rate density of sources as $r(t_j)=\sum_i I_i^X(t_j,t_j+\tau)/[\tau V_{\mathrm{box}}]$, with $t_j = t(z_j)$. $I_i^X(t_j, t_j + \tau)$ indicates if the $i$-th system in the sample undergoes an event $X$ at redshift $z_j$ between times $t_j$ and $t_j + \tau$, where $\tau$ is a chosen time interval significantly shorter than the timescale on which the rate density evolves.
For practical purposes, $\tau$ can be thought of as a chosen bin width in time.
Eq.~\eqref{eqn:short_sum_final} allows for an efficient computation of $N_{\mathrm{short}}$.

\subsubsection{Detection Weights} \label{sec:weight_calc}

We apply the formalism developed in the previous subsections to assign relative detection weights to all binaries in our sample. 
Specifically, the sums over redshift in Eq.~\eqref{eqn:long_sum_final} and Eq.~\eqref{eqn:short_sum_final} are the expressions we use to define the weight of source $i$ in state $X$. 
As we consider relative weights, we can ignore the constant volume factor $V_{\mathrm{box}}$ and the time factors $T$ and $\tau$.
We use the formalism for long-lived sources to calculate detectable \changed{\ac{HMXB}} weights, and that for short-lived sources to calculate detectable \ac{GW} weights.

For \ac{HMXB} sources that emit for long times, this weighting is the sum over redshift in Eq.~\eqref{eqn:long_sum_final}, which becomes
\begin{equation}\label{eqn:xrb_weights}
    \mathcal{W}_{\mathrm{xrb}, i} \propto \sum_j I_i^{\mathrm{HMXB}_{\mathrm{obs}}}(z_j) \frac{\mathrm{d}V_\mathrm{c}}{\mathrm{d}z_j}\Delta z_j ,
\end{equation}
where the indicator $I^{\mathrm{HMXB}_{\mathrm{obs}}}(z_j)$ has value equal to unity if the \ac{HMXB} exceeds a given X-ray detection threshold at redshift $z_j$, and has value zero otherwise. 
Thus, sources that emit above the detection threshold for longer periods of time will contribute more in the summation and be given a higher relative detection weighting. 

The weights for merging \ac{BBH} sources are similar, but now the \ac{GW} detection probability is included with the indicator function to account for selection effects,
\begin{equation}
    \mathcal{W}_{\mathrm{gw}, i} \propto \sum_{j} I_i^{\mathrm{BBH}_\mathrm{m}}(z_j) p_{\mathrm{det}, i}\frac{\mathrm{d}V_\mathrm{c}}{\mathrm{d}z_j} \frac{1}{1+z_j}\Delta z_j,
\end{equation}
where the indicator $I_i^{\mathrm{BBH}_\mathrm{m}}(z_j)$ has value unity if the \ac{BBH} coalesces at redshift $z_j$ and zero otherwise. 
Thus, for \acp{BBH} merging at redshift $z_m$, each system has a relative weight of
\begin{equation} \label{eqn:gw_weights}
    \mathcal{W}_{\mathrm{gw}, i} \propto p_{\mathrm{det},i}\frac{\mathrm{d}V_\mathrm{c}}{\mathrm{d}z_m} \frac{1}{1+z_m}\Delta z_m .
\end{equation}
Building this formalism for relative detection weights allows us to easily compare \changed{detectable} systems and their progenitors in Section~\ref{sec:results}.

\subsubsection{Probability Expressions}

We use the counting expressions in Section~\ref{sec:longlived} and Section~\ref{sec:shortlived} to calculate probabilities in the form of Eq.~\eqref{eqn:probs_general}. 
For \changed{detectable} \acp{HMXB}, this equation becomes
\begin{equation} 
    P(\mathrm{HMXB}_{\mathrm{obs}}) = \frac{N_{\mathrm{HMXB}_{\mathrm{obs}}}}{N_{\mathrm{total}}},
\end{equation}
which, written in summation form, is
\begin{equation} \label{eqn:pHMXB_obs}
   P(\mathrm{HMXB}_{\mathrm{obs}})   = \frac{\sum_i \sum_j I_i^{\mathrm{HMXB}_{\mathrm{obs}}}(z_j)[\mathrm{d}V_\mathrm{c}/\mathrm{d}z_j]\Delta z_j}{\sum_{i'} \sum_{j'} I_{i'}^\mathrm{ZAMS}(z_{j'})[\mathrm{d}V_\mathrm{c}/\mathrm{d}z_{j'}]\Delta z_{j'}},
\end{equation}
where $I_{i'}^{\mathrm{ZAMS}}(z_{j'})$ indicates if the $i$-th system in the sample has been initialized at \ac{ZAMS} and exists as any form of bound binary at redshift $z_{j'}$. 
These are the systems we count as members of the $N_{\mathrm{total}}$ sample of binaries.

For merging \acp{BBH}, the probability of detecting a source is
\begin{multline} \label{eqn:pBBHm_obs}
    P(\mathrm{BBHm}_{\mathrm{obs}}) = \\ \frac{T}{\tau}\frac{\sum_i \sum_jI_i^\mathrm{BBHm}(t_j, t_j + \tau)p_{\mathrm{det},i}[{\mathrm{d}V_\mathrm{c}}/{\mathrm{d}z_j}][{1+z_j}]^{-1}\Delta z_j }{\sum_{i'} \sum_{j'}I_{i'}^\mathrm{ZAMS}(z_{j'})[{\mathrm{d}V_\mathrm{c}}/{\mathrm{d}z_{j'}}]\Delta z_{j'}}.
\end{multline}
This probability heavily depends on choices for the observing window $T$ \changed{and the detector sensitivity described by $p_{\mathrm{det}}$,  so it is primarily determined by survey details. 
The time $\tau$ is a choice of bin width in time for approximating time integrals. 
The probability $P(\mathrm{BBHm}_{\mathrm{obs}})$ is only loosely dependent on $\tau$, because $I_i^{\mathrm{BBHm}} (t_j, t_j+\tau)$ scales approximately linearly with $\tau$, which counters the $\tau$ in the denominator.}

We also consider \emph{conditional} probabilities for sources in our sample. 
For example, the probability a \changed{detectable} \ac{HMXB} will become a \ac{BBH} that merges within a Hubble time can be written as
\begin{equation}
    P(\mathrm{BBHm}_\mathrm{H} | \mathrm{HMXB}_{\mathrm{obs}}) = \frac{P(\mathrm{BBHm}_\mathrm{H} \cap \mathrm{HMXB}_{\mathrm{obs}})}{P(\mathrm{HMXB}_{\mathrm{obs}})},
\end{equation}
which, in summation form, is
\begin{multline}
    P(\mathrm{BBHm}_\mathrm{H} | \mathrm{HMXB}_{\mathrm{obs}}) = \\ \frac{\sum_i \sum_j I_i^{\mathrm{HMXB}_{\mathrm{obs}} \cap \mathrm{BBHm}_\mathrm{H}}(z_j)[{\mathrm{d}V_\mathrm{c}}/{\mathrm{d}z_j}]\Delta z_j}{\sum_{i'} \sum_{j'}  I_{i'}^{\mathrm{HMXB}_{\mathrm{obs}}}(z_{j'})[{\mathrm{d}V_\mathrm{c}}/{\mathrm{d}z_{j'}}]\Delta z_{j'}},
\end{multline}
where $I_i^{\mathrm{HMXB}_{\mathrm{obs}} \cap \mathrm{BBHm}_\mathrm{H}}(z_j)$ indicates if the $i$-th system in the sample is an \ac{HMXB} emitting above a given \changed{detection} flux threshold at redshift $z_j$ that will \emph{also} become a merging \ac{BBH} within a Hubble time.

These probability expressions, along with those for the relative detection weights in Eq.~\eqref{eqn:xrb_weights} and Eq.~\eqref{eqn:gw_weights}, provide all the mathematical tools necessary to quantify the relationships between \changed{detectable} sources and their progenitors.

\section{Results} \label{sec:results}

To determine the impact of X-ray and \ac{GW} selection effects on their observed populations, we apply detection weights as calculated in Section~\ref{sec:weight_calc} to our sampled binaries from \texttt{COSMIC}.
To fully understand these results, we must also understand the emission behavior of our X-ray and \ac{GW} populations.
This provides context for their evolutionary behavior and how it relates to their \changed{detectability}, especially in the \ac{HMXB} case where emission behavior has not been well-characterized for large theoretical populations.
We also examine the mass distributions of detected \ac{HMXB} and \ac{BBH} populations and calculate probabilities describing the relationships of sources to one another.
Through this analysis, we illuminate possible causes of the observational discrepancies in \ac{BH} masses and evolutionary predictions between X-ray and \ac{GW} sources.

\subsection{HMXB Emission Properties} \label{sec:xrb_results}

\begin{figure}
\includegraphics[width=0.5\textwidth]{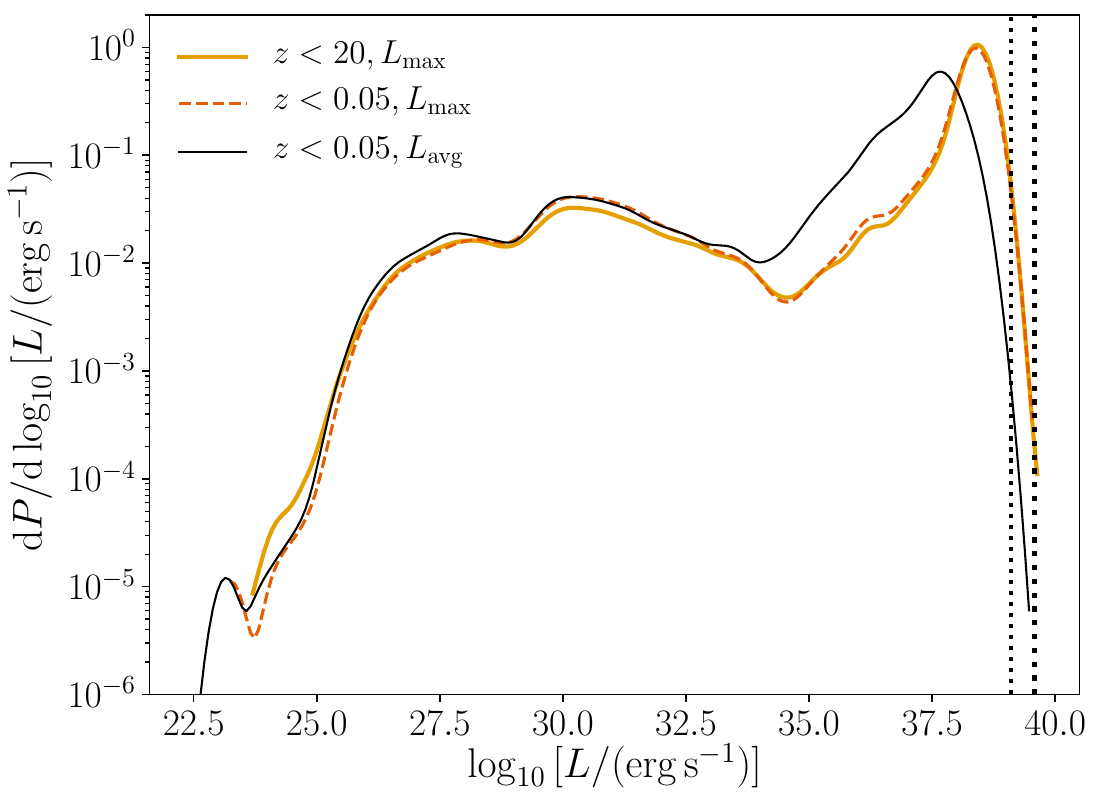}
\caption{Distributions of the maximum X-ray luminosity $L_{\mathrm{max}}$ reached by all \acp{HMXB} in the $z_{<20}$ (solid gold line) and $z_{<0.05}$ (dashed red line) populations, along with the time-averaged X-ray luminosity $L_{\mathrm{avg}}$ of binaries during the \changed{\ac{XRB}} phase for the $z_{<0.05}$ population (solid black line). 
The left and right vertical dotted lines mark the Eddington luminosity limits for $10 M_{\odot}$ and $30 M_{\odot}$ \ac{BH} accretors, respectively.}
\label{fig:max_lum_kde}
\end{figure}

The emission properties of our \ac{HMXB} populations, as calculated in Section \ref{sec:xrb_calc}, provide insight into the nature of their detectability. 
Figure~\ref{fig:max_lum_kde} shows the maximum X-ray luminosity $L_{\mathrm{max}}$ reached by all \acp{HMXB} in our $z_{<0.05}$ and $z_{<20}$ samples, as well as the time-averaged X-ray luminosity $L_{\mathrm{avg}}$ of \acp{HMXB} during the \changed{\ac{XRB}} phase for the $z_{<0.05}$ sample. 
We mark the Eddington luminosities for $10~M_{\odot}$ and $30~M_{\odot}$ \ac{BH} accretors with vertical dotted lines.
Almost $70\%$ of the $z_{<0.05}$ \acp{HMXB} achieve peak luminosities between $10^{38}$ and $10^{39}~\mathrm{erg\, s^{-1}}$ during their \ac{XRB} phase, which is roughly within one order of magnitude of their Eddington luminosity. 
The overall maximum emission behavior of the $z_{<0.05}$ and $z_{<20}$ \acp{HMXB} is similar across all luminosities.
Both exhibit a steep drop in the distribution close to the Eddington limit along with a tail to lower X-ray luminosities.

The time-averaged luminosity during the \ac{XRB} phase follows approximately the same distribution as the maximum luminosity for values $< 10^{33}~\mathrm{erg\, s^{-1}}$.
At higher luminosities, however, these distributions diverge, with a larger discrepancy between the average and maximum luminosities reached during the \changed{\ac{XRB}} phase.
This is because in our models, most systems emitting at higher luminosities only reach their peak emission for a short time and emit at luminosities a few orders of magnitude lower for most of the \changed{\ac{XRB}} phase.

\begin{figure}
\includegraphics[width=0.5\textwidth]{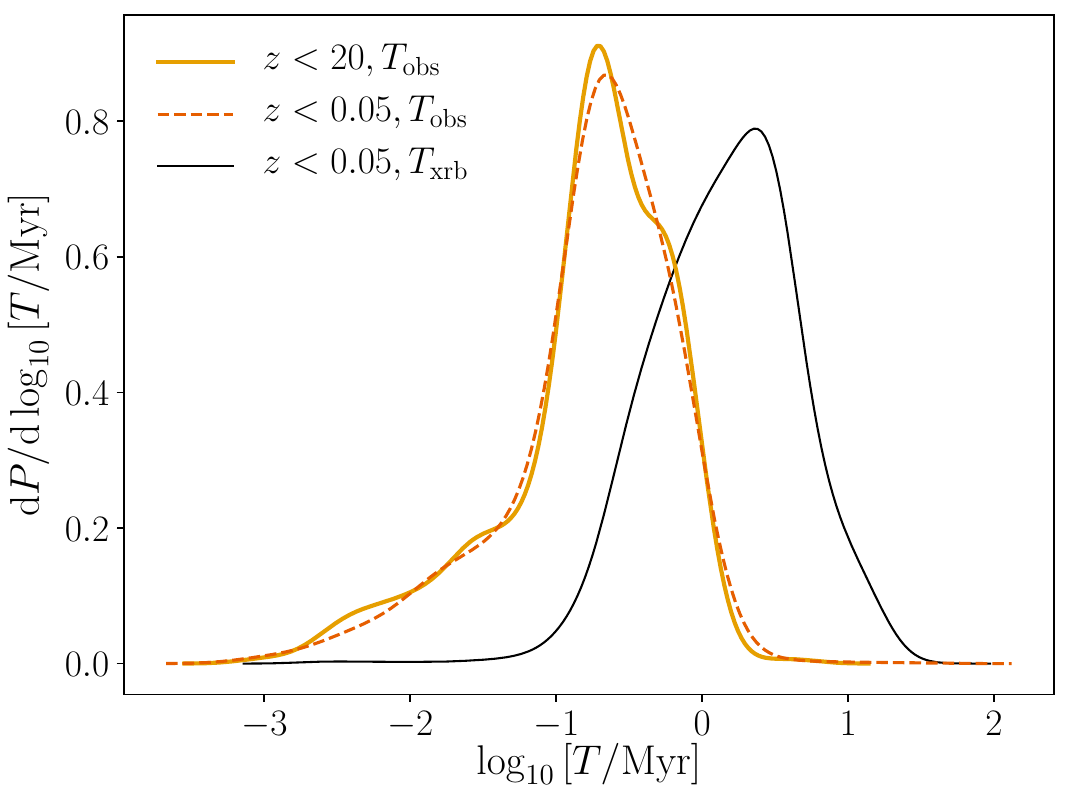}
\caption{Distributions of the duration of \changed{detectable} X-ray emission $T_{\mathrm{obs}}$ for all \acp{HMXB} in the $z_{<20}$ (solid gold line) and $z_{<0.05}$ (dashed red line) populations that emit above the minimum X-ray detection threshold for a nonzero period of time.
For the $z_{<0.05}$ population, we also show the total duration of the \changed{\ac{XRB}} phase $T_{\mathrm{xrb}}$ (solid black line).
\changed{This distribution is nearly identical to that for the $z_{<20}$ population.}}
\label{fig:emit_duration_kde}
\end{figure}

The detectability of \acp{HMXB} is influenced by the time for which they maintain high luminosities, which varies between binaries. 
\changed{\ac{HMXB} detectability may also be correlated with accretor mass, as indicated in Eq.~\eqref{eq:lum} through Eq.~\eqref{eq:mdot_BH}, and we explore this relationship in conjunction with \ac{HMXB} emission properties in Section~\ref{sec:mass_results}}.
Figure~\ref{fig:emit_duration_kde} shows the duration of \changed{detectable} emission $T_{\mathrm{obs}}$ for \acp{HMXB} in our $z_{<20}$ and $z_{<0.05}$ samples that emit above the minimum X-ray detection threshold of $5\times10^{-15}~\mathrm{erg\, s^{-1}\, cm^{-2}}$ for a nonzero period of time. 
We also show the total duration of the \changed{\ac{XRB}} phase $T_{\mathrm{xrb}}$ for the same binaries in our $z_{<0.05}$ sample.
Once again, the emission behavior between the $z_{<20}$ and $z_{<0.05}$ samples is consistent.
More than $98\%$ of \acp{HMXB} emit \changed{above the X-ray detection threshold} for less than $1~\mathrm{Myr}$, with the $T_{\mathrm{obs}}$ distributions peaking near $0.3~\mathrm{Myr}$. 
For the majority of binaries, this time is less than half of the duration of the \changed{\ac{XRB}} phase, which is governed by the remaining lifetime of the donor star after the formation of the first \ac{CO}.

\subsection{Comparing Selection Effects}

Here we compare the detectable \ac{HMXB} and detectable merging \ac{BBH} populations in order to understand the existing observational discrepancies between \acp{BH} detected in X-ray versus \ac{GW} sources.
Namely, we investigate the lack of observed \acp{HMXB} that are predicted to become \ac{BBH} mergers, as well as the lack of high-mass \acp{BH} found in \changed{\acp{HMXB}}.
We use Eq.~\eqref{eqn:xrb_weights} and Eq.~\eqref{eqn:gw_weights} to calculate detectability weights for \acp{HMXB} ($\mathcal{W}_{\mathrm{xrb}}$) and \acp{GW} ($\mathcal{W}_{\mathrm{gw}}$), respectively.

In all of these comparisons, we use the $z_{<0.05}$ \ac{HMXB} population and the $z_{<20}$ \ac{BBH} population. 
This is because \changed{the effective sample size of \changed{detectable} \acp{HMXB} in the $z_{<0.05}$ population is $\simeq50$ times larger, and thus allows for more robust statistics. 
The detectable \acp{HMXB} in both populations have comparable properties (in addition to the emission properties in Figure~\ref{fig:max_lum_kde} and Figure~\ref{fig:emit_duration_kde}, their redshift, metallicity, and mass distributions are consistent).}
We find that more than $99.9\%$ of \changed{all} \acp{HMXB} in the $z_{<20}$ sample are too distant to be observed, even if they achieve high luminosities.
In addition, the maximum \ac{ZAMS} formation redshift of \changed{detectable} \acp{HMXB} in our $z_{<0.05}$ sample is $\simeq8 \times 10^{-3}$, which is much smaller than the upper sampling bound of $z=0.05$.
Thus, even if we sampled a $z_{<20}$ population with one billion systems, we still would not find any \changed{detectable} \acp{HMXB} beyond $z \sim 0.01$.

\subsubsection{Redshifts and Metallicities}

\begin{figure}
\includegraphics[width=0.5\textwidth]{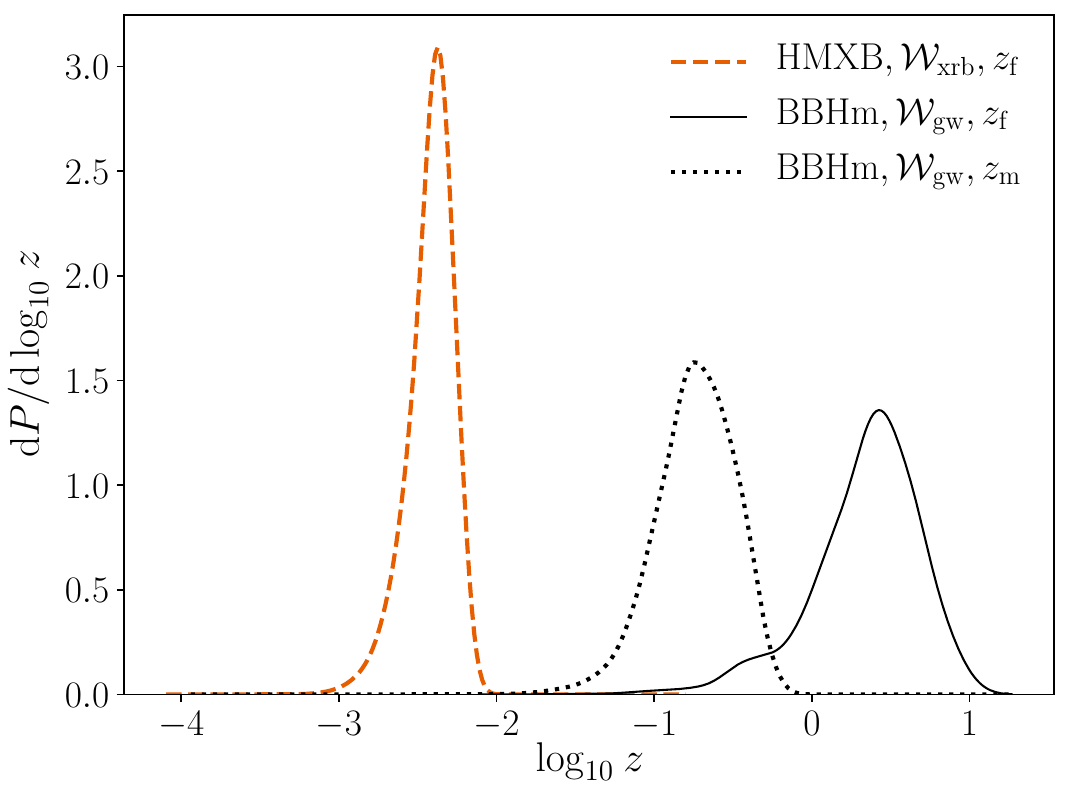}
\caption{Distributions of the \ac{ZAMS} formation redshifts $z_\mathrm{f}$ for $z_{<0.05}$ \ac{HMXB} and  $z_{<20}$ merging \ac{BBH} progenitors, as well as the merger redshifts $z_\mathrm{m}$ for $z_{<20}$ merging \acp{BBH}, weighted by their X-ray and \ac{GW} detectability, respectively.}
\label{fig:zf_comp_kde}
\end{figure}

\begin{figure}
\includegraphics[width=0.5\textwidth]{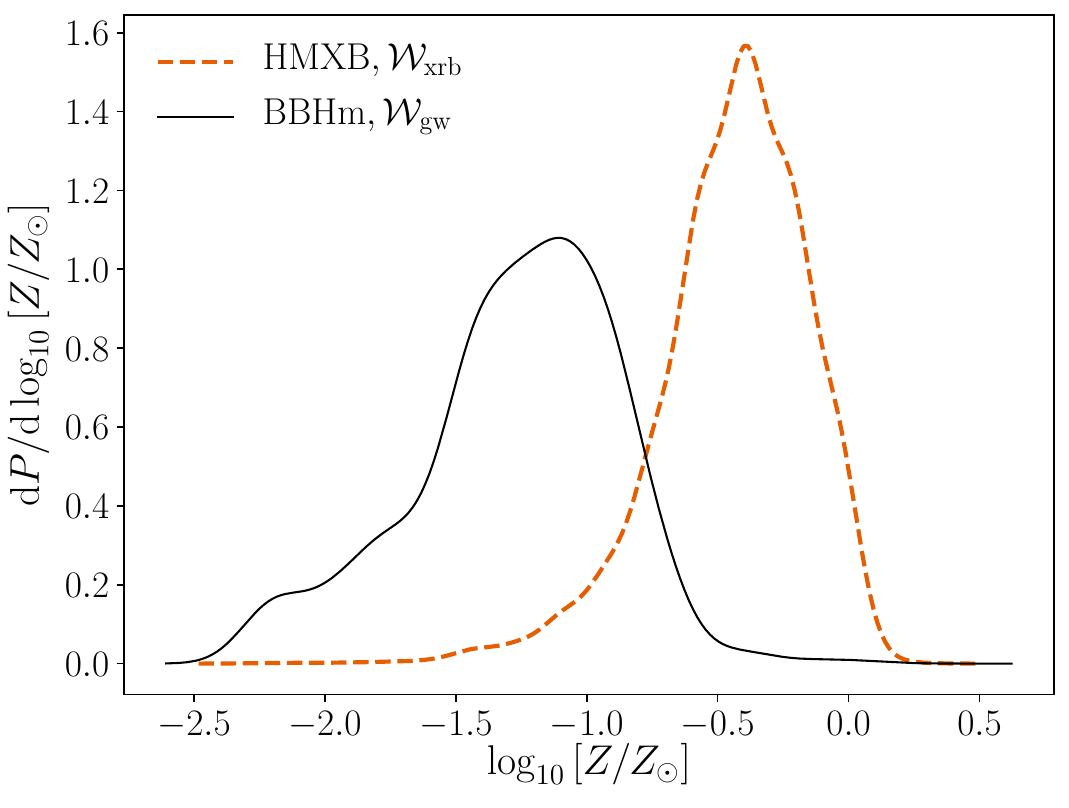}
\caption{Distributions of progenitor metallicities for $z_{<0.05}$ \acp{HMXB} and $z_{<20}$ merging \acp{BBH}, weighted by their X-ray and \ac{GW} detectability, respectively.}
\label{fig:met_comp_kde}
\end{figure}

In Figure~\ref{fig:zf_comp_kde}, we compare the redshift distributions of \acp{HMXB} formed by today (HMXB$_{z0}$) and \acp{BBH} merged by today (BBHm$_{z0}$), weighted by their respective detectability
In Figure~\ref{fig:met_comp_kde}, we do the same for the metallicity distributions. 
As expected, the progenitors of \changed{detectable} \acp{HMXB} form at very low redshifts ($z_\mathrm{f} \lesssim 0.01$), whereas the progenitors of \changed{detectable} \acp{BBH} mergers form at much higher redshifts ($z_\mathrm{f} \gtrsim 1$), leading to detectable \ac{BBH} mergers in the redshift range $0.01 \lesssim z_\mathrm{m} \lesssim 1$. 

As a consequence of the differing redshift distributions of \changed{detectable} \ac{HMXB} and \ac{BBH}-merger progenitors, there are also differences in the population metallicities, with the detection-weighted \acp{HMXB} having higher typical metallicities than the merging \acp{BBH}. 
The metallicity distribution of detectable \acp{HMXB} peaks near $0.5 Z/Z_{\odot}$ while that for detectable merging \acp{BBH} peaks near $0.08 Z/Z_{\odot}$.
There is almost no support for detectable \ac{HMXB} formation below $0.1 Z/Z_{\odot}$, as fewer than $0.05\%$ of \changed{detectable} \acp{HMXB} have metallicities below this limit.
This is expected because the distance at which \ac{GW} detectors are sensitive to merging \acp{BBH} is much larger than the distance to which we can observe \acp{HMXB}, and thus \acp{GW} probe higher redshifts and lower metallicites.

\subsubsection{Binary Component Masses}\label{sec:mass_results}

\begin{figure}
\includegraphics[width=0.5\textwidth]{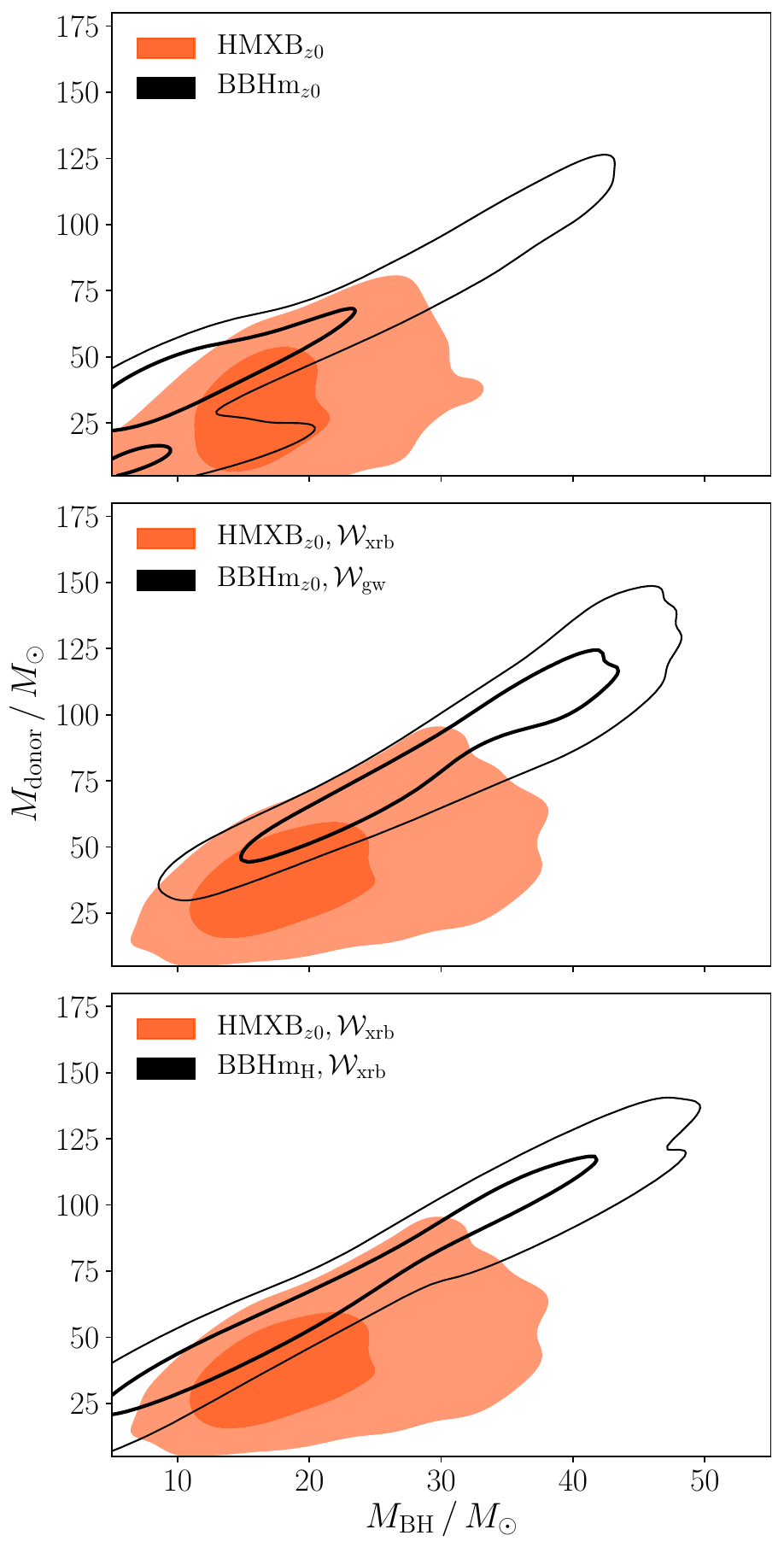}
\caption{Distributions of \ac{HMXB} component masses at the start of the \ac{XRB} phase, weighted by different detector selection effects. 
We plot the $50\%$ and $90\%$ probability levels. 
The \emph{top} panel shows the \ac{HMXB} component masses for the full underlying $z_{<0.05}$ \ac{HMXB} population (HMXB$_{z0}$) and the subpopulation of $z_{<20}$ \acp{HMXB} that will form merging \acp{BBH} by $z=0$ (BBHm$_{z0}$).
The \emph{central} panel shows these populations weighted by X-ray detector ($\mathcal{W}_{\mathrm{xrb}}$) and \ac{GW} detector ($\mathcal{W}_{\mathrm{gw}}$) selection effects. 
The \emph{bottom} panel shows the same HMXB$_{z0}$ population compared with the supopulation of $z_{<0.05}$ \acp{HMXB} that will form merging \acp{BBH} within a Hubble time (BBHm$_\mathrm{H}$), both weighted by X-ray detector selection effects ($\mathcal{W}_{\mathrm{xrb}}$).
Changing the \ac{BBH} sample from $z_{<20}$ in the central panel to $z_{<0.05}$ in the bottom panel creates extra support for low-mass \ac{BBH} mergers due to the $z_{<0.05}$ population residing at higher metallicities.}
\label{fig:hmxb_masses}
\end{figure}

Next, we examine the mass distributions of the $z_{<0.05}$ \ac{HMXB} and $z_{<20}$ merging \ac{BBH} populations. 
In Figure~\ref{fig:hmxb_masses}, we show distributions of the binary component masses at the beginning of the \ac{HMXB} phase for all $z_{<0.05}$ \acp{HMXB} formed by today (HMXB$_{z0}$) and all $z_{<20}$ \acp{HMXB} that will merge as \acp{BBH} by today (BBHm$_{z0}$). 
The horizontal axis shows the \ac{BH} accretor mass, $M_\mathrm{BH}/M_{\odot}$, and the vertical axis shows the donor star mass, $M_\mathrm{donor}/M_{\odot}$, at the start of the \ac{HMXB} phase.
In the top panel of Figure~\ref{fig:hmxb_masses}, we show the underlying mass distributions for the HMXB$_{z0}$ and BBHm$_{z0}$ populations. 
While the distributions mostly overlap, the BBHm$_{z0}$ distribution is centered at lower \ac{BH} masses ($< 20 M_{\odot}$) because the majority ($\sim70$\%) of \acp{HMXB} with high-mass \ac{BH} accretors form \ac{BBH} systems with long delay times that do not merge by $z=0$.
Finding \acp{BH} with longer delay times at higher redshifts could be an artifact of various population synthesis prescriptions for binary evolution, such as those for \ac{BH} kicks, common-envelope evolution, mass transfer, etc.~\citep[e.g.,][]{van_son_redshift_2022}.
However, because we are only interested in obtaining a fiducial model for binary evolution in this paper, we do not investigate these effects in detail.

In the central panel of Figure~\ref{fig:hmxb_masses}, we show the same distributions weighted by their respective selection effects. 
We weight the HMXB$_{z0}$ population by their X-ray detection weights $\mathcal{W}_{\mathrm{xrb}}$ and the BBHm$_{z0}$ population by their \ac{GW} detection weights $\mathcal{W}_{\mathrm{gw}}$, defined in Eq.~\eqref{eqn:xrb_weights} and Eq.~\eqref{eqn:gw_weights}, respectively. 
While the distributions still overlap, the detectable BBHm$_{z0}$ distribution is pulled to higher masses ($>20 M_{\odot}$), which is the result of heavier \ac{BH} mergers having higher \ac{GW} detection probabilities. 
The detectable HMXB$_{z0}$ distribution remains at masses similar to the underlying distribution, as we do not find a strong correlation of $\mathcal{W}_{\mathrm{xrb}}$ with \ac{BH} or donor mass. 
\changed{If all \acp{HMXB} were emitting at the same distance for the exact same duration and had Eddington-limited accretion, the \ac{HMXB} detectability would scale with the mass of the \ac{BH} accretor according to Eq.~\eqref{eq:lum} through Eq.~\eqref{eq:mdot_BH}. 
However, since we do not find a discernible correlation between \ac{BH} mass and detectability, it is clear that the duration of detectable emission plays a critical role in the detectability of \acp{HMXB}.}

There are very few detectable \acp{HMXB} that have \ac{BH} accretor masses $>35 M_{\odot}$, while there is significant support for these massive systems in the \ac{GW} population: fewer than $3\%$ of detectable \acp{HMXB} host \acp{BH} $>35 M_{\odot}$, while $\simeq 20\%$ of detectable merging \ac{BBH} progenitors have primary \ac{BH} masses that exceed this limit.
\changed{However, there is significant overlap between the \ac{BH} masses in the detectable \ac{HMXB} and detectable \ac{BBH} below  $35~M_{\odot}$. 
This falls in accordance with observations: the \ac{BH} mass distribution inferred from \ac{GW} observations peaks near $10~M_{\odot}$~\citep{abbott_population_2022}, and all observed \acp{HMXB} have \ac{BH} accretors with masses well below $35~M_{\odot}$.}

In the bottom panel of Figure~\ref{fig:hmxb_masses}, we plot the same HMXB$_{z0}$ distribution along with the subpopulation of $z_{<0.05}$ \acp{HMXB} that will form merging \acp{BBH} within a Hubble time of \ac{ZAMS} (BBHm$_{\mathrm{H}}$). 
Here, the distributions are \emph{both} weighted by their X-ray detectability. 
This compares the full \changed{detectable} population of \acp{HMXB} to the subpopulation that will become merging \acp{BBH} within a Hubble time. 
Changing the (black) distribution of the \ac{BBH} sample from $z_{<20}$ in the central panel to $z_{<0.05}$ in the bottom panel creates extra support for low-mass \ac{BBH} mergers that is independent of changing the weights from $\mathcal{W}_{\mathrm{gw}}$ to $\mathcal{W}_{\mathrm{xrb}}$.
This spread to lower masses is a result of the $z_{<0.05}$ population residing at higher metallicities.

We find that the X-ray detection-weighted BBHm$_{\mathrm{H}}$ progenitors have higher donor masses relative to the full detection-weighted \ac{HMXB} population. 
This is because \changed{detectable} \acp{HMXB} are unlikely to form \acp{BBH} in our sample; in addition to the progenitors of these stars having lower \ac{ZAMS} masses, we find that many lose additional mass during the \ac{HMXB} phase, making it more likely that they will ultimately form a \ac{NS} or \ac{WD}. 
In fact, only $\simeq 20\%$ of \acp{HMXB} that emit \changed{above the X-ray detection threshold} for $1~\mathrm{Myr}$ or longer (the upper tail of the emission distribution in Figure~\ref{fig:emit_duration_kde}) contain donor stars that will form \acp{BH}.
Of these \acp{HMXB} that emit for longer times \textit{and} form \acp{BBH}, over $80\%$ are too wide to merge within a Hubble time.

\begin{figure}
\includegraphics[width=0.5\textwidth]{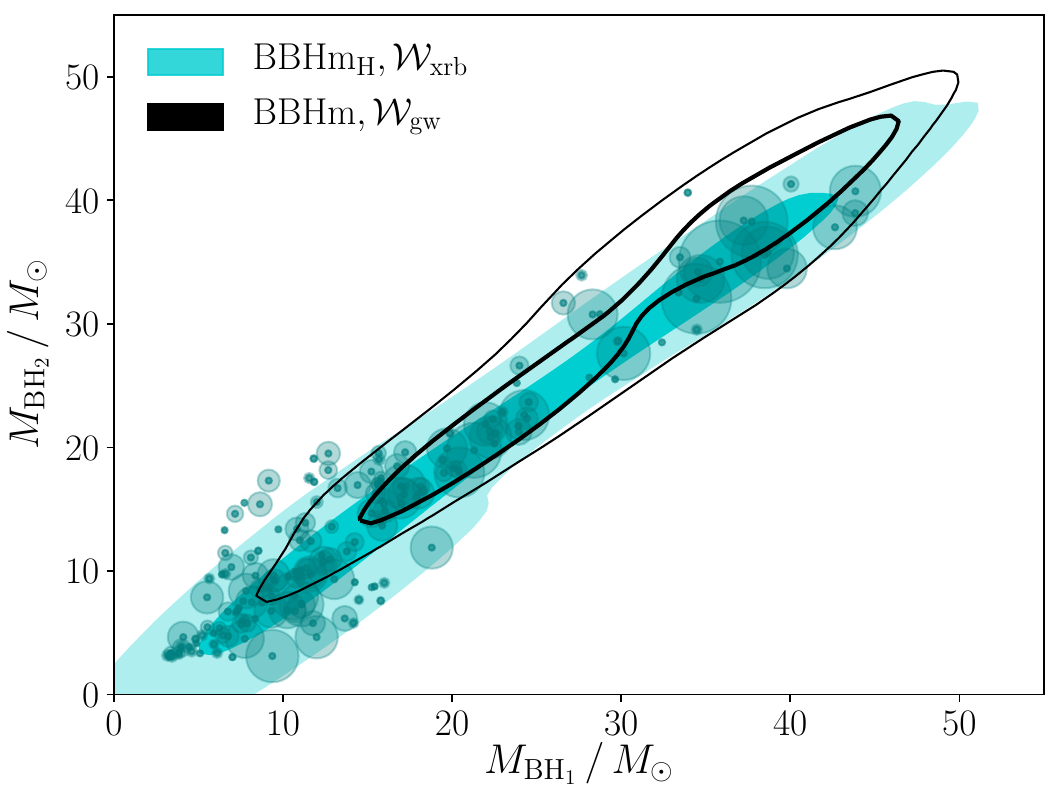}
\caption{Distributions of the final \ac{BH} component masses for the subpopulation of $z_{<0.05}$ \acp{HMXB} that will become merging \acp{BBH} in a Hubble time (BBHm$_\mathrm{H}$) and the full population of $z_{<20}$ \acp{BBH} that merge by today (BBHm$_{z0}$). 
We plot the $50\%$ and $90\%$ probability levels.
The BBHm$_\mathrm{H}$ distribution is weighted by X-ray detector selection effects ($\mathcal{W}_{\mathrm{xrb}}$) and the BBHm$_{z0}$ distribution is weighted by \ac{GW} detector selection effects ($\mathcal{W}_{\mathrm{gw}}$). 
As the number of \acp{HMXB} with $\mathcal{W}_{\mathrm{xrb}} >0$ that become BBHm$_\mathrm{H}$ in our population is small, we plot the points for individual systems, with the point size scaling with $\mathcal{W}_{\mathrm{xrb}}$.}
\label{fig:BH_weighted_masses}
\end{figure}

Finally, in Figure~\ref{fig:BH_weighted_masses}, we plot the component \ac{BH} masses for \acp{HMXB} in the $z_{<0.05}$ sample that become merging \acp{BBH} in a Hubble time (BBHm$_\mathrm{H}$) and the component \ac{BH} masses for the population of \acp{BBH} in the $z_{<20}$ sample that merge by today (BBHm$_{z0}$).
We weight the BBHm$_\mathrm{H}$ population by their X-ray detectability and the BBHm$_{z0}$ population by their \ac{GW} detectability.
This allows for comparison of the component \ac{BH} mass distributions for \changed{detectable} \acp{HMXB} that become \acp{BBH} and \changed{detectable} \ac{BBH} mergers. 
The population of \acp{HMXB} that become merging \acp{BBH} within a Hubble time of formation accounts for $<1\%$ of the full \ac{HMXB}$_{z0}$ population, and the number of these systems that are detectable in X-ray is even smaller ($<0.05\%$ of the \ac{HMXB}$_{z0}$ population).
Consequently, we also plot the individual points for these systems, with the point size scaling with their X-ray detection weighting.

We find that the detection-weighted \ac{BH} mass distributions overlap significantly.
This is primarily because there are a handful of \acp{HMXB} with large X-ray detection weights that will form high-mass \ac{BBH} mergers, even though the majority of detectable \acp{HMXB} in this population form lower-mass \acp{BBH} that are unlikely to be detected using \acp{GW}.
Thus, though \changed{detectable} \acp{HMXB} that result in \ac{BBH} mergers have lower typical masses than \ac{BBH} mergers detected via \acp{GW}, it is plausible that X-ray detectors could find a high-mass \ac{BH} in a \changed{detectable} \ac{HMXB} system that is predicted to become a \ac{BBH} \ac{GW} source.

\subsubsection{Population Probability Results}\label{sec:prob_results}

\begin{deluxetable}{c c}
\tablecaption{Probabilities of obtaining various sources in our sampled populations as calculated using the methods described in Section~\ref{sec:probs}.
\changed{Here, $P(\mathrm{HMXB}_\mathrm{obs} | \mathrm{HMXB}_{z0})$ is the probability of detecting an \ac{HMXB} in our sample; 
$P(\mathrm{BBHm}_\mathrm{obs} | \mathrm{BBHm}_{T})$ is the probability of detecting a merging \ac{BBH} in our sample within an observing window of $T = 100~\mathrm{yr}$;
$P(\mathrm{BBHm}_\mathrm{H} | \mathrm{HMXB}_\mathrm{obs})$ is the probability that a detectable \ac{HMXB} becomes a merging \ac{BBH} within a Hubble time of formation, and
$P(\mathrm{HMXB}_\mathrm{obs} | \mathrm{BBHm}_\mathrm{H})$ is the probability that a \ac{BBH} that merges within a Hubble time of formation underwent a detectable \ac{HMXB} phase in the past.}}
\label{table:probs}
\tablehead{
\colhead{Probability} &  \colhead{Approximate Value} 
}
\startdata
$P(\mathrm{HMXB}_\mathrm{obs} | \mathrm{HMXB}_{z0})$ & $2.6 \times 10^{-6}$ \\
\changed{$P(\mathrm{BBHm}_\mathrm{obs} | \mathrm{BBHm}_{T})$} & \changed{$4.8 \times 10^{-3}$} \\
$P(\mathrm{BBHm}_\mathrm{H} | \mathrm{HMXB}_\mathrm{obs})$ & $6.2 \times 10^{-3}$ \\
$P(\mathrm{HMXB}_\mathrm{obs} | \mathrm{BBHm}_\mathrm{H})$ & $9.3 \times 10^{-7}$ \\
\enddata
\end{deluxetable}

To further quantify our results, we calculate the probabilities of obtaining sources in our sampled populations using the method described in Section~\ref{sec:probs}. 
These quantities are defined in the context of our sampled populations, which consist of binaries that have survived the first \ac{SN} with primary \ac{NS} or \ac{BH} progenitor stars and companion stars that are $\geq 5 M_{\odot}$ at first \ac{CO} formation.
The probability results are summarized in Table~\ref{table:probs}.

We find that the probability of detecting an \ac{HMXB} in our $z_{<0.05}$ sample is $P(\mathrm{HMXB}_\mathrm{obs} | \mathrm{HMXB}_{z0}) \simeq 2.6 \times 10^{-6}$. 
This means that most \acp{HMXB} in our sample are not probable to be detected via their X-ray emission.
We calculate the analogous probability for the merging \ac{BBH} population in our $z_{<20}$ sample, and find the probability of detecting a \ac{BBH} within an observing window of $T=100$ years is \changed{$P(\mathrm{BBHm}_\mathrm{obs} | \mathrm{BBHm}_{T}) \simeq 4.8 \times 10^{-3}$}.
\changed{Here, $\mathrm{BBHm}_T$ is the population of \acp{BBH} that merge within our past light cone for the observing window $T$.
This population can be thought of as the \ac{BBH} mergers we would detect in time $T$ if with a perfect detector ($p_{\mathrm{det}} = 1$).}
\changed{Our result for this probability indicates that most \acp{BBH} in our sample are not detectable as \ac{GW} sources, although this probability is higher than that for the \changed{detectable} \acp{HMXB}}. 
Thus, we conclude that both \changed{detectable} \acp{HMXB} and \changed{detectable} \acp{BBH} are rare outcomes from binaries across the Universe.

The probability that a detectable \ac{HMXB} becomes a merging \ac{BBH} in a Hubble time is $P(\mathrm{BBHm}_\mathrm{H} | \mathrm{HMXB}_\mathrm{obs}) \simeq 6.2 \times 10^{-3}$.
This implies that even if we detect an \ac{HMXB}, which itself is improbable, the detected binary will most probably not merge as a \ac{BBH} within a Hubble time.
This is consistent with predictions for the fates of Cyg X-1, LMC X-1, and Cyg X-3, as discussed in Section~\ref{sec:intro}.
Conversely, we calculate the probability that a \ac{BBH} that merges within a Hubble time of formation underwent a \changed{detectable} \ac{HMXB} phase in the past. 
We find this to be $P(\mathrm{HMXB}_{\mathrm{obs}} | \mathrm{BBHm}_{\mathrm{H}})\simeq 9.3 \times 10^{-7}$ for the $z_{<0.05}$ population. 
This probability is considerably smaller than $P(\mathrm{BBHm}_{\mathrm{H}} | \mathrm{HMXB}_{\mathrm{obs}})$, which indicates that for the population of $\mathrm{BBHm}_{\mathrm{H}}$, far fewer will have undergone a \changed{detectable} \ac{HMXB} phase compared to the amount of \changed{detectable} \acp{HMXB} that will become merging \acp{BBH} within a Hubble time. 
This is expected, as only binaries formed at low redshifts can become \changed{detectable} \acp{HMXB}. 
Thus, although $>97\%$ of \ac{BBH} mergers experience an \ac{XRB} phase during their evolution, it is unsurprising that the \emph{observed} \acp{HMXB} are not the progenitors of merging \acp{BBH}.

The probability of a detectable \ac{HMXB} forming a detectable \ac{BBH} merger, $P(\mathrm{BBHm}_{\mathrm{obs}} | \mathrm{HMXB}_{\mathrm{obs}}$), is essentially zero.
This is because the delay time between the \ac{HMXB} phase and the \ac{GW} merger is much longer than the observing window $T$.
This probability is also highly sensitive to the choice of $T$, as one could theoretically choose a very long observing window and force this probability to be non-zero.
Thus, $P(\mathrm{BBHm}_{\mathrm{obs}} | \mathrm{HMXB}_{\mathrm{obs}})$, and similarly $P(\mathrm{HMXB}_{\mathrm{obs}} | \mathrm{BBHm}_{\mathrm{obs}})$, do not have strong physical meanings, as they are largely determined by the choice of observing period.

\section{Discussion} \label{sec:discussion}

We investigate the major current discrepancies in \ac{HMXB} and \ac{GW} \ac{BH} observations, namely the lack of observed \acp{HMXB} that are predicted to become \ac{BBH} mergers as well as the lack of high-mass \acp{BH} found in \acp{XRB}, and find that X-ray and \ac{GW} detector selection effects can explain them. 
To arrive at this conclusion, we applied X-ray and \ac{GW} selection effects to simulated binaries from \texttt{COSMIC} and examined their impact on population parameters. 
Using our new probability formalism, we quantified the probability of obtaining \changed{detectable} sources in our samples of binaries. 
We discuss the main conclusions of our study in detail in Section~\ref{sec:conclusions}, and summarise caveats and areas for future work in Section~\ref{sec:future}.

\subsection{Main Conclusions}\label{sec:conclusions}

From our population synthesis analysis, we conclude that:
\begin{enumerate}

    \item \textbf{\changed{Detectable} \acp{HMXB} are not likely to host \acp{BH} $\mathbf{>35 \emph{M}_{\odot}}$ relative to 
    \changed{detectable} \ac{BBH} mergers.} 
The mass distributions in the central panel of Figure~\ref{fig:hmxb_masses} show that \ac{GW} detection-weighted merging \ac{BBH} progenitors pull to higher donor and \ac{BH} masses compared to the full \changed{detectable} \ac{HMXB} population. 
This is expected because the \ac{GW} detection probability increases with mass, while we find no strong correlation between mass and X-ray \changed{detectability}. 
Fewer than $3\%$ of \changed{detectable} \acp{HMXB} host \ac{BH} accretors $>35 M_{\odot}$, while $\simeq 20\%$ of \changed{detectable} \ac{BBH} mergers have primary \ac{BH} masses that exceed this limit.
These results indicate that \ac{GW} detectors will preferentially see more binaries of higher mass than X-ray surveys.

    \item \textbf{It is highly unlikely that \changed{detectable} \acp{HMXB} will form merging \acp{BBH} within a Hubble time.} 
We calculate that the probability a detectable \ac{HMXB} will merge as a \ac{BBH} within a Hubble time is $\simeq 0.6\%$.  
This result is within one order of magnitude of predictions for the fates of Cyg X-1, Cyg X-3, and LMC X-1 (Section~\ref{sec:intro}), even though population synthesis prescriptions for binary evolution vary between all studies~\citep{belczynski_high_2012, belczynski_cyg_2013,neijssel_wind_2021}. 
This rarity implies that the population of \acp{HMXB} that we do observe is unlikely to form merging \acp{BBH}. 
Since the sample size of observed \ac{HMXB} sources with well-constrained binary properties is small, our results confirm that it is unsurprising none of these systems are likely to form merging \acp{BBH}.
    
    \item \textbf{X-ray and \ac{GW} selection effects probe different redshifts and metallicities.} 
The \ac{ZAMS} formation redshift distributions for detectable \ac{HMXB} and detectable merging \ac{BBH} sources in Figure~\ref{fig:zf_comp_kde} show that 
\changed{detectable} \acp{HMXB} form locally (around $13~\mathrm{Mpc}$, if these systems were in the Hubble flow), while \changed{detectable} merging \ac{BBH} sources form and merge at much farther distances.
As a result, \changed{detectable} \acp{HMXB} also form at higher metallicities near $0.5 Z/Z_{\odot}$ while \changed{detectable} \ac{BBH} mergers form at lower metallicities below $0.1 Z/Z_{\odot}$.
This behavior is expected, as it mirrors what we see in observations: most \acp{HMXB} with well-constrained binary parameters are in the Milky Way or nearby Local Group galaxies that have typical metallicities of $0.1 \lesssim Z/Z_{\odot} \lesssim 1$, whereas \ac{GW} sources are detected at farther distances where there are more environments with low metallicity~\citep[e.g.,][]{evans_chandra_2010, rosen_xmm-newton_2016, krivonos_15_2021, abbott_gwtc-1_2019, abbott_gwtc-2_2021, abbott_gwtc-3_2021, abbott_gwtc-21_2022}. 

\end{enumerate}
The differences between the observational samples of \acp{HMXB} and \acp{GW} sources means that these measurements are complementary, each providing different probes of the evolution of massive stars and the formation of \acp{CO}. 


\subsection{Caveats \& Future Work}\label{sec:future}

In performing our calculations, we made several simplifying approximations. 
First, we assume that the local universe is homogeneous, and we do not model individual galaxies or changes in metallicity within those galaxies. 
Variations in metallicity within galaxies can be significant \citep[e.g.,][]{williams_two-dimensional_2021, taibi_stellar_2022}, having potential effects on binary stellar evolution. 
Our results should therefore be taken as approximate estimates for the $z_{<0.05}$ \ac{HMXB} population. 

We have also approximated the selection function of X-ray sources with a single flux threshold, but in reality these detected sources may not be resolvable as \ac{BH} X-ray binaries~\citep[e.g.,][]{lutovinov_population_2013, clavel_nustar_2019}.
This discrepancy can be seen through the difference in redshifts between our sampled \ac{HMXB} distribution in Figure~\ref{fig:zf_comp_kde} and those found in observations~\citep[e.g.,][]{evans_chandra_2010, arnason_distances_2021, krivonos_15_2021}.
Most observed \acp{HMXB} with well-constrained binary parameters are ${< 50~\mathrm{kpc}}$ away, which is significantly closer than the peak of our redshift distribution. 
To accurately model these sources, separate simulations of the Milky Way and local galaxies are required. 
However, as we do not attempt to reproduce existing observations of individual \acp{HMXB}, our results give a good sense of how selection effects play a role in detecting larger X-ray populations.

\changed{In addition, the Roche lobe filling factor may be important in determining \ac{HMXB} \changed{detectability}. 
Recent results from~\citet{hirai_conditions_2021} show that focused accretion streams necessary for \ac{BH} disk formation can only form in wind-fed binaries when the Roche lobe filling factor is $\gtrsim 0.8$.
We check whether this condition for detectability affects our results by cutting our detectable \ac{HMXB} population to include only wind-fed systems with Roche lobe filling factors $>0.8$.
We find that these cuts do not significantly change our main qualitative conclusions, but they reduce the size of our detectable \ac{HMXB} population \changed{by 95\%}.
This reduction in population size will change the probabilities reported in Section~\ref{sec:prob_results}, though it does not strongly alter the distributions in our plots; the \ac{HMXB} redshift and metallicity distributions in Figure~\ref{fig:zf_comp_kde} and Figure~\ref{fig:met_comp_kde} are nearly identical to those of the population cut based on Roche lobe filling factor, and the \ac{HMXB} mass distributions in the top and center panels of Figure~\ref{fig:hmxb_masses} shift to slightly lower masses with the cut population.
Thus, if anything, incorporating this additional condition for detectability will further emphasize the differences in the detected binary populations due to selection effects.}

While we sample our populations with redshift--metallicity evolution from the \texttt{Illustris-TNG} simulations~\citep{nelson_illustristng_2021}, the true distribution of $P(Z|z)$ is uncertain~\citep[e.g.,][]{neijssel_effect_2019}.
For example, though sampling with a truncated log-normal distribution with a spread of $\sigma = 0.5~\mathrm{dex}$ for $P(Z|z)$ gives similar fundamental results to sampling with our method using \texttt{Illustris-TNG}, using a tighter log-normal distribution with $\sigma = 0.1~\mathrm{dex}$ significantly alters our results. 
We choose to sample with \texttt{Illustris-TNG} because this redshift--metallicity evolution is more physically motivated than simple analytic alternatives.
Uncertainties in the redshift--metallicity evolution can affect \ac{GW} populations by changing the initial conditions and evolutionary properties of their progenitor populations~\citep[e.g.,][]{chruslinska_chemical_2022}.
However, features in the \ac{BBH} mass spectrum are relatively robust to variations in the joint redshift and metallicity evolution~\citep{van_son_locations_2022}.

Rapid population synthesis codes like \texttt{COSMIC} necessarily use approximate prescriptions for stellar and binary evolution. 
Therefore, they are known to have systematic effects on their produced binary populations~\citep[e.g.,][]{shao_population_2019, gallegos-garcia_binary_2021, marchant_role_2021}.
Most of the shortcomings associated with these approximations can be well-addressed with state-of-the-art population synthesis codes like \texttt{POSYDON}~\citep{fragos_posydon_2022}, which use detailed \texttt{MESA} simulations~\citep{paxton_modules_2011} to model the full evolution of binary systems. 
Currently, such codes only cover a limited range of metallicities, and so cannot yet be used for studies such as ours. 
However, our analysis framework can be adapted to use updated population synthesis results as they become available.

We only examine isolated binary evolution scenarios in this project.
In reality, we expect \ac{GW} sources to form through a number of channels~\citep[e.g.,][]{zevin_one_2021}, and these must be considered as well in order to fully understand \ac{GW} sources.
Considering additional formation channels may further differentiate the evolution of \ac{HMXB} sources from \ac{BBH} sources, which exemplifies the need to understand the selection effects (both in terms of source formation and detectability) unique to each of these populations.

\changedtwo{Once we have more sophisticated population models that accurately track mass transfer and stellar structure, and once we include all potential formation channels, it will be possible to make more accurate forecasts of the diverse populations of detectable \acp{HMXB} and detectable \acp{BBH}. 
These predictions could then be compared to observations to jointly constrain uncertainties in the physics of these systems' evolution. 
While we only consider masses in our work, including properties such as \ac{BH} spin would provide further insights into the formation of \acp{HMXB} and \acp{BBH}, and how these populations are related.}

As the LIGO--Virgo--KAGRA detector network prepares for its fourth observing run~\citep{abbott_prospects_2020}, new \ac{GW} data will better resolve the mass distribution of \acp{BBH}.
In addition, as \ac{HMXB} measurements continue to improve with upcoming missions like the \textit{eROSITA} survey set to complete in 2023~\citep{basu-zych_next_2020}, more binaries will be detected with resolved component masses and more accurate \ac{BH} mass measurements will be attained for previously observed systems. 
Combining \ac{GW} and X-ray observations using studies like ours can help us build a more complete concordance model of binary evolution.

\section*{Acknowledgements}

The authors thank Scott Coughlin and Katie Breivik for their assistance with \texttt{COSMIC}. 
We thank Jeff Andrews, Pablo Marchant, Ilya Mandel, and Neta Bahcall for insightful conversations and input on our results. 
C.L.\ acknowledges support from CIERA and Northwestern University.
Support for M.Z.\ was provided by NASA through the NASA Hubble Fellowship grant HST-HF2-51474.001-A awarded by the Space Telescope Science Institute, which is operated by the Association of Universities for Research in Astronomy, Incorporated, under NASA contract NAS5-26555. 
C.P.L.B.\ acknowledges support from the CIERA Board of Visitors Research Professorship and from STFC grant ST/V005634/1. 
Z.D.\ also acknowledges support from the CIERA Board of Visitors Research Professorship. 
V.K.\ was partially supported through a CIFAR Senior Fellowship, a Guggenheim Fellowship, and the Gordon and Betty Moore Foundation (grant award GBMF8477). 
This work utilized the computing resources at CIERA provided by the Quest high performance computing facility at Northwestern University, which is jointly supported by the Office of the Provost, the Office for Research, and Northwestern University Information Technology, and used computing resources at CIERA funded by NSF PHY-1726951. 
\changed{The data that support the findings of this study are openly available from Zenodo~\href{https://doi.org/10.5281/zenodo.7216270}{doi.org/10.5281/zenodo.7216270}.}

\textit{Software:} \texttt{COSMIC}~\citep{breivik_cosmic_2020}; \texttt{Matplotlib}~\citep{hunter_matplotlib_2007}; \texttt{NumPy}~\citep{van_der_walt_numpy_2011}; \texttt{Pandas}~\citep{mckinney_data_2010}; \texttt{seaborn}~\citep{waskom_seaborn_2021}.

\bibliography{references}{}

\begin{thebibliography}{}
\expandafter\ifx\csname natexlab\endcsname\relax\def\natexlab#1{#1}\fi
\providecommand{\url}[1]{\href{#1}{#1}}
\providecommand{\dodoi}[1]{doi:~\href{http://doi.org/#1}{\nolinkurl{#1}}}
\providecommand{\doeprint}[1]{\href{http://ascl.net/#1}{\nolinkurl{http://ascl.net/#1}}}
\providecommand{\doarXiv}[1]{\href{https://arxiv.org/abs/#1}{\nolinkurl{https://arxiv.org/abs/#1}}}

\bibitem[{Aasi {et~al.}(2015)Aasi, Abbott, Abbott, Abbott, Abernathy, Ackley,
  Adams, Adams, Addesso, Adhikari, Adya, Affeldt, Aggarwal, Aguiar, Ain, Ajith,
  Alemic, Allen, Amariutei, Anderson, Anderson, Arai, Araya, Arceneaux, Areeda,
  Ashton, Ast, Aston, Aufmuth, Aulbert, Aylott, Babak, Baker, Ballmer,
  Barayoga, Barbet, Barclay, Barish, Barker, Barr, Barsotti, Bartlett, Barton,
  Bartos, Bassiri, Batch, Baune, Behnke, Bell, Bell, Benacquista, Bergman,
  Bergmann, Berry, Betzwieser, Bhagwat, Bhandare, Bilenko, Billingsley, Birch,
  Biscans, Biwer, Blackburn, Blackburn, Blair, Blair, Bock, Bodiya, Bojtos,
  Bond, Bork, Born, Bose, Brady, Braginsky, Brau, Bridges, Brinkmann, Brooks,
  Brown, Brown, Brown, Buchman, Buikema, Buonanno, Cadonati, Bustillo, Camp,
  Cannon, Cao, Capano, Caride, Caudill, Cavaglià, Cepeda, Chakraborty,
  Chalermsongsak, Chamberlin, Chao, Charlton, Chen, Cho, Cho, Chow,
  Christensen, Chu, Chung, Ciani, Clara, Clark, Collette, Cominsky, Constancio,
  Cook, Corbitt, Cornish, Corsi, Costa, Coughlin, Countryman, Couvares, Coward,
  Cowart, Coyne, Coyne, Craig, Creighton, Creighton, Cripe, Crowder, Cumming,
  Cunningham, Cutler, Dahl, Canton, Damjanic, Danilishin, Danzmann, Dartez,
  Dave, Daveloza, Davies, Daw, DeBra, Pozzo, Denker, Dent, Dergachev, DeRosa,
  DeSalvo, Dhurandhar, D´ıaz, Palma, Dojcinoski, Dominguez, Donovan, Dooley,
  Doravari, Douglas, Downes, Driggers, Du, Dwyer, Eberle, Edo, Edwards,
  Edwards, Effler, Eggenstein, Ehrens, Eichholz, Eikenberry, Essick, Etzel,
  Evans, Evans, Factourovich, Fairhurst, Fan, Fang, Farr, Farr, Favata, Fays,
  Fehrmann, Fejer, Feldbaum, Ferreira, Fisher, Frei, Freise, Frey, Fricke,
  Fritschel, Frolov, Fuentes-Tapia, Fulda, Fyffe, Gair, Gaonkar, Gehrels,
  Gergely´, Giaime, Giardina, Gleason, Goetz, Goetz, Gondan, González,
  Gordon, Gorodetsky, Gossan, Goßler, Gräf, Graff, Grant, Gras, Gray,
  Greenhalgh, Gretarsson, Grote, Grunewald, Guido, Guo, Gushwa, Gustafson,
  Gustafson, Hacker, Hall, Hammond, Hanke, Hanks, Hanna, Hannam, Hanson,
  Hardwick, Harry, Harry, Hart, Hartman, Haster, Haughian, Hee, Heintze,
  Heinzel, Hendry, Heng, Heptonstall, Heurs, Hewitson, Hild, Hoak, Hodge,
  Hollitt, Holt, Hopkins, Hosken, Hough, Houston, Howell, Hu, Huerta, Hughey,
  Husa, Huttner, Huynh, Huynh-Dinh, Idrisy, Indik, Ingram, Inta, Islas, Isler,
  Isogai, Iyer, Izumi, Jacobson, Jang, Jawahar, Ji, Jiménez-Forteza, Johnson,
  Jones, Jones, Ju, Haris, Kalogera, Kandhasamy, Kang, Kanner, Katsavounidis,
  Katzman, Kaufer, Kaufer, Kaur, Kawabe, Kawazoe, Keiser, Keitel, Kelley,
  Kells, Keppel, Key, Khalaidovski, Khalili, Khazanov, Kim, Kim, Kim, Kim, Kim,
  King, King, Kinzel, Kissel, Klimenko, Kline, Koehlenbeck, Kokeyama,
  Kondrashov, Korobko, Korth, Kozak, Kringel, Krishnan, Krueger, Kuehn, Kumar,
  Kumar, Kuo, Landry, Lantz, Larson, Lasky, Lazzarini, Lazzaro, Le, Leaci,
  Leavey, Lebigot, Lee, Lee, Lee, Leong, Levin, Levine, Lewis, Li, Libbrecht,
  Libson, Lin, Littenberg, Lockerbie, Lockett, Logue, Lombardi, Lormand, Lough,
  Lubinski, Lück, Lundgren, Lynch, Ma, Macarthur, MacDonald, Machenschalk,
  MacInnis, Macleod, Magaña-Sandoval, Magee, Mageswaran, Maglione, Mailand,
  Mandel, Mandic, Mangano, Mansell, Márka, Márka, Markosyan, Maros, Martin,
  Martin, Martynov, Marx, Mason, Massinger, Matichard, Matone, Mavalvala,
  Mazumder, Mazzolo, McCarthy, McClelland, McCormick, McGuire, McIntyre,
  McIver, McLin, McWilliams, Meadors, Meinders, Melatos, Mendell, Mercer,
  Meshkov, Messenger, Meyers, Miao, Middleton, Mikhailov, Miller, Miller,
  Millhouse, Ming, Mirshekari, Mishra, Mitra, Mitrofanov, Mitselmakher,
  Mittleman, Moe, Mohanty, Mohapatra, Moore, Moraru, Moreno, Morriss, Mossavi,
  Mow-Lowry, Mueller, Mueller, Mukherjee, Mullavey, Munch, Murphy, Murray,
  Mytidis, Nash, Nayak, Necula, Nedkova, Newton, Nguyen, Nielsen, Nissanke,
  Nitz, Nolting, Normandin, Nuttall, Ochsner, O’Dell, Oelker, Ogin, Oh, Oh,
  Ohme, Oppermann, Oram, O’Reilly, Ortega, O’Shaughnessy, Osthelder, Ott,
  Ottaway, Ottens, Overmier, Owen, Padilla, Pai, Pai, Palashov, Pal-Singh, Pan,
  Pankow, Pannarale, Pant, Papa, Paris, Patrick, Pedraza, Pekowsky, Pele, Penn,
  Perreca, Phelps, Pierro, Pinto, Pitkin, Poeld, Post, Poteomkin, Powell,
  Prasad, Predoi, Premachandra, Prestegard, Price, Principe, Privitera, Prix,
  Prokhorov, Puncken, Pürrer, Qin, Quetschke, Quintero, Quiroga,
  Quitzow-James, Raab, Rabeling, Radkins, Raffai, Raja, Rajalakshmi, Rakhmanov,
  Ramirez, Raymond, Reed, Reid, Reitze, Reula, Riles, Robertson, Robie,
  Rollins, Roma, Romano, Romanov, Romie, Rowan, Rüdiger, Ryan, Sachdev,
  Sadecki, Sadeghian, Saleem, Salemi, Sammut, Sandberg, Sanders, Sannibale,
  Santiago-Prieto, Sathyaprakash, Saulson, Savage, Sawadsky, Scheuer,
  Schilling, Schmidt, Schnabel, Schofield, Schreiber, Schuette, Schutz, Scott,
  Scott, Sellers, Sengupta, Sergeev, Serna, Sevigny, Shaddock, Shahriar,
  Shaltev, Shao, Shapiro, Shawhan, Shoemaker, Sidery, Siemens, Sigg, Silva,
  Simakov, Singer, Singer, Singh, Sintes, Slagmolen, Smith, Smith, Smith,
  Smith-Lefebvre, Son, Sorazu, Souradeep, Staley, Stebbins, Steinke,
  Steinlechner, Steinlechner, Steinmeyer, Stephens, Steplewski, Stevenson,
  Stone, Strain, Strigin, Sturani, Stuver, Summerscales, Sutton, Szczepanczyk,
  Szeifert, Talukder, Tanner, Tápai, Tarabrin, Taracchini, Taylor, Tellez,
  Theeg, Thirugnanasambandam, Thomas, Thomas, Thorne, Thorne, Thrane, Tiwari,
  Tomlinson, Torres, Torrie, Traylor, Tse, Tshilumba, Ugolini, Unnikrishnan,
  Urban, Usman, Vahlbruch, Vajente, Valdes, Vallisneri, Veggel, Vass, Vaulin,
  Vecchio, Veitch, Veitch, Venkateswara, Vincent-Finley, Vitale, Vo, Vorvick,
  Vousden, Vyatchanin, Wade, Wade, Wade, Walker, Wallace, Walsh, Wang, Wang,
  Wang, Ward, Warner, Was, Weaver, Weinert, Weinstein, Weiss, Welborn, Wen,
  Wessels, Westphal, Wette, Whelan, Whitcomb, White, Whiting, Wilkinson,
  Williams, Williams, Williamson, Willis, Willke, Wimmer, Winkler, Wipf,
  Wittel, Woan, Worden, Xie, Yablon, Yakushin, Yam, Yamamoto, Yancey, Yang,
  Zanolin, Zhang, Zhang, Zhang, Zhang, Zhao, Zhou, Zhu, Zucker, Zuraw, \&
  Zweizig}]{aasi_advanced_2015}
Aasi, J., Abbott, B.~P., Abbott, R., {et~al.} 2015, Classical and Quantum
  Gravity, 32, 074001, \dodoi{10.1088/0264-9381/32/7/074001}

\bibitem[{Abbott {et~al.}(2016{\natexlab{a}})Abbott, Abbott, Abbott, Abernathy,
  Acernese, Ackley, Adams, Adams, Addesso, Adhikari, Adya, Affeldt, Agathos,
  Agatsuma, Aggarwal, Aguiar, Aiello, Ain, Ajith, Allen, Allocca, Altin,
  Anderson, Anderson, Arai, Arain, Araya, Arceneaux, Areeda, Arnaud, Arun,
  Ascenzi, Ashton, Ast, Aston, Astone, Aufmuth, Aulbert, Babak, Bacon, Bader,
  Baker, Baldaccini, Ballardin, Ballmer, Barayoga, Barclay, Barish, Barker,
  Barone, Barr, Barsotti, Barsuglia, Barta, Bartlett, Barton, Bartos, Bassiri,
  Basti, Batch, Baune, Bavigadda, Bazzan, Behnke, Bejger, Belczynski, Bell,
  Bell, Berger, Bergman, Bergmann, Berry, Bersanetti, Bertolini, Betzwieser,
  Bhagwat, Bhandare, Bilenko, Billingsley, Birch, Birney, Birnholtz, Biscans,
  Bisht, Bitossi, Biwer, Bizouard, Blackburn, Blair, Blair, Blair, Bloemen,
  Bock, Bodiya, Boer, Bogaert, Bogan, Bohe, Bojtos, Bond, Bondu, Bonnand, Boom,
  Bork, Boschi, Bose, Bouffanais, Bozzi, Bradaschia, Brady, Braginsky,
  Branchesi, Brau, Briant, Brillet, Brinkmann, Brisson, Brockill, Brooks,
  Brown, Brown, Brown, Buchanan, Buikema, Bulik, Bulten, Buonanno, Buskulic,
  Buy, Byer, Cabero, Cadonati, Cagnoli, Cahillane, Bustillo, Callister,
  Calloni, Camp, Cannon, Cao, Capano, Capocasa, Carbognani, Caride, Diaz,
  Casentini, Caudill, Cavaglià, Cavalier, Cavalieri, Cella, Cepeda, Baiardi,
  Cerretani, Cesarini, Chakraborty, Chalermsongsak, Chamberlin, Chan, Chao,
  Charlton, Chassande-Mottin, Chen, Chen, Cheng, Chincarini, Chiummo, Cho, Cho,
  Chow, Christensen, Chu, Chua, Chung, Ciani, Clara, Clark, Cleva, Coccia,
  Cohadon, Colla, Collette, Cominsky, Constancio, Conte, Conti, Cook, Corbitt,
  Cornish, Corsi, Cortese, Costa, Coughlin, Coughlin, Coulon, Countryman,
  Couvares, Cowan, Coward, Cowart, Coyne, Coyne, Craig, Creighton, Creighton,
  Cripe, Crowder, Cruise, Cumming, Cunningham, Cuoco, Canton, Danilishin,
  D’Antonio, Danzmann, Darman, Da~Silva~Costa, Dattilo, Dave, Daveloza,
  Davier, Davies, Daw, Day, De, DeBra, Debreczeni, Degallaix, De~Laurentis,
  Deléglise, Del~Pozzo, Denker, Dent, Dereli, Dergachev, DeRosa, De~Rosa,
  DeSalvo, Dhurandhar, Díaz, Di~Fiore, Di~Giovanni, Di~Lieto, Di~Pace,
  Di~Palma, Di~Virgilio, Dojcinoski, Dolique, Donovan, Dooley, Doravari,
  Douglas, Downes, Drago, Drever, Driggers, Du, Ducrot, Dwyer, Edo, Edwards,
  Effler, Eggenstein, Ehrens, Eichholz, Eikenberry, Engels, Essick, Etzel,
  Evans, Evans, Everett, Factourovich, Fafone, Fair, Fairhurst, Fan, Fang,
  Farinon, Farr, Farr, Favata, Fays, Fehrmann, Fejer, Feldbaum, Ferrante,
  Ferreira, Ferrini, Fidecaro, Finn, Fiori, Fiorucci, Fisher, Flaminio,
  Fletcher, Fong, Fournier, Franco, Frasca, Frasconi, Frede, Frei, Freise,
  Frey, Frey, Fricke, Fritschel, Frolov, Fulda, Fyffe, Gabbard, Gair,
  Gammaitoni, Gaonkar, Garufi, Gatto, Gaur, Gehrels, Gemme, Gendre, Genin,
  Gennai, George, Gergely, Germain, Ghosh, Ghosh, Ghosh, Giaime, Giardina,
  Giazotto, Gill, Glaefke, Gleason, Goetz, Goetz, Gondan, González, Castro,
  Gopakumar, Gordon, Gorodetsky, Gossan, Gosselin, Gouaty, Graef, Graff,
  Granata, Grant, Gras, Gray, Greco, Green, Greenhalgh, Groot, Grote,
  Grunewald, Guidi, Guo, Gupta, Gupta, Gushwa, Gustafson, Gustafson, Hacker,
  Hall, Hall, Hammond, Haney, Hanke, Hanks, Hanna, Hannam, Hanson, Hardwick,
  Harms, Harry, Harry, Hart, Hartman, Haster, Haughian, Healy, Heefner,
  Heidmann, Heintze, Heinzel, Heitmann, Hello, Hemming, Hendry, Heng, Hennig,
  Heptonstall, Heurs, Hild, Hoak, Hodge, Hofman, Hollitt, Holt, Holz, Hopkins,
  Hosken, Hough, Houston, Howell, Hu, Huang, Huerta, Huet, Hughey, Husa,
  Huttner, Huynh-Dinh, Idrisy, Indik, Ingram, Inta, Isa, Isac, Isi, Islas,
  Isogai, Iyer, Izumi, Jacobson, Jacqmin, Jang, Jani, Jaranowski, Jawahar,
  Jiménez-Forteza, Johnson, Johnson-McDaniel, Jones, Jones, Jonker, Ju, Haris,
  Kalaghatgi, Kalogera, Kandhasamy, Kang, Kanner, Karki, Kasprzack,
  Katsavounidis, Katzman, Kaufer, Kaur, Kawabe, Kawazoe, Kéfélian, Kehl,
  Keitel, Kelley, Kells, Kennedy, Keppel, Key, Khalaidovski, Khalili, Khan,
  Khan, Khan, Khazanov, Kijbunchoo, Kim, Kim, Kim, Kim, Kim, Kim, King, King,
  Kinzel, Kissel, Kleybolte, Klimenko, Koehlenbeck, Kokeyama, Koley,
  Kondrashov, Kontos, Koranda, Korobko, Korth, Kowalska, Kozak, Kringel,
  Krishnan, Królak, Krueger, Kuehn, Kumar, Kumar, Kuo, Kutynia, Kwee, Lackey,
  Landry, Lange, Lantz, Lasky, Lazzarini, Lazzaro, Leaci, Leavey, Lebigot, Lee,
  Lee, Lee, Lee, Lenon, Leonardi, Leong, Leroy, Letendre, Levin, Levine, Li,
  Libson, Littenberg, Lockerbie, Logue, Lombardi, London, Lord, Lorenzini,
  Loriette, Lormand, Losurdo, Lough, Lousto, Lovelace, Lück, Lundgren, Luo,
  Lynch, Ma, MacDonald, Machenschalk, MacInnis, Macleod, Magaña-Sandoval,
  Magee, Mageswaran, Majorana, Maksimovic, Malvezzi, Man, Mandel, Mandic,
  Mangano, Mansell, Manske, Mantovani, Marchesoni, Marion, Márka, Márka,
  Markosyan, Maros, Martelli, Martellini, Martin, Martin, Martynov, Marx,
  Mason, Masserot, Massinger, Masso-Reid, Matichard, Matone, Mavalvala,
  Mazumder, Mazzolo, McCarthy, McClelland, McCormick, McGuire, McIntyre,
  McIver, McManus, McWilliams, Meacher, Meadors, Meidam, Melatos, Mendell,
  Mendoza-Gandara, Mercer, Merilh, Merzougui, Meshkov, Messenger, Messick,
  Meyers, Mezzani, Miao, Michel, Middleton, Mikhailov, Milano, Miller,
  Millhouse, Minenkov, Ming, Mirshekari, Mishra, Mitra, Mitrofanov,
  Mitselmakher, Mittleman, Moggi, Mohan, Mohapatra, Montani, Moore, Moore,
  Moraru, Moreno, Morriss, Mossavi, Mours, Mow-Lowry, Mueller, Mueller, Muir,
  Mukherjee, Mukherjee, Mukherjee, Mukund, Mullavey, Munch, Murphy, Murray,
  Mytidis, Nardecchia, Naticchioni, Nayak, Necula, Nedkova, Nelemans, Neri,
  Neunzert, Newton, Nguyen, Nielsen, Nissanke, Nitz, Nocera, Nolting,
  Normandin, Nuttall, Oberling, Ochsner, O’Dell, Oelker, Ogin, Oh, Oh, Ohme,
  Oliver, Oppermann, Oram, O’Reilly, O’Shaughnessy, Ott, Ottaway, Ottens,
  Overmier, Owen, Pai, Pai, Palamos, Palashov, Palomba, Pal-Singh, Pan, Pan,
  Pankow, Pannarale, Pant, Paoletti, Paoli, Papa, Paris, Parker, Pascucci,
  Pasqualetti, Passaquieti, Passuello, Patricelli, Patrick, Pearlstone,
  Pedraza, Pedurand, Pekowsky, Pele, Penn, Perreca, Pfeiffer, Phelps, Piccinni,
  Pichot, Pickenpack, Piergiovanni, Pierro, Pillant, Pinard, Pinto, Pitkin,
  Poeld, Poggiani, Popolizio, Post, Powell, Prasad, Predoi, Premachandra,
  Prestegard, Price, Prijatelj, Principe, Privitera, Prix, Prodi, Prokhorov,
  Puncken, Punturo, Puppo, Pürrer, Qi, Qin, Quetschke, Quintero,
  Quitzow-James, Raab, Rabeling, Radkins, Raffai, Raja, Rakhmanov, Ramet,
  Rapagnani, Raymond, Razzano, Re, Read, Reed, Regimbau, Rei, Reid, Reitze,
  Rew, Reyes, Ricci, Riles, Robertson, Robie, Robinet, Rocchi, Rolland,
  Rollins, Roma, Romano, Romano, Romanov, Romie, Rosińska, Rowan, Rüdiger,
  Ruggi, Ryan, Sachdev, Sadecki, Sadeghian, Salconi, Saleem, Salemi, Samajdar,
  Sammut, Sampson, Sanchez, Sandberg, Sandeen, Sanders, Sanders, Sassolas,
  Sathyaprakash, Saulson, Sauter, Savage, Sawadsky, Schale, Schilling, Schmidt,
  Schmidt, Schnabel, Schofield, Schönbeck, Schreiber, Schuette, Schutz, Scott,
  Scott, Sellers, Sengupta, Sentenac, Sequino, Sergeev, Serna, Setyawati,
  Sevigny, Shaddock, Shaffer, Shah, Shahriar, Shaltev, Shao, Shapiro, Shawhan,
  Sheperd, Shoemaker, Shoemaker, Siellez, Siemens, Sigg, Silva, Simakov,
  Singer, Singer, Singh, Singh, Singhal, Sintes, Slagmolen, Smith, Smith,
  Smith, Smith, Son, Sorazu, Sorrentino, Souradeep, Srivastava, Staley,
  Steinke, Steinlechner, Steinlechner, Steinmeyer, Stephens, Stevenson, Stone,
  Strain, Straniero, Stratta, Strauss, Strigin, Sturani, Stuver, Summerscales,
  Sun, Sutton, Swinkels, Szczepańczyk, Tacca, Talukder, Tanner, Tápai,
  Tarabrin, Taracchini, Taylor, Theeg, Thirugnanasambandam, Thomas, Thomas,
  Thomas, Thorne, Thorne, Thrane, Tiwari, Tiwari, Tokmakov, Tomlinson, Tonelli,
  Torres, Torrie, Töyrä, Travasso, Traylor, Trifirò, Tringali, Trozzo, Tse,
  Turconi, Tuyenbayev, Ugolini, Unnikrishnan, Urban, Usman, Vahlbruch, Vajente,
  Valdes, Vallisneri, van Bakel, van Beuzekom, van~den Brand, Van Den~Broeck,
  Vander-Hyde, van~der Schaaf, van Heijningen, van Veggel, Vardaro, Vass,
  Vasúth, Vaulin, Vecchio, Vedovato, Veitch, Veitch, Venkateswara, Verkindt,
  Vetrano, Viceré, Vinciguerra, Vine, Vinet, Vitale, Vo, Vocca, Vorvick, Voss,
  Vousden, Vyatchanin, Wade, Wade, Wade, Waldman, Walker, Wallace, Walsh, Wang,
  Wang, Wang, Wang, Wang, Ward, Ward, Warner, Was, Weaver, Wei, Weinert,
  Weinstein, Weiss, Welborn, Wen, Weßels, Westphal, Wette, Whelan, Whitcomb,
  White, Whiting, Wiesner, Wilkinson, Willems, Williams, Williams, Williamson,
  Willis, Willke, Wimmer, Winkelmann, Winkler, Wipf, Wiseman, Wittel, Woan,
  Worden, Wright, Wu, Yablon, Yakushin, Yam, Yamamoto, Yancey, Yap, Yu, Yvert,
  Zadrożny, Zangrando, Zanolin, Zendri, Zevin, Zhang, Zhang, Zhang, Zhang,
  Zhao, Zhou, Zhou, Zhu, Zucker, Zuraw, Zweizig, \& {LIGO Scientific
  Collaboration and Virgo Collaboration}}]{abbott_observation_2016}
Abbott, B., Abbott, R., Abbott, T., {et~al.} 2016{\natexlab{a}}, Physical
  Review Letters, 116, 061102, \dodoi{10.1103/PhysRevLett.116.061102}

\bibitem[{Abbott {et~al.}(2016{\natexlab{b}})Abbott, Abbott, Abbott, Abernathy,
  Acernese, Ackley, Adams, Adams, Addesso, Adhikari, Adya, Affeldt, Agathos,
  Agatsuma, Aggarwal, Aguiar, Aiello, Ain, Ajith, Allen, Allocca, Altin,
  Anderson, Anderson, Arai, Araya, Arceneaux, Areeda, Arnaud, Arun, Ascenzi,
  Ashton, Ast, Aston, Astone, Aufmuth, Aulbert, Babak, Bacon, Bader, Baker,
  Baldaccini, Ballardin, Ballmer, Barayoga, Barclay, Barish, Barker, Barone,
  Barr, Barsotti, Barsuglia, Barta, Bartlett, Bartos, Bassiri, Basti, Batch,
  Baune, Bavigadda, Bazzan, Behnke, Bejger, Belczynski, Bell, Bell, Berger,
  Bergman, Bergmann, Berry, Bersanetti, Bertolini, Betzwieser, Bhagwat,
  Bhandare, Bilenko, Billingsley, Birch, Birney, Biscans, Bisht, Bitossi,
  Biwer, Bizouard, Blackburn, Blair, Blair, Blair, Bloemen, Bock, Bodiya, Boer,
  Bogaert, Bogan, Bohe, Bojtos, Bond, Bondu, Bonnand, Boom, Bork, Boschi, Bose,
  Bouffanais, Bozzi, Bradaschia, Brady, Braginsky, Branchesi, Brau, Briant,
  Brillet, Brinkmann, Brisson, Brockill, Brooks, Brown, Brown, Brown, Buchanan,
  Buikema, Bulik, Bulten, Buonanno, Buskulic, Buy, Byer, Cadonati, Cagnoli,
  Cahillane, Bustillo, Callister, Calloni, Camp, Cannon, Cao, Capano, Capocasa,
  Carbognani, Caride, Diaz, Casentini, Caudill, Cavaglià, Cavalier, Cavalieri,
  Cella, Cepeda, Baiardi, Cerretani, Cesarini, Chakraborty, Chalermsongsak,
  Chamberlin, Chan, Chao, Charlton, Chassande-Mottin, Chen, Chen, Cheng,
  Chincarini, Chiummo, Cho, Cho, Chow, Christensen, Chu, Chua, Chung, Ciani,
  Clara, Clark, Cleva, Coccia, Cohadon, Colla, Collette, Cominsky, Jr, Conte,
  Conti, Cook, Corbitt, Cornish, Corsi, Cortese, Costa, Coughlin, Coughlin,
  Coulon, Countryman, Couvares, Cowan, Coward, Cowart, Coyne, Coyne, Craig,
  Creighton, Cripe, Crowder, Cumming, Cunningham, Cuoco, Canton, Danilishin,
  D'Antonio, Danzmann, Darman, Dattilo, Dave, Daveloza, Davier, Davies, Daw,
  Day, DeBra, Debreczeni, Degallaix, Laurentis, Deléglise, Pozzo, Denker,
  Dent, Dereli, Dergachev, DeRosa, DeRosa, DeSalvo, Dhurandhar, Díaz, Fiore,
  Giovanni, Lieto, Pace, Palma, Virgilio, Dojcinoski, Dolique, Donovan, Dooley,
  Doravari, Douglas, Downes, Drago, Drever, Driggers, Du, Ducrot, Dwyer, Edo,
  Edwards, Effler, Eggenstein, Ehrens, Eichholz, Eikenberry, Engels, Essick,
  Etzel, Evans, Evans, Everett, Factourovich, Fafone, Fair, Fairhurst, Fan,
  Fang, Farinon, Farr, Farr, Favata, Fays, Fehrmann, Fejer, Ferrante, Ferreira,
  Ferrini, Fidecaro, Fiori, Fiorucci, Fisher, Flaminio, Fletcher, Fournier,
  Franco, Frasca, Frasconi, Frei, Freise, Frey, Frey, Fricke, Fritschel,
  Frolov, Fulda, Fyffe, Gabbard, Gair, Gammaitoni, Gaonkar, Garufi, Gatto,
  Gaur, Gehrels, Gemme, Gendre, Genin, Gennai, George, Gergely, Germain, Ghosh,
  Ghosh, Giaime, Giardina, Giazotto, Gill, Glaefke, Goetz, Goetz, Gondan,
  González, Castro, Gopakumar, Gordon, Gorodetsky, Gossan, Gosselin, Gouaty,
  Graef, Graff, Granata, Grant, Gras, Gray, Greco, Green, Groot, Grote,
  Grunewald, Guidi, Guo, Gupta, Gupta, Gushwa, Gustafson, Gustafson, Hacker,
  Hall, Hall, Hammond, Haney, Hanke, Hanks, Hanna, Hannam, Hanson, Hardwick,
  Harms, Harry, Harry, Hart, Hartman, Haster, Haughian, Heidmann, Heintze,
  Heitmann, Hello, Hemming, Hendry, Heng, Hennig, Heptonstall, Heurs, Hild,
  Hoak, Hodge, Hofman, Hollitt, Holt, Holz, Hopkins, Hosken, Hough, Houston,
  Howell, Hu, Huang, Huerta, Huet, Hughey, Husa, Huttner, Huynh-Dinh, Idrisy,
  Indik, Ingram, Inta, Isa, Isac, Isi, Islas, Isogai, Iyer, Izumi, Jacqmin,
  Jang, Jani, Jaranowski, Jawahar, Jiménez-Forteza, Johnson, Jones, Jones,
  Jonker, Ju, K, Kalaghatgi, Kalogera, Kandhasamy, Kang, Kanner, Karki,
  Kasprzack, Katsavounidis, Katzman, Kaufer, Kaur, Kawabe, Kawazoe, Kéfélian,
  Kehl, Keitel, Kelley, Kells, Kennedy, Key, Khalaidovski, Khalili, Khan, Khan,
  Khan, Khazanov, Kijbunchoo, Kim, Kim, Kim, Kim, Kim, Kim, King, King, Kinzel,
  Kissel, Kleybolte, Klimenko, Koehlenbeck, Kokeyama, Koley, Kondrashov,
  Kontos, Korobko, Korth, Kowalska, Kozak, Kringel, Krishnan, Królak, Krueger,
  Kuehn, Kumar, Kuo, Kutynia, Lackey, Landry, Lange, Lantz, Lasky, Lazzarini,
  Lazzaro, Leaci, Leavey, Lebigot, Lee, Lee, Lee, Lee, Lenon, Leonardi, Leong,
  Leroy, Letendre, Levin, Levine, Li, Libson, Littenberg, Lockerbie, Logue,
  Lombardi, Lord, Lorenzini, Loriette, Lormand, Losurdo, Lough, Lück,
  Lundgren, Luo, Lynch, Ma, MacDonald, Machenschalk, MacInnis, Macleod,
  Magaña-Sandoval, Magee, Mageswaran, Majorana, Maksimovic, Malvezzi, Man,
  Mandel, Mandic, Mangano, Mansell, Manske, Mantovani, Marchesoni, Marion,
  Márka, Márka, Markosyan, Maros, Martelli, Martellini, Martin, Martin,
  Martynov, Marx, Mason, Masserot, Massinger, Masso-Reid, Matichard, Matone,
  Mavalvala, Mazumder, Mazzolo, McCarthy, McClelland, McCormick, McGuire,
  McIntyre, McIver, McManus, McWilliams, Meacher, Meadors, Meidam, Melatos,
  Mendell, Mendoza-Gandara, Mercer, Merilh, Merzougui, Meshkov, Messenger,
  Messick, Meyers, Mezzani, Miao, Michel, Middleton, Mikhailov, Milano, Miller,
  Millhouse, Minenkov, Ming, Mirshekari, Mishra, Mitra, Mitrofanov,
  Mitselmakher, Mittleman, Moggi, Mohan, Mohapatra, Montani, Moore, Moore,
  Moraru, Moreno, Morriss, Mossavi, Mours, Mow-Lowry, Mueller, Mueller, Muir,
  Mukherjee, Mukherjee, Mukherjee, Mukund, Mullavey, Munch, Murphy, Murray,
  Mytidis, Nardecchia, Naticchioni, Nayak, Necula, Nedkova, Nelemans, Neri,
  Neunzert, Newton, Nguyen, Nielsen, Nissanke, Nitz, Nocera, Nolting,
  Normandin, Nuttall, Oberling, Ochsner, O'Dell, Oelker, Ogin, Oh, Oh, Ohme,
  Oliver, Oppermann, Oram, O'Reilly, O'Shaughnessy, Ottaway, Ottens, Overmier,
  Owen, Pai, Pai, Palamos, Palashov, Palomba, Pal-Singh, Pan, Pankow,
  Pannarale, Pant, Paoletti, Paoli, Papa, Paris, Parker, Pascucci, Pasqualetti,
  Passaquieti, Passuello, Patricelli, Patrick, Pearlstone, Pedraza, Pedurand,
  Pekowsky, Pele, Penn, Perreca, Phelps, Piccinni, Pichot, Piergiovanni,
  Pierro, Pillant, Pinard, Pinto, Pitkin, Poggiani, Popolizio, Post, Powell,
  Prasad, Predoi, Premachandra, Prestegard, Price, Prijatelj, Principe,
  Privitera, Prix, Prodi, Prokhorov, Puncken, Punturo, Puppo, Pürrer, Qi, Qin,
  Quetschke, Quintero, Quitzow-James, Raab, Rabeling, Radkins, Raffai, Raja,
  Rakhmanov, Rapagnani, Raymond, Razzano, Re, Read, Reed, Regimbau, Rei, Reid,
  Reitze, Rew, Reyes, Ricci, Riles, Robertson, Robie, Robinet, Rocchi, Rolland,
  Rollins, Roma, Romano, Romano, Romanov, Romie, Rosińska, Rowan, Rüdiger,
  Ruggi, Ryan, Sachdev, Sadecki, Sadeghian, Salconi, Saleem, Salemi, Samajdar,
  Sammut, Sanchez, Sandberg, Sandeen, Sanders, Sassolas, Sathyaprakash,
  Saulson, Sauter, Savage, Sawadsky, Schale, Schilling, Schmidt, Schmidt,
  Schnabel, Schofield, Schönbeck, Schreiber, Schuette, Schutz, Scott, Scott,
  Sellers, Sentenac, Sequino, Sergeev, Serna, Setyawati, Sevigny, Shaddock,
  Shah, Shahriar, Shaltev, Shao, Shapiro, Shawhan, Sheperd, Shoemaker,
  Shoemaker, Siellez, Siemens, Sigg, Silva, Simakov, Singer, Singer, Singh,
  Singh, Singhal, Sintes, Slagmolen, Smith, Smith, Smith, Son, Sorazu,
  Sorrentino, Souradeep, Srivastava, Staley, Steinke, Steinlechner,
  Steinlechner, Steinmeyer, Stephens, Stevenson, Stone, Strain, Straniero,
  Stratta, Strauss, Strigin, Sturani, Stuver, Summerscales, Sun, Sutton,
  Swinkels, Szczepańczyk, Tacca, Talukder, Tanner, Tápai, Tarabrin,
  Taracchini, Taylor, Theeg, Thirugnanasambandam, Thomas, Thomas, Thomas,
  Thorne, Thorne, Thrane, Tiwari, Tiwari, Tokmakov, Tomlinson, Tonelli, Torres,
  Torrie, Töyrä, Travasso, Traylor, Trifirò, Tringali, Trozzo, Tse, Turconi,
  Tuyenbayev, Ugolini, Unnikrishnan, Urban, Usman, Vahlbruch, Vajente, Valdes,
  Bakel, Beuzekom, Brand, Broeck, Vander-Hyde, Schaaf, Heijningen, Veggel,
  Vardaro, Vass, Vasúth, Vaulin, Vecchio, Vedovato, Veitch, Veitch,
  Venkateswara, Verkindt, Vetrano, Viceré, Vinciguerra, Vine, Vinet, Vitale,
  Vo, Vocca, Vorvick, Voss, Vousden, Vyatchanin, Wade, Wade, Wade, Walker,
  Wallace, Walsh, Wang, Wang, Wang, Wang, Wang, Ward, Warner, Was, Weaver, Wei,
  Weinert, Weinstein, Weiss, Welborn, Wen, Weßels, Westphal, Wette, Whelan,
  White, Whiting, Williams, Williamson, Willis, Willke, Wimmer, Winkler, Wipf,
  Wittel, Woan, Worden, Wright, Wu, Yablon, Yam, Yamamoto, Yancey, Yap, Yu,
  Yvert, Zadro{\textbackslash}.zny, Zangrando, Zanolin, Zendri, Zevin, Zhang,
  Zhang, Zhang, Zhang, Zhao, Zhou, Zhou, Zhu, Zucker, Zuraw, {and}, \&
  and}]{abbott_astrophysical_2016}
Abbott, B.~P., Abbott, R., Abbott, T.~D., {et~al.} 2016{\natexlab{b}}, 818,
  L22, \dodoi{10.3847/2041-8205/818/2/L22}

\bibitem[{Abbott {et~al.}(2017)Abbott, Abbott, Abbott, Acernese, Ackley, Adams,
  Adams, Addesso, Adhikari, Adya, Affeldt, Afrough, Agarwal, Agathos, Agatsuma,
  Aggarwal, Aguiar, Aiello, Ain, Ajith, Allen, Allen, Allocca, Altin, Amato,
  Ananyeva, Anderson, Anderson, Angelova, Antier, Appert, Arai, Araya, Areeda,
  Arnaud, Arun, Ascenzi, Ashton, Ast, Aston, Astone, Atallah, Aufmuth, Aulbert,
  AultONeal, Austin, Avila-Alvarez, Babak, Bacon, Bader, Bae, Bailes, Baker,
  Baldaccini, Ballardin, Ballmer, Banagiri, Barayoga, Barclay, Barish, Barker,
  Barkett, Barone, Barr, Barsotti, Barsuglia, Barta, Barthelmy, Bartlett,
  Bartos, Bassiri, Basti, Batch, Bawaj, Bayley, Bazzan, Bécsy, Beer, Bejger,
  Belahcene, Bell, Berger, Bergmann, Bernuzzi, Bero, Berry, Bersanetti,
  Bertolini, Betzwieser, Bhagwat, Bhandare, Bilenko, Billingsley, Billman,
  Birch, Birney, Birnholtz, Biscans, Biscoveanu, Bisht, Bitossi, Biwer,
  Bizouard, Blackburn, Blackman, Blair, Blair, Blair, Bloemen, Bock, Bode,
  Boer, Bogaert, Bohe, Bondu, Bonilla, Bonnand, Boom, Bork, Boschi, Bose,
  Bossie, Bouffanais, Bozzi, Bradaschia, Brady, Branchesi, Brau, Briant,
  Brillet, Brinkmann, Brisson, Brockill, Broida, Brooks, Brown, Brown, Brunett,
  Buchanan, Buikema, Bulik, Bulten, Buonanno, Buskulic, Buy, Byer, Cabero,
  Cadonati, Cagnoli, Cahillane, Calderón~Bustillo, Callister, Calloni, Camp,
  Canepa, Canizares, Cannon, Cao, Cao, Capano, Capocasa, Carbognani, Caride,
  Carney, Carullo, Casanueva~Diaz, Casentini, Caudill, Cavaglià, Cavalier,
  Cavalieri, Cella, Cepeda, Cerdá-Durán, Cerretani, Cesarini, Chamberlin,
  Chan, Chao, Charlton, Chase, Chassande-Mottin, Chatterjee, Chatziioannou,
  Cheeseboro, Chen, Chen, Chen, Cheng, Chia, Chincarini, Chiummo, Chmiel, Cho,
  Cho, Chow, Christensen, Chu, Chua, Chua, Chung, Chung, Ciani, Ciolfi,
  Cirelli, Cirone, Clara, Clark, Clearwater, Cleva, Cocchieri, Coccia, Cohadon,
  Cohen, Colla, Collette, Cominsky, Constancio, Conti, Cooper, Corban, Corbitt,
  Cordero-Carrión, Corley, Cornish, Corsi, Cortese, Costa, Coughlin, Coughlin,
  Coulon, Countryman, Couvares, Covas, Cowan, Coward, Cowart, Coyne, Coyne,
  Creighton, Creighton, Cripe, Crowder, Cullen, Cumming, Cunningham, Cuoco,
  Dal~Canton, Dálya, Danilishin, D'Antonio, Danzmann, Dasgupta,
  Da~Silva~Costa, Dattilo, Dave, Davier, Davis, Daw, Day, De, DeBra, Degallaix,
  De~Laurentis, Deléglise, Del~Pozzo, Demos, Denker, Dent, De~Pietri,
  Dergachev, De~Rosa, DeRosa, De~Rossi, DeSalvo, de~Varona, Devenson,
  Dhurandhar, Díaz, Dietrich, Di~Fiore, Di~Giovanni, Di~Girolamo, Di~Lieto,
  Di~Pace, Di~Palma, Di~Renzo, Doctor, Dolique, Donovan, Dooley, Doravari,
  Dorrington, Douglas, Dovale~Álvarez, Downes, Drago, Dreissigacker, Driggers,
  Du, Ducrot, Dudi, Dupej, Dwyer, Edo, Edwards, Effler, Eggenstein, Ehrens,
  Eichholz, Eikenberry, Eisenstein, Essick, Estevez, Etienne, Etzel, Evans,
  Evans, Factourovich, Fafone, Fair, Fairhurst, Fan, Farinon, Farr, Farr,
  Fauchon-Jones, Favata, Fays, Fee, Fehrmann, Feicht, Fejer, Fernandez-Galiana,
  Ferrante, Ferreira, Ferrini, Fidecaro, Finstad, Fiori, Fiorucci, Fishbach,
  Fisher, Fitz-Axen, Flaminio, Fletcher, Fong, Font, Forsyth, Forsyth,
  Fournier, Frasca, Frasconi, Frei, Freise, Frey, Frey, Fries, Fritschel,
  Frolov, Fulda, Fyffe, Gabbard, Gadre, Gaebel, Gair, Gammaitoni, Ganija,
  Gaonkar, Garcia-Quiros, Garufi, Gateley, Gaudio, Gaur, Gayathri, Gehrels,
  Gemme, Genin, Gennai, George, George, Gergely, Germain, Ghonge, Ghosh, Ghosh,
  Ghosh, Giaime, Giardina, Giazotto, Gill, Glover, Goetz, Goetz, Gomes,
  Goncharov, González, Gonzalez~Castro, Gopakumar, Gorodetsky, Gossan,
  Gosselin, Gouaty, Grado, Graef, Granata, Grant, Gras, Gray, Greco, Green,
  Gretarsson, Groot, Grote, Grunewald, Gruning, Guidi, Guo, Gupta, Gupta,
  Gushwa, Gustafson, Gustafson, Halim, Hall, Hall, Hamilton, Hammond, Haney,
  Hanke, Hanks, Hanna, Hannam, Hannuksela, Hanson, Hardwick, Harms, Harry,
  Harry, Hart, Haster, Haughian, Healy, Heidmann, Heintze, Heitmann, Hello,
  Hemming, Hendry, Heng, Hennig, Heptonstall, Heurs, Hild, Hinderer, Ho, Hoak,
  Hofman, Holt, Holz, Hopkins, Horst, Hough, Houston, Howell, Hreibi, Hu,
  Huerta, Huet, Hughey, Husa, Huttner, Huynh-Dinh, Indik, Inta, Intini, Isa,
  Isac, Isi, Iyer, Izumi, Jacqmin, Jani, Jaranowski, Jawahar, Jiménez-Forteza,
  Johnson, Johnson-McDaniel, Jones, Jones, Jonker, Ju, Junker, Kalaghatgi,
  Kalogera, Kamai, Kandhasamy, Kang, Kanner, Kapadia, Karki, Karvinen,
  Kasprzack, Kastaun, Katolik, Katsavounidis, Katzman, Kaufer, Kawabe,
  Kéfélian, Keitel, Kemball, Kennedy, Kent, Key, Khalili, Khan, Khan, Khan,
  Khazanov, Kijbunchoo, Kim, Kim, Kim, Kim, Kim, Kim, Kimbrell, King, King,
  Kinley-Hanlon, Kirchhoff, Kissel, Kleybolte, Klimenko, Knowles, Koch,
  Koehlenbeck, Koley, Kondrashov, Kontos, Korobko, Korth, Kowalska, Kozak,
  Krämer, Kringel, Krishnan, Królak, Kuehn, Kumar, Kumar, Kumar, Kuo,
  Kutynia, Kwang, Lackey, Lai, Landry, Lang, Lange, Lantz, Lanza, Larson,
  Lartaux-Vollard, Lasky, Laxen, Lazzarini, Lazzaro, Leaci, Leavey, Lee, Lee,
  Lee, Lee, Lee, Lehmann, Lenon, Leon, Leonardi, Leroy, Letendre, Levin, Li,
  Linker, Littenberg, Liu, Liu, Lo, Lockerbie, London, Lord, Lorenzini,
  Loriette, Lormand, Losurdo, Lough, Lousto, Lovelace, Lück, Lumaca, Lundgren,
  Lynch, Ma, Macas, Macfoy, Machenschalk, MacInnis, Macleod, Magaña~Hernandez,
  Magaña-Sandoval, Magaña~Zertuche, Magee, Majorana, Maksimovic, Man, Mandic,
  Mangano, Mansell, Manske, Mantovani, Marchesoni, Marion, Márka, Márka,
  Markakis, Markosyan, Markowitz, Maros, Marquina, Marsh, Martelli, Martellini,
  Martin, Martin, Martynov, Marx, Mason, Massera, Masserot, Massinger,
  Masso-Reid, Mastrogiovanni, Matas, Matichard, Matone, Mavalvala, Mazumder,
  McCarthy, McClelland, McCormick, McCuller, McGuire, McIntyre, McIver,
  McManus, McNeill, McRae, McWilliams, Meacher, Meadors, Mehmet, Meidam,
  Mejuto-Villa, Melatos, Mendell, Mercer, Merilh, Merzougui, Meshkov,
  Messenger, Messick, Metzdorff, Meyers, Miao, Michel, Middleton, Mikhailov,
  Milano, Miller, Miller, Miller, Millhouse, Milovich-Goff, Minazzoli,
  Minenkov, Ming, Mishra, Mitra, Mitrofanov, Mitselmakher, Mittleman, Moffa,
  Moggi, Mogushi, Mohan, Mohapatra, Molina, Montani, Moore, Moraru, Moreno,
  Morisaki, Morriss, Mours, Mow-Lowry, Mueller, Muir, Mukherjee, Mukherjee,
  Mukherjee, Mukund, Mullavey, Munch, Muñiz, Muratore, Murray, Nagar, Napier,
  Nardecchia, Naticchioni, Nayak, Neilson, Nelemans, Nelson, Nery, Neunzert,
  Nevin, Newport, Newton, Ng, Nguyen, Nguyen, Nichols, Nielsen, Nissanke, Nitz,
  Noack, Nocera, Nolting, North, Nuttall, Oberling, O'Dea, Ogin, Oh, Oh, Ohme,
  Okada, Oliver, Oppermann, Oram, O'Reilly, Ormiston, Ortega, O'Shaughnessy,
  Ossokine, Ottaway, Overmier, Owen, Pace, Page, Page, Pai, Pai, Palamos,
  Palashov, Palomba, Pal-Singh, Pan, Pan, Pang, Pang, Pankow, Pannarale, Pant,
  Paoletti, Paoli, Papa, Parida, Parker, Pascucci, Pasqualetti, Passaquieti,
  Passuello, Patil, Patricelli, Pearlstone, Pedraza, Pedurand, Pekowsky, Pele,
  Penn, Perez, Perreca, Perri, Pfeiffer, Phelps, Piccinni, Pichot,
  Piergiovanni, Pierro, Pillant, Pinard, Pinto, Pirello, Pitkin, Poe, Poggiani,
  Popolizio, Porter, Post, Powell, Prasad, Pratt, Pratten, Predoi, Prestegard,
  Prijatelj, Principe, Privitera, Prix, Prodi, Prokhorov, Puncken, Punturo,
  Puppo, Pürrer, Qi, Quetschke, Quintero, Quitzow-James, Raab, Rabeling,
  Radkins, Raffai, Raja, Rajan, Rajbhandari, Rakhmanov, Ramirez, Ramos-Buades,
  Rapagnani, Raymond, Razzano, Read, Regimbau, Rei, Reid, Reitze, Ren, Reyes,
  Ricci, Ricker, Rieger, Riles, Rizzo, Robertson, Robie, Robinet, Rocchi,
  Rolland, Rollins, Roma, Romano, Romano, Romel, Romie, Rosińska, Ross, Rowan,
  Rüdiger, Ruggi, Rutins, Ryan, Sachdev, Sadecki, Sadeghian, Sakellariadou,
  Salconi, Saleem, Salemi, Samajdar, Sammut, Sampson, Sanchez, Sanchez,
  Sanchis-Gual, Sandberg, Sanders, Sassolas, Sathyaprakash, Saulson, Sauter,
  Savage, Sawadsky, Schale, Scheel, Scheuer, Schmidt, Schmidt, Schnabel,
  Schofield, Schönbeck, Schreiber, Schuette, Schulte, Schutz, Schwalbe, Scott,
  Scott, Seidel, Sellers, Sengupta, Sentenac, Sequino, Sergeev, Shaddock,
  Shaffer, Shah, Shahriar, Shaner, Shao, Shapiro, Shawhan, Sheperd, Shoemaker,
  Shoemaker, Siellez, Siemens, Sieniawska, Sigg, Silva, Singer, Singh, Singhal,
  Sintes, Slagmolen, Smith, Smith, Smith, Somala, Son, Sonnenberg, Sorazu,
  Sorrentino, Souradeep, Spencer, Srivastava, Staats, Staley, Steinke,
  Steinlechner, Steinlechner, Steinmeyer, Stevenson, Stone, Stops, Strain,
  Stratta, Strigin, Strunk, Sturani, Stuver, Summerscales, Sun, Sunil, Suresh,
  Sutton, Swinkels, Szczepańczyk, Tacca, Tait, Talbot, Talukder, Tanner,
  Tápai, Taracchini, Tasson, Taylor, Taylor, Tewari, Theeg, Thies, Thomas,
  Thomas, Thomas, Thorne, Thorne, Thrane, Tiwari, Tiwari, Tokmakov, Toland,
  Tonelli, Tornasi, Torres-Forné, Torrie, Töyrä, Travasso, Traylor,
  Trinastic, Tringali, Trozzo, Tsang, Tse, Tso, Tsukada, Tsuna, Tuyenbayev,
  Ueno, Ugolini, Unnikrishnan, Urban, Usman, Vahlbruch, Vajente, Valdes,
  Vallisneri, van Bakel, van Beuzekom, van~den Brand, Van Den~Broeck,
  Vander-Hyde, van~der Schaaf, van Heijningen, van Veggel, Vardaro, Varma,
  Vass, Vasúth, Vecchio, Vedovato, Veitch, Veitch, Venkateswara, Venugopalan,
  Verkindt, Vetrano, Viceré, Viets, Vinciguerra, Vine, Vinet, Vitale, Vo,
  Vocca, Vorvick, Vyatchanin, Wade, Wade, Wade, Walet, Walker, Wallace, Walsh,
  Wang, Wang, Wang, Wang, Wang, Ward, Warner, Was, Watchi, Weaver, Wei,
  Weinert, Weinstein, Weiss, Wen, Wessel, Weßels, Westerweck, Westphal, Wette,
  Whelan, Whitcomb, Whiting, Whittle, Wilken, Williams, Williams, Williamson,
  Willis, Willke, Wimmer, Winkler, Wipf, Wittel, Woan, Woehler, Wofford, Wong,
  Worden, Wright, Wu, Wysocki, Xiao, Yamamoto, Yancey, Yang, Yap, Yazback, Yu,
  Yu, Yvert, ZadroŻny, Zanolin, Zelenova, Zendri, Zevin, Zhang, Zhang, Zhang,
  Zhang, Zhao, Zhou, Zhou, Zhu, Zhu, Zimmerman, Zucker, Zweizig, {LIGO
  Scientific Collaboration}, \& {Virgo Collaboration}}]{abbott_gw170817_2017}
---. 2017, Physical Review Letters, 119, 161101,
  \dodoi{10.1103/PhysRevLett.119.161101}

\bibitem[{Abbott {et~al.}(2019)Abbott, Abbott, Abbott, Abraham, Acernese,
  Ackley, Adams, Adhikari, Adya, Affeldt, Agathos, Agatsuma, Aggarwal, Aguiar,
  Aiello, Ain, Ajith, Allen, Allocca, Aloy, Altin, Amato, Ananyeva, Anderson,
  Anderson, Angelova, Antier, Appert, Arai, Araya, Areeda, Arène, Arnaud,
  Arun, Ascenzi, Ashton, Aston, Astone, Aubin, Aufmuth, AultONeal, Austin,
  Avendano, Avila-Alvarez, Babak, Bacon, Badaracco, Bader, Bae, Baker,
  Baldaccini, Ballardin, Ballmer, Banagiri, Barayoga, Barclay, Barish, Barker,
  Barkett, Barnum, Barone, Barr, Barsotti, Barsuglia, Barta, Bartlett, Bartos,
  Bassiri, Basti, Bawaj, Bayley, Bazzan, Bécsy, Bejger, Belahcene, Bell,
  Beniwal, Berger, Bergmann, Bernuzzi, Bero, Berry, Bersanetti, Bertolini,
  Betzwieser, Bhandare, Bidler, Bilenko, Bilgili, Billingsley, Birch, Birney,
  Birnholtz, Biscans, Biscoveanu, Bisht, Bitossi, Bizouard, Blackburn,
  Blackman, Blair, Blair, Blair, Bloemen, Bode, Boer, Boetzel, Bogaert, Bondu,
  Bonilla, Bonnand, Booker, Boom, Booth, Bork, Boschi, Bose, Bossie, Bossilkov,
  Bosveld, Bouffanais, Bozzi, Bradaschia, Brady, Bramley, Branchesi, Brau,
  Briant, Briggs, Brighenti, Brillet, Brinkmann, Brisson, Brockill, Brooks,
  Brown, Brunett, Buikema, Bulik, Bulten, Buonanno, Buskulic,
  Bustamante~Rosell, Buy, Byer, Cabero, Cadonati, Cagnoli, Cahillane,
  Calderón~Bustillo, Callister, Calloni, Camp, Campbell, Canepa, Cannon, Cao,
  Cao, Capocasa, Carbognani, Caride, Carney, Carullo, Casanueva~Diaz,
  Casentini, Caudill, Cavaglià, Cavalier, Cavalieri, Cella, Cerdá-Durán,
  Cerretani, Cesarini, Chaibi, Chakravarti, Chamberlin, Chan, Chao, Charlton,
  Chase, Chassande-Mottin, Chatterjee, Chaturvedi, Chatziioannou, Cheeseboro,
  Chen, Chen, Chen, Cheng, Cheong, Chia, Chincarini, Chiummo, Cho, Cho, Cho,
  Christensen, Chu, Chua, Chung, Chung, Ciani, Ciobanu, Ciolfi, Cipriano,
  Cirone, Clara, Clark, Clearwater, Cleva, Cocchieri, Coccia, Cohadon, Cohen,
  Colgan, Colleoni, Collette, Collins, Cominsky, Constancio, Conti, Cooper,
  Corban, Corbitt, Cordero-Carrión, Corley, Cornish, Corsi, Cortese, Costa,
  Cotesta, Coughlin, Coughlin, Coulon, Countryman, Couvares, Covas, Cowan,
  Coward, Cowart, Coyne, Coyne, Creighton, Creighton, Cripe, Croquette,
  Crowder, Cullen, Cumming, Cunningham, Cuoco, Canton, Dálya, Danilishin,
  D'Antonio, Danzmann, Dasgupta, Da~Silva~Costa, Datrier, Dattilo, Dave,
  Davier, Davis, Daw, DeBra, Deenadayalan, Degallaix, De~Laurentis, Deléglise,
  Del~Pozzo, DeMarchi, Demos, Dent, De~Pietri, Derby, De~Rosa, De~Rossi,
  DeSalvo, de~Varona, Dhurandhar, Díaz, Dietrich, Di~Fiore, Di~Giovanni,
  Di~Girolamo, Di~Lieto, Ding, Di~Pace, Di~Palma, Di~Renzo, Dmitriev, Doctor,
  Donovan, Dooley, Doravari, Dorrington, Downes, Drago, Driggers, Du, Ducoin,
  Dupej, Dwyer, Easter, Edo, Edwards, Effler, Ehrens, Eichholz, Eikenberry,
  Eisenmann, Eisenstein, Essick, Estelles, Estevez, Etienne, Etzel, Evans,
  Evans, Fafone, Fair, Fairhurst, Fan, Farinon, Farr, Farr, Fauchon-Jones,
  Favata, Fays, Fazio, Fee, Feicht, Fejer, Feng, Fernandez-Galiana, Ferrante,
  Ferreira, Ferreira, Ferrini, Fidecaro, Fiori, Fiorucci, Fishbach, Fisher,
  Fishner, Fitz-Axen, Flaminio, Fletcher, Flynn, Fong, Font, Forsyth, Fournier,
  Frasca, Frasconi, Frei, Freise, Frey, Frey, Fritschel, Frolov, Fulda, Fyffe,
  Gabbard, Gadre, Gaebel, Gair, Gammaitoni, Ganija, Gaonkar, Garcia,
  García-Quirós, Garufi, Gateley, Gaudio, Gaur, Gayathri, Gemme, Genin,
  Gennai, George, George, Gergely, Germain, Ghonge, Ghosh, Ghosh, Ghosh,
  Giacomazzo, Giaime, Giardina, Giazotto, Gill, Giordano, Glover, Godwin,
  Goetz, Goetz, Goncharov, González, Gonzalez~Castro, Gopakumar, Gorodetsky,
  Gossan, Gosselin, Gouaty, Grado, Graef, Granata, Grant, Gras, Grassia, Gray,
  Gray, Greco, Green, Green, Gretarsson, Groot, Grote, Grunewald, Gruning,
  Guidi, Gulati, Guo, Gupta, Gupta, Gustafson, Gustafson, Haegel, Halim, Hall,
  Hall, Hamilton, Hammond, Haney, Hanke, Hanks, Hanna, Hannam, Hannuksela,
  Hanson, Hardwick, Haris, Harms, Harry, Harry, Haster, Haughian, Hayes, Healy,
  Heidmann, Heintze, Heitmann, Hello, Hemming, Hendry, Heng, Hennig,
  Heptonstall, Hernandez~Vivanco, Heurs, Hild, Hinderer, Hoak, Hochheim,
  Hofman, Holgado, Holland, Holt, Holz, Hopkins, Horst, Hough, Howell, Hoy,
  Hreibi, Huang, Huerta, Huet, Hughey, Hulko, Husa, Huttner, Huynh-Dinh,
  Idzkowski, Iess, Ingram, Inta, Intini, Irwin, Isa, Isac, Isi, Iyer, Izumi,
  Jacqmin, Jadhav, Jani, Janthalur, Jaranowski, Jenkins, Jiang, Johnson,
  Johnson-McDaniel, Jones, Jones, Jones, Jonker, Ju, Junker, Kalaghatgi,
  Kalogera, Kamai, Kandhasamy, Kang, Kanner, Kapadia, Karki, Karvinen, Kashyap,
  Kasprzack, Katsanevas, Katsavounidis, Katzman, Kaufer, Kawabe, Keerthana,
  Kéfélian, Keitel, Kennedy, Key, Khalili, Khan, Khan, Khan, Khan, Khazanov,
  Khursheed, Kijbunchoo, Kim, Kim, Kim, Kim, Kim, Kim, Kimball, King, King,
  Kinley-Hanlon, Kirchhoff, Kissel, Kleybolte, Klika, Klimenko, Knowles, Koch,
  Koehlenbeck, Koekoek, Koley, Kondrashov, Kontos, Koper, Korobko, Korth,
  Kowalska, Kozak, Kringel, Krishnendu, Królak, Kuehn, Kumar, Kumar, Kumar,
  Kumar, Kuo, Kutynia, Kwang, Lackey, Lai, Lam, Landry, Lane, Lang, Lange,
  Lantz, Lanza, Lartaux-Vollard, Lasky, Laxen, Lazzarini, Lazzaro, Leaci,
  Leavey, Lecoeuche, Lee, Lee, Lee, Lee, Lee, Lee, Lehmann, Lenon, Leroy,
  Letendre, Levin, Li, Li, Li, Li, Lin, Linde, Linker, Littenberg, Liu, Liu,
  Lo, Lockerbie, London, Longo, Lorenzini, Loriette, Lormand, Losurdo, Lough,
  Lousto, Lovelace, Lower, Lück, Lumaca, Lundgren, Lynch, Ma, Macas, Macfoy,
  MacInnis, Macleod, Macquet, Magaña-Sandoval, Magaña~Zertuche, Magee,
  Majorana, Maksimovic, Malik, Man, Mandic, Mangano, Mansell, Manske,
  Mantovani, Marchesoni, Marion, Márka, Márka, Markakis, Markosyan,
  Markowitz, Maros, Marquina, Marsat, Martelli, Martin, Martin, Martynov,
  Mason, Massera, Masserot, Massinger, Masso-Reid, Mastrogiovanni, Matas,
  Matichard, Matone, Mavalvala, Mazumder, McCann, McCarthy, McClelland,
  McCormick, McCuller, McGuire, McIver, McManus, McRae, McWilliams, Meacher,
  Meadors, Mehmet, Mehta, Meidam, Melatos, Mendell, Mercer, Mereni, Merilh,
  Merzougui, Meshkov, Messenger, Messick, Metzdorff, Meyers, Miao, Michel,
  Middleton, Mikhailov, Milano, Miller, Miller, Millhouse, Mills,
  Milovich-Goff, Minazzoli, Minenkov, Mishkin, Mishra, Mistry, Mitra,
  Mitrofanov, Mitselmakher, Mittleman, Mo, Moffa, Mogushi, Mohapatra, Montani,
  Moore, Moraru, Moreno, Morisaki, Mours, Mow-Lowry, Mukherjee, Mukherjee,
  Mukherjee, Mukund, Mullavey, Munch, Muñiz, Muratore, Murray, Nagar,
  Nardecchia, Naticchioni, Nayak, Neilson, Nelemans, Nelson, Nery, Neunzert,
  Ng, Ng, Nguyen, Nichols, Nielsen, Nissanke, Nitz, Nocera, North, Nuttall,
  Obergaulinger, Oberling, O'Brien, O'Dea, Ogin, Oh, Oh, Ohme, Ohta, Okada,
  Oliver, Oppermann, Oram, O'Reilly, Ormiston, Ortega, O'Shaughnessy, Ossokine,
  Ottaway, Overmier, Owen, Pace, Pagano, Page, Pai, Pai, Palamos, Palashov,
  Palomba, Pal-Singh, Pan, Pang, Pang, Pankow, Pannarale, Pant, Paoletti,
  Paoli, Papa, Parida, Parker, Pascucci, Pasqualetti, Passaquieti, Passuello,
  Patil, Patricelli, Pearlstone, Pedersen, Pedraza, Pedurand, Pele, Penn,
  Perego, Perez, Perreca, Pfeiffer, Phelps, Phukon, Piccinni, Pichot,
  Piergiovanni, Pillant, Pinard, Pirello, Pitkin, Poggiani, Pong, Ponrathnam,
  Popolizio, Porter, Powell, Prajapati, Prasad, Prasai, Prasanna, Pratten,
  Prestegard, Privitera, Prodi, Prokhorov, Puncken, Punturo, Puppo, Pürrer,
  Qi, Quetschke, Quinonez, Quintero, Quitzow-James, Raab, Radkins, Radulescu,
  Raffai, Raja, Rajan, Rajbhandari, Rakhmanov, Ramirez, Ramos-Buades, Rana,
  Rao, Rapagnani, Raymond, Razzano, Read, Regimbau, Rei, Reid, Reitze, Ren,
  Ricci, Richardson, Richardson, Ricker, Riemenschneider, Riles, Rizzo,
  Robertson, Robie, Robinet, Rocchi, Rolland, Rollins, Roma, Romanelli, Romano,
  Romel, Romie, Rose, Rosińska, Rosofsky, Ross, Rowan, Rüdiger, Ruggi,
  Rutins, Ryan, Sachdev, Sadecki, Sakellariadou, Salafia, Salconi, Saleem,
  Salemi, Samajdar, Sammut, Sanchez, Sanchez, Sanchis-Gual, Sandberg, Sanders,
  Santiago, Sarin, Sassolas, Sathyaprakash, Saulson, Sauter, Savage, Schale,
  Scheel, Scheuer, Schmidt, Schnabel, Schofield, Schönbeck, Schreiber,
  Schulte, Schutz, Schwalbe, Scott, Scott, Seidel, Sellers, Sengupta, Sennett,
  Sentenac, Sequino, Sergeev, Setyawati, Shaddock, Shaffer, Shahriar, Shaner,
  Shao, Sharma, Shawhan, Shen, Shink, Shoemaker, Shoemaker, ShyamSundar,
  Siellez, Sieniawska, Sigg, Silva, Singer, Singh, Singhal, Sintes,
  Sitmukhambetov, Skliris, Slagmolen, Slaven-Blair, Smith, Smith, Somala, Son,
  Sorazu, Sorrentino, Souradeep, Sowell, Spencer, Srivastava, Srivastava,
  Staats, Stachie, Standke, Steer, Steinke, Steinlechner, Steinlechner,
  Steinmeyer, Stevenson, Stocks, Stone, Stops, Strain, Stratta, Strigin,
  Strunk, Sturani, Stuver, Sudhir, Summerscales, Sun, Sunil, Suresh, Sutton,
  Swinkels, Szczepańczyk, Tacca, Tait, Talbot, Talukder, Tanner, Tápai,
  Taracchini, Tasson, Taylor, Thies, Thomas, Thomas, Thondapu, Thorne, Thrane,
  Tiwari, Tiwari, Tiwari, Toland, Tonelli, Tornasi, Torres-Forné, Torrie,
  Töyrä, Travasso, Traylor, Tringali, Trovato, Trozzo, Trudeau, Tsang, Tse,
  Tso, Tsukada, Tsuna, Tuyenbayev, Ueno, Ugolini, Unnikrishnan, Urban, Usman,
  Vahlbruch, Vajente, Valdes, van Bakel, van Beuzekom, van~den Brand, Van
  Den~Broeck, Vander-Hyde, van Heijningen, van~der Schaaf, van Veggel, Vardaro,
  Varma, Vass, Vasúth, Vecchio, Vedovato, Veitch, Veitch, Venkateswara,
  Venugopalan, Verkindt, Vetrano, Viceré, Viets, Vine, Vinet, Vitale, Vo,
  Vocca, Vorvick, Vyatchanin, Wade, Wade, Wade, Walet, Walker, Wallace, Walsh,
  Wang, Wang, Wang, Wang, Wang, Ward, Warden, Warner, Was, Watchi, Weaver, Wei,
  Weinert, Weinstein, Weiss, Wellmann, Wen, Wessel, Weßels, Westhouse, Wette,
  Whelan, White, Whiting, Whittle, Wilken, Williams, Williamson, Willis,
  Willke, Wimmer, Winkler, Wipf, Wittel, Woan, Woehler, Wofford, Worden,
  Wright, Wu, Wysocki, Xiao, Yamamoto, Yancey, Yang, Yap, Yazback, Yeeles, Yu,
  Yu, Yuen, Yvert, ZadroŻny, Zanolin, Zappa, Zelenova, Zendri, Zevin, Zhang,
  Zhang, Zhang, Zhao, Zhou, Zhou, Zhu, Zimmerman, Zlochower, Zucker, Zweizig,
  {LIGO Scientific Collaboration}, \& {Virgo
  Collaboration}}]{abbott_gwtc-1_2019}
---. 2019, Physical Review X, 9, 031040, \dodoi{10.1103/PhysRevX.9.031040}

\bibitem[{Abbott {et~al.}(2020{\natexlab{a}})Abbott, Abbott, Abbott, Abraham,
  Acernese, Ackley, Adams, Adhikari, Adya, Affeldt, Agathos, Agatsuma,
  Aggarwal, Aguiar, Aiello, Ain, Ajith, Allen, Allocca, Aloy, Altin, Amato,
  Anand, Ananyeva, Anderson, Anderson, Angelova, Antier, Appert, Arai, Araya,
  Areeda, Arène, Arnaud, Aronson, Arun, Ascenzi, Ashton, Aston, Astone, Aubin,
  Aufmuth, AultONeal, Austin, Avendano, Avila-Alvarez, Babak, Bacon, Badaracco,
  Bader, Bae, Baird, Baker, Baldaccini, Ballardin, Ballmer, Bals, Banagiri,
  Barayoga, Barbieri, Barclay, Barish, Barker, Barkett, Barnum, Barone, Barr,
  Barsotti, Barsuglia, Barta, Bartlett, Bartos, Bassiri, Basti, Bawaj, Bayley,
  Baylor, Bazzan, Bécsy, Bejger, Belahcene, Bell, Beniwal, Benjamin, Berger,
  Bergmann, Bernuzzi, Berry, Bersanetti, Bertolini, Betzwieser, Bhandare,
  Bidler, Biggs, Bilenko, Bilgili, Billingsley, Birney, Birnholtz, Biscans,
  Bischi, Biscoveanu, Bisht, Bitossi, Bizouard, Blackburn, Blackman, Blair,
  Blair, Blair, Bloemen, Bobba, Bode, Boer, Boetzel, Bogaert, Bondu, Bonnand,
  Booker, Boom, Bork, Boschi, Bose, Bossilkov, Bosveld, Bouffanais, Bozzi,
  Bradaschia, Brady, Bramley, Branchesi, Brau, Breschi, Briant, Briggs,
  Brighenti, Brillet, Brinkmann, Brockill, Brooks, Brooks, Brown, Brunett,
  Buikema, Bulik, Bulten, Buonanno, Buskulic, Buy, Byer, Cabero, Cadonati,
  Cagnoli, Cahillane, Bustillo, Callister, Calloni, Camp, Campbell, Canepa,
  Cannon, Cao, Cao, Carapella, Carbognani, Caride, Carney, Carullo, Diaz,
  Casentini, Caudill, Cavaglià, Cavalier, Cavalieri, Cella, Cerdá-Durán,
  Cesarini, Chaibi, Chakravarti, Chamberlin, Chan, Chao, Charlton, Chase,
  Chassande-Mottin, Chatterjee, Chaturvedi, Chatziioannou, Cheeseboro, Chen,
  Chen, Chen, Cheng, Cheong, Chia, Chiadini, Chincarini, Chiummo, Cho, Cho,
  Cho, Christensen, Chu, Chua, Chung, Chung, Ciani, Cieślar, Ciobanu, Ciolfi,
  Cipriano, Cirone, Clara, Clark, Clearwater, Cleva, Coccia, Cohadon, Cohen,
  Colleoni, Collette, Collins, Colpi, Cominsky, Constancio, Conti, Cooper,
  Corban, Corbitt, Cordero-Carrión, Corezzi, Corley, Cornish, Corre, Corsi,
  Cortese, Costa, Cotesta, Coughlin, Coughlin, Coulon, Countryman, Couvares,
  Covas, Cowan, Coward, Cowart, Coyne, Coyne, Creighton, Creighton, Cripe,
  Croquette, Crowder, Cullen, Cumming, Cunningham, Cuoco, Canton, Dálya,
  D'Angelo, Danilishin, D'Antonio, Danzmann, Dasgupta, Costa, Datrier, Dattilo,
  Dave, Davier, Davis, Daw, DeBra, Deenadayalan, Degallaix, Laurentis,
  Deléglise, Lillo, Pozzo, DeMarchi, Demos, Dent, Pietri, Rosa, Rossi,
  DeSalvo, Varona, Dhurandhar, Díaz, Dietrich, Fiore, DiFronzo, Giorgio,
  Giovanni, Giovanni, Girolamo, Lieto, Ding, Pace, Palma, Renzo, Divakarla,
  Dmitriev, Doctor, Donovan, Dooley, Doravari, Dorrington, Downes, Drago,
  Driggers, Du, Ducoin, Dudi, Dupej, Durante, Dwyer, Easter, Eddolls, Edo,
  Effler, Ehrens, Eichholz, Eikenberry, Eisenmann, Eisenstein, Errico, Essick,
  Estelles, Estevez, Etienne, Etzel, Evans, Evans, Fafone, Fairhurst, Fan,
  Farinon, Farr, Farr, Fauchon-Jones, Favata, Fays, Fazio, Fee, Feicht, Fejer,
  Feng, Fernandez-Galiana, Ferrante, Ferreira, Ferreira, Fidecaro, Fiori,
  Fiorucci, Fishbach, Fisher, Fishner, Fittipaldi, Fitz-Axen, Fiumara,
  Flaminio, Fletcher, Floden, Flynn, Fong, Font, Forsyth, Fournier, Vivanco,
  Frasca, Frasconi, Frei, Freise, Frey, Frey, Fritschel, Frolov, Fronzè,
  Fulda, Fyffe, Gabbard, Gadre, Gaebel, Gair, Gamba, Gammaitoni, Gaonkar,
  García-Quirós, Garufi, Gateley, Gaudio, Gaur, Gayathri, Gemme, Genin,
  Gennai, George, George, George, Gergely, Ghonge, Ghosh, Ghosh, Ghosh,
  Giacomazzo, Giaime, Giardina, Gibson, Gill, Glover, Gniesmer, Godwin, Goetz,
  Goetz, Goncharov, González, Castro, Gopakumar, Gossan, Gosselin, Gouaty,
  Grace, Grado, Granata, Grant, Gras, Grassia, Gray, Gray, Greco, Green, Green,
  Gretarsson, Grimaldi, Grimm, Groot, Grote, Grunewald, Gruning, Guidi, Gulati,
  Guo, Gupta, Gupta, Gupta, Gustafson, Gustafson, Haegel, Halim, Hall, Hall,
  Hamilton, Hammond, Haney, Hanke, Hanks, Hanna, Hannam, Hannuksela, Hansen,
  Hanson, Harder, Hardwick, Haris, Harms, Harry, Harry, Hasskew, Haster,
  Haughian, Hayes, Healy, Heidmann, Heintze, Heitmann, Hellman, Hello, Hemming,
  Hendry, Heng, Hennig, Heurs, Hild, Hinderer, Ho, Hochheim, Hofman, Holgado,
  Holland, Holt, Holz, Hopkins, Horst, Hough, Howell, Hoy, Huang, Hübner,
  Huerta, Huet, Hughey, Hui, Husa, Huttner, Huynh-Dinh, Idzkowski, Iess,
  Inchauspe, Ingram, Inta, Intini, Irwin, Isa, Isac, Isi, Iyer, Jacqmin,
  Jadhav, Jani, Janthalur, Jaranowski, Jariwala, Jenkins, Jiang, Johnson,
  Johnson-McDaniel, Jones, Jones, Jones, Jones, Jonker, Ju, Junker, Kalaghatgi,
  Kalogera, Kamai, Kandhasamy, Kang, Kanner, Kapadia, Karki, Kashyap,
  Kasprzack, Kastaun, Katsanevas, Katsavounidis, Katzman, Kaufer, Kawabe,
  Keerthana, Kéfélian, Keitel, Kennedy, Key, Khalili, Khan, Khan, Khazanov,
  Khetan, Khursheed, Kijbunchoo, Kim, Kim, Kim, Kim, Kim, Kim, Kimball, King,
  Kinley-Hanlon, Kirchhoff, Kissel, Kleybolte, Klika, Klimenko, Knowles, Koch,
  Koehlenbeck, Koekoek, Koley, Kondrashov, Kontos, Koper, Korobko, Korth,
  Kovalam, Kozak, Krämer, Kringel, Krishnendu, Królak, Krupinski, Kuehn,
  Kumar, Kumar, Kumar, Kumar, Kuo, Kutynia, Kwang, Lackey, Laghi, Lai, Lam,
  Landry, Landry, Lane, Lang, Lange, Lantz, Lanza, Lartaux-Vollard, Lasky,
  Laxen, Lazzarini, Lazzaro, Leaci, Leavey, Lecoeuche, Lee, Lee, Lee, Lee, Lee,
  Lee, Lehmann, Lenon, Leroy, Letendre, Levin, Li, Li, Li, Li, Li, Lin, Linde,
  Linker, Littenberg, Liu, Liu, Llorens-Monteagudo, Lo, London, Longo,
  Lorenzini, Loriette, Lormand, Losurdo, Lough, Lousto, Lovelace, Lower,
  Lucaccioni, Lück, Lumaca, Lundgren, Lynch, Ma, Macas, Macfoy, MacInnis,
  Macleod, Macquet, Hernandez, Magaña-Sandoval, Magee, Majorana, Maksimovic,
  Malik, Man, Mandic, Mangano, Mansell, Manske, Mantovani, Mapelli, Marchesoni,
  Marion, Márka, Márka, Markakis, Markosyan, Markowitz, Maros, Marquina,
  Marsat, Martelli, Martin, Martin, Martinez, Martynov, Masalehdan, Mason,
  Massera, Masserot, Massinger, Masso-Reid, Mastrogiovanni, Matas, Matichard,
  Matone, Mavalvala, McCann, McCarthy, McClelland, McCormick, McCuller,
  McGuire, McIsaac, McIver, McManus, McRae, McWilliams, Meacher, Meadors,
  Mehmet, Mehta, Meidam, Villa, Melatos, Mendell, Mercer, Mereni, Merfeld,
  Merilh, Merzougui, Meshkov, Messenger, Messick, Messina, Metzdorff, Meyers,
  Meylahn, Miani, Miao, Michel, Middleton, Milano, Miller, Millhouse, Mills,
  Milovich-Goff, Minazzoli, Minenkov, Mishkin, Mishra, Mistry, Mitra,
  Mitrofanov, Mitselmakher, Mittleman, Mo, Moffa, Mogushi, Mohapatra,
  Molina-Ruiz, Mondin, Montani, Moore, Moraru, Morawski, Moreno, Morisaki,
  Mours, Mow-Lowry, Muciaccia, Mukherjee, Mukherjee, Mukherjee, Mukherjee,
  Mukund, Mullavey, Munch, Muñiz, Muratore, Murray, Nagar, Nardecchia,
  Naticchioni, Nayak, Neil, Neilson, Nelemans, Nelson, Nery, Neunzert, Nevin,
  Ng, Ng, Nguyen, Nguyen, Nichols, Nichols, Nissanke, Nocera, North, Nuttall,
  Obergaulinger, Oberling, O'Brien, Oganesyan, Ogin, Oh, Oh, Ohme, Ohta, Okada,
  Oliver, Oppermann, Oram, O'Reilly, Ormiston, Ortega, O'Shaughnessy, Ossokine,
  Ottaway, Overmier, Owen, Pace, Pagano, Page, Pagliaroli, Pai, Pai, Palamos,
  Palashov, Palomba, Pan, Panda, Pang, Pankow, Pannarale, Pant, Paoletti,
  Paoli, Parida, Parker, Pascucci, Pasqualetti, Passaquieti, Passuello, Patil,
  Patricelli, Payne, Pearlstone, Pechsiri, Pedersen, Pedraza, Pedurand, Pele,
  Penn, Perego, Perez, Périgois, Perreca, Petermann, Pfeiffer, Phelps, Phukon,
  Piccinni, Pichot, Piergiovanni, Pierro, Pillant, Pinard, Pinto, Pirello,
  Pitkin, Plastino, Poggiani, Pong, Ponrathnam, Popolizio, Porter, Powell,
  Prajapati, Prasad, Prasai, Prasanna, Pratten, Prestegard, Principe, Prodi,
  Prokhorov, Punturo, Puppo, Pürrer, Qi, Quetschke, Quinonez, Raab,
  Raaijmakers, Radkins, Radulesco, Raffai, Raja, Rajan, Rajbhandari, Rakhmanov,
  Ramirez, Ramos-Buades, Rana, Rao, Rapagnani, Raymond, Razzano, Read,
  Regimbau, Rei, Reid, Reitze, Rettegno, Ricci, Richardson, Richardson, Ricker,
  Riemenschneider, Riles, Rizzo, Robertson, Robinet, Rocchi, Rolland, Rollins,
  Roma, Romanelli, Romano, Romel, Romie, Rose, Rose, Rose, Rosell, Rosińska,
  Rosofsky, Ross, Rowan, Roy, Rüdiger, Ruggi, Rutins, Ryan, Sachdev, Sadecki,
  Sakellariadou, Salafia, Salconi, Saleem, Samajdar, Sammut, Sanchez, Sanchez,
  Sanchis-Gual, Sanders, Santiago, Santos, Sarin, Sassolas, Sathyaprakash,
  Sauter, Savage, Schale, Scheel, Scheuer, Schmidt, Schnabel, Schofield,
  Schönbeck, Schreiber, Schulte, Schutz, Scott, Scott, Seidel, Sellers,
  Sengupta, Sennett, Sentenac, Sequino, Sergeev, Setyawati, Shaddock, Shaffer,
  Shahriar, Shaner, Sharma, Sharma, Shawhan, Shen, Shink, Shoemaker, Shoemaker,
  Shukla, ShyamSundar, Siellez, Sieniawska, Sigg, Singer, Singh, Singh,
  Singhal, Sintes, Sitmukhambetov, Skliris, Slagmolen, Slaven-Blair, Smith,
  Smith, Somala, Son, Soni, Sorazu, Sorrentino, Souradeep, Sowell, Spencer,
  Spera, Srivastava, Srivastava, Staats, Stachie, Standke, Steer, Steinke,
  Steinlechner, Steinlechner, Steinmeyer, Stevenson, Stocks, Stone, Stops,
  Strain, Stratta, Strigin, Strunk, Sturani, Stuver, Sudhir, Summerscales, Sun,
  Sunil, Sur, Suresh, Sutton, Swinkels, Szczepańczyk, Tacca, Tait, Talbot,
  Tanner, Tao, Tápai, Tapia, Tasson, Taylor, Tenorio, Terkowski, Thomas,
  Thomas, Thondapu, Thorne, Thrane, Tiwari, Tiwari, Tiwari, Toland, Tonelli,
  Tornasi, Torres-Forné, Torrie, Töyrä, Travasso, Traylor, Tringali,
  Tripathee, Trovato, Trozzo, Tsang, Tse, Tso, Tsukada, Tsuna, Tsutsui,
  Tuyenbayev, Ueno, Ugolini, Unnikrishnan, Urban, Usman, Vahlbruch, Vajente,
  Valdes, Valentini, Bakel, Beuzekom, Brand, Broeck, Vander-Hyde, Schaaf,
  VanHeijningen, Veggel, Vardaro, Varma, Vass, Vasúth, Vecchio, Vedovato,
  Veitch, Veitch, Venkateswara, Venugopalan, Verkindt, Vetrano, Viceré, Viets,
  Vinciguerra, Vine, Vinet, Vitale, Vo, Vocca, Vorvick, Vyatchanin, Wade, Wade,
  Wade, Walet, Walker, Wallace, Walsh, Wang, Wang, Wang, Wang, Ward, Warden,
  Warner, Was, Watchi, Weaver, Wei, Weinert, Weinstein, Weiss, Wellmann, Wen,
  Wessel, Weßels, Westhouse, Wette, Whelan, White, Whiting, Whittle, Wilken,
  Williams, Williamson, Willis, Willke, Winkler, Wipf, Wittel, Woan, Woehler,
  Wofford, Wright, Wu, Wysocki, Xiao, Xu, Yamamoto, Yancey, Yang, Yang, Yang,
  Yap, Yazback, Yeeles, Yu, Yu, Yuen, Zadro{\textbackslash}.zny,
  Zadro{\textbackslash}.zny, Zanolin, Zelenova, Zendri, Zevin, Zhang, Zhang,
  Zhang, Zhao, Zhao, Zhou, Zhou, Zhu, Zimmerman, Zucker, \&
  Zweizig}]{abbott_gw190425_2020}
---. 2020{\natexlab{a}}, The Astrophysical Journal Letters, 892, L3,
  \dodoi{10.3847/2041-8213/ab75f5}

\bibitem[{Abbott {et~al.}(2020{\natexlab{b}})Abbott, Abbott, Abbott, Abraham,
  Acernese, Ackley, Adams, Adya, Affeldt, Agathos, Agatsuma, Aggarwal, Aguiar,
  Aiello, Ain, Ajith, Akutsu, Allen, Allocca, Aloy, Altin, Amato, Ananyeva,
  Anderson, Anderson, Ando, Angelova, Antier, Appert, Arai, Arai, Arai, Araki,
  Araya, Araya, Areeda, Arène, Aritomi, Arnaud, Arun, Ascenzi, Ashton, Aso,
  Aston, Astone, Aubin, Aufmuth, AultONeal, Austin, Avendano, Avila-Alvarez,
  Babak, Bacon, Badaracco, Bader, Bae, Bae, Baiotti, Bajpai, Baker, Baldaccini,
  Ballardin, Ballmer, Banagiri, Barayoga, Barclay, Barish, Barker, Barkett,
  Barnum, Barone, Barr, Barsotti, Barsuglia, Barta, Bartlett, Barton, Bartos,
  Bassiri, Basti, Bawaj, Bayley, Bazzan, Bécsy, Bejger, Belahcene, Bell,
  Beniwal, Berger, Bergmann, Bernuzzi, Bero, Berry, Bersanetti, Bertolini,
  Betzwieser, Bhandare, Bidler, Bilenko, Bilgili, Billingsley, Birch, Birney,
  Birnholtz, Biscans, Biscoveanu, Bisht, Bitossi, Bizouard, Blackburn, Blair,
  Blair, Blair, Bloemen, Bode, Boer, Boetzel, Bogaert, Bondu, Bonilla, Bonnand,
  Booker, Boom, Booth, Bork, Boschi, Bose, Bossie, Bossilkov, Bosveld,
  Bouffanais, Bozzi, Bradaschia, Brady, Bramley, Branchesi, Brau, Briant,
  Briggs, Brighenti, Brillet, Brinkmann, Brisson, Brockill, Brooks, Brown,
  Brown, Brunett, Buikema, Bulik, Bulten, Buonanno, Buskulic, Buy, Byer,
  Cabero, Cadonati, Cagnoli, Cahillane, Bustillo, Callister, Calloni, Camp,
  Campbell, Canepa, Cannon, Cannon, Cao, Cao, Capocasa, Carbognani, Caride,
  Carney, Carullo, Diaz, Casentini, Caudill, Cavaglià, Cavalier, Cavalieri,
  Cella, Cerdá-Durán, Cerretani, Cesarini, Chaibi, Chakravarti, Chamberlin,
  Chan, Chan, Chao, Charlton, Chase, Chassande-Mottin, Chatterjee, Chaturvedi,
  Chatziioannou, Cheeseboro, Chen, Chen, Chen, Chen, Chen, Chen, Cheng, Cheong,
  Chia, Chincarini, Chiummo, Cho, Cho, Cho, Christensen, Chu, Chu, Chu, Chua,
  Chung, Chung, Ciani, Ciobanu, Ciolfi, Cipriano, Cirone, Clara, Clark,
  Clearwater, Cleva, Cocchieri, Coccia, Cohadon, Cohen, Colgan, Colleoni,
  Collette, Collins, Cominsky, Constancio, Conti, Cooper, Corban, Corbitt,
  Cordero-Carrión, Corley, Cornish, Corsi, Cortese, Costa, Cotesta, Coughlin,
  Coughlin, Coulon, Countryman, Couvares, Covas, Cowan, Coward, Cowart, Coyne,
  Coyne, Creighton, Creighton, Cripe, Croquette, Crowder, Cullen, Cumming,
  Cunningham, Cuoco, Canton, Dálya, Danilishin, D’Antonio, Danzmann,
  Dasgupta, Da~Silva~Costa, Datrier, Dattilo, Dave, Davier, Davis, Daw, DeBra,
  Deenadayalan, Degallaix, De~Laurentis, Deléglise, Pozzo, DeMarchi, Demos,
  Dent, De~Pietri, Derby, De~Rosa, De~Rossi, DeSalvo, de~Varona, Dhurandhar,
  Díaz, Dietrich, \& Fiore}]{abbott_prospects_2020}
---. 2020{\natexlab{b}}, Living Reviews in Relativity, 23, 3,
  \dodoi{10.1007/s41114-020-00026-9}

\bibitem[{Abbott {et~al.}(2020{\natexlab{c}})Abbott, Abbott, Abraham, Acernese,
  Ackley, Adams, Adhikari, Adya, Affeldt, Agathos, Agatsuma, Aggarwal, Aguiar,
  Aich, Aiello, Ain, Ajith, Akcay, Allen, Allocca, Altin, Amato, Anand,
  Ananyeva, Anderson, Anderson, Angelova, Ansoldi, Antier, Appert, Arai, Araya,
  Areeda, Arène, Arnaud, Aronson, Arun, Asali, Ascenzi, Ashton, Aston, Astone,
  Aubin, Aufmuth, AultONeal, Austin, Avendano, Babak, Bacon, Badaracco, Bader,
  Bae, Baer, Baird, Baldaccini, Ballardin, Ballmer, Bals, Balsamo, Baltus,
  Banagiri, Bankar, Bankar, Barayoga, Barbieri, Barish, Barker, Barkett,
  Barneo, Barone, Barr, Barsotti, Barsuglia, Barta, Bartlett, Bartos, Bassiri,
  Basti, Bawaj, Bayley, Bazzan, Bécsy, Bejger, Belahcene, Bell, Beniwal,
  Benjamin, Benkel, Bentley, Bergamin, Berger, Bergmann, Bernuzzi, Berry,
  Bersanetti, Bertolini, Betzwieser, Bhandare, Bhandari, Bidler, Biggs,
  Bilenko, Billingsley, Birney, Birnholtz, Biscans, Bischi, Biscoveanu, Bisht,
  Bissenbayeva, Bitossi, Bizouard, Blackburn, Blackman, Blair, Blair, Blair,
  Bobba, Bode, Boer, Boetzel, Bogaert, Bondu, Bonilla, Bonnand, Booker, Boom,
  Bork, Boschi, Bose, Bossilkov, Bosveld, Bouffanais, Bozzi, Bradaschia, Brady,
  Bramley, Branchesi, Brau, Breschi, Briant, Briggs, Brighenti, Brillet,
  Brinkmann, Brito, Brockill, Brooks, Brooks, Brown, Brunett, Bruno, Bruntz,
  Buikema, Bulik, Bulten, Buonanno, Buskulic, Byer, Cabero, Cadonati, Cagnoli,
  Cahillane, Bustillo, Callaghan, Callister, Calloni, Camp, Canepa, Cannon,
  Cao, Cao, Carapella, Carbognani, Caride, Carney, Carullo, Diaz, Casentini,
  Castañeda, Caudill, Cavaglià, Cavalier, Cavalieri, Cella, Cerdá-Durán,
  Cesarini, Chaibi, Chakravarti, Chan, Chan, Chao, Charlton, Chase,
  Chassande-Mottin, Chatterjee, Chaturvedi, Chatziioannou, Chen, Chen, Chen,
  Cheng, Cheong, Chia, Chiadini, Chierici, Chincarini, Chiummo, Cho, Cho, Cho,
  Christensen, Chu, Chua, Chung, Chung, Ciani, Ciecielag, Cieślar, Ciobanu,
  Ciolfi, Cipriano, Cirone, Clara, Clark, Clearwater, Clesse, Cleva, Coccia,
  Cohadon, Cohen, Colleoni, Collette, Collins, Colpi, Constancio, Conti,
  Cooper, Corban, Corbitt, Cordero-Carrión, Corezzi, Corley, Cornish, Corre,
  Corsi, Cortese, Costa, Cotesta, Coughlin, Coughlin, Coulon, Countryman,
  Couvares, Covas, Coward, Cowart, Coyne, Coyne, Creighton, Creighton, Cripe,
  Croquette, Crowder, Cudell, Cullen, Cumming, Cummings, Cunningham, Cuoco,
  Curylo, Canton, Dálya, Dana, Daneshgaran-Bajastani, D'Angelo, Danilishin,
  D'Antonio, Danzmann, Darsow-Fromm, Dasgupta, Datrier, Dattilo, Dave, Davier,
  Davies, Davis, Daw, DeBra, Deenadayalan, Degallaix, Laurentis, Deléglise,
  Delfavero, Lillo, Pozzo, DeMarchi, D'Emilio, Demos, Dent, Pietri, Rosa,
  Rossi, DeSalvo, Varona, Dhurandhar, Díaz, Diaz-Ortiz, Dietrich, Fiore,
  Fronzo, Giorgio, Giovanni, Giovanni, Girolamo, Lieto, Ding, Pace, Palma,
  Renzo, Divakarla, Dmitriev, Doctor, Donovan, Dooley, Doravari, Dorrington,
  Downes, Drago, Driggers, Du, Ducoin, Dupej, Durante, D'Urso, Dwyer, Easter,
  Eddolls, Edelman, Edo, Edy, Effler, Ehrens, Eichholz, Eikenberry, Eisenmann,
  Eisenstein, Ejlli, Errico, Essick, Estelles, Estevez, Etienne, Etzel, Evans,
  Evans, Ewing, Fafone, Fairhurst, Fan, Farinon, Farr, Farr, Fauchon-Jones,
  Favata, Fays, Fazio, Feicht, Fejer, Feng, Fenyvesi, Ferguson,
  Fernandez-Galiana, Ferrante, Ferreira, Ferreira, Fidecaro, Fiori, Fiorucci,
  Fishbach, Fisher, Fittipaldi, Fitz-Axen, Fiumara, Flaminio, Floden, Flynn,
  Fong, Font, Forsyth, Fournier, Frasca, Frasconi, Frei, Freise, Frey, Frey,
  Fritschel, Frolov, Fronzè, Fulda, Fyffe, Gabbard, Gadre, Gaebel, Gair,
  Galaudage, Ganapathy, Ganguly, Gaonkar, García-Quirós, Garufi, Gateley,
  Gaudio, Gayathri, Gemme, Genin, Gennai, George, George, Gergely, Ghonge,
  Ghosh, Ghosh, Ghosh, Giacomazzo, Giaime, Giardina, Gibson, Gier, Gill,
  Glanzer, Gniesmer, Godwin, Goetz, Goetz, Gohlke, Goncharov, González,
  Gopakumar, Gossan, Gosselin, Gouaty, Grace, Grado, Granata, Grant, Gras,
  Grassia, Gray, Gray, Greco, Green, Green, Gretarsson, Griggs, Grignani,
  Grimaldi, Grimm, Grote, Grunewald, Gruning, Guidi, Guimaraes, Guixé, Gulati,
  Guo, Gupta, Gupta, Gupta, Gustafson, Gustafson, Haegel, Halim, Hall,
  Hamilton, Hammond, Haney, Hanke, Hanks, Hanna, Hannam, Hannuksela, Hansen,
  Hanson, Harder, Hardwick, Haris, Harms, Harry, Harry, Hasskew, Haster,
  Haughian, Hayes, Healy, Heidmann, Heintze, Heinze, Heitmann, Hellman, Hello,
  Hemming, Hendry, Heng, Hennes, Hennig, Heurs, Hild, Hinderer, Hoback,
  Hochheim, Hofgard, Hofman, Holgado, Holland, Holt, Holz, Hopkins, Horst,
  Hough, Howell, Hoy, Huang, Hübner, Huerta, Huet, Hughey, Hui, Husa, Huttner,
  Huxford, Huynh-Dinh, Idzkowski, Iess, Inchauspe, Ingram, Intini, Isac, Isi,
  Iyer, Jacqmin, Jadhav, Jadhav, James, Jani, Janthalur, Jaranowski, Jariwala,
  Jaume, Jenkins, Jiang, Johns, Johnson-McDaniel, Jones, Jones, Jones, Jones,
  Jones, Jonker, Ju, Junker, Kalaghatgi, Kalogera, Kamai, Kandhasamy, Kang,
  Kanner, Kapadia, Karki, Kashyap, Kasprzack, Kastaun, Katsanevas,
  Katsavounidis, Katzman, Kaufer, Kawabe, Kéfélian, Keitel, Keivani, Kennedy,
  Key, Khadka, Khalili, Khan, Khan, Khan, Khazanov, Khetan, Khursheed,
  Kijbunchoo, Kim, Kim, Kim, Kim, Kim, Kim, Kim, Kimball, King, Kinley-Hanlon,
  Kirchhoff, Kissel, Kleybolte, Klimenko, Knowles, Knyazev, Koch, Koehlenbeck,
  Koekoek, Koley, Kondrashov, Kontos, Koper, Korobko, Korth, Kovalam, Kozak,
  Kringel, Krishnendu, Królak, Krupinski, Kuehn, Kumar, Kumar, Kumar, Kumar,
  Kumar, Kuo, Kutynia, Lackey, Laghi, Lalande, Lam, Lamberts, Landry, Landry,
  Lane, Lang, Lange, Lantz, Lanza, Rosa, Lartaux-Vollard, Lasky, Laxen,
  Lazzarini, Lazzaro, Leaci, Leavey, Lecoeuche, Lee, Lee, Lee, Lee, Lee,
  Lehmann, Leroy, Letendre, Levin, Li, Li, li, Li, Li, Linde, Linker, Linley,
  Littenberg, Liu, Liu, Llorens-Monteagudo, Lo, Lockwood, London, Longo,
  Lorenzini, Loriette, Lormand, Losurdo, Lough, Lousto, Lovelace, Lück,
  Lumaca, Lundgren, Ma, Macas, Macfoy, MacInnis, Macleod, MacMillan, Macquet,
  Hernandez, Magaña-Sandoval, Magee, Majorana, Maksimovic, Malik, Man, Mandic,
  Mangano, Mansell, Manske, Mantovani, Mapelli, Marchesoni, Marion, Márka,
  Márka, Markakis, Markosyan, Markowitz, Maros, Marquina, Marsat, Martelli,
  Martin, Martin, Martinez, Martynov, Masalehdan, Mason, Massera, Masserot,
  Massinger, Masso-Reid, Mastrogiovanni, Matas, Matichard, Mavalvala, Maynard,
  McCann, McCarthy, McClelland, McCormick, McCuller, McGuire, McIsaac, McIver,
  McManus, McRae, McWilliams, Meacher, Meadors, Mehmet, Mehta, Villa, Melatos,
  Mendell, Mercer, Mereni, Merfeld, Merilh, Merritt, Merzougui, Meshkov,
  Messenger, Messick, Metzdorff, Meyers, Meylahn, Mhaske, Miani, Miao,
  Michaloliakos, Michel, Middleton, Milano, Miller, Millhouse, Mills, Milotti,
  Milovich-Goff, Minazzoli, Minenkov, Mishkin, Mishra, Mistry, Mitra,
  Mitrofanov, Mitselmakher, Mittleman, Mo, Mogushi, Mohapatra, Mohite,
  Molina-Ruiz, Mondin, Montani, Moore, Moraru, Morawski, Moreno, Morisaki,
  Mours, Mow-Lowry, Mozzon, Muciaccia, Mukherjee, Mukherjee, Mukherjee,
  Mukherjee, Mukund, Mullavey, Munch, Muñiz, Murray, Nagar, Nardecchia,
  Naticchioni, Nayak, Neil, Neilson, Nelemans, Nelson, Nery, Neunzert, Ng, Ng,
  Nguyen, Nguyen, Nichols, Nichols, Nissanke, Nocera, Noh, North, Nothard,
  Nuttall, Oberling, O'Brien, Oganesyan, Ogin, Oh, Oh, Ohme, Ohta, Okada,
  Oliver, Olivetto, Oppermann, Oram, O'Reilly, Ormiston, Ortega, O'Shaughnessy,
  Ossokine, Osthelder, Ottaway, Overmier, Owen, Pace, Pagano, Page, Pagliaroli,
  Pai, Pai, Palamos, Palashov, Palomba, Pan, Panda, Pang, Pankow, Pannarale,
  Pant, Paoletti, Paoli, Parida, Parker, Pascucci, Pasqualetti, Passaquieti,
  Passuello, Patricelli, Payne, Pearlstone, Pechsiri, Pedersen, Pedraza, Pele,
  Penn, Perego, Perez, Périgois, Perreca, Perriès, Petermann, Pfeiffer,
  Phelps, Phukon, Piccinni, Pichot, Piendibene, Piergiovanni, Pierro, Pillant,
  Pinard, Pinto, Piotrzkowski, Pirello, Pitkin, Plastino, Poggiani, Pong,
  Ponrathnam, Popolizio, Porter, Powell, Prajapati, Prasai, Prasanna, Pratten,
  Prestegard, Principe, Prodi, Prokhorov, Punturo, Puppo, Pürrer, Qi,
  Quetschke, Quinonez, Raab, Raaijmakers, Radkins, Radulesco, Raffai, Rafferty,
  Raja, Rajan, Rajbhandari, Rakhmanov, Ramirez, Ramos-Buades, Rana, Rao,
  Rapagnani, Raymond, Razzano, Read, Regimbau, Rei, Reid, Reitze, Rettegno,
  Ricci, Richardson, Richardson, Ricker, Riemenschneider, Riles, Rizzo,
  Robertson, Robinet, Rocchi, Rodriguez-Soto, Rolland, Rollins, Roma,
  Romanelli, Romano, Romel, Romero-Shaw, Romie, Rose, Rose, Rose, Rosińska,
  Rosofsky, Ross, Rowan, Rowlinson, Roy, Roy, Roy, Ruggi, Rutins, Ryan,
  Sachdev, Sadecki, Sakellariadou, Salafia, Salconi, Saleem, Salemi, Samajdar,
  Sanchez, Sanchez, Sanchis-Gual, Sanders, Santiago, Santos, Sarin, Sassolas,
  Sathyaprakash, Sauter, Savage, Savant, Sawant, Sayah, Schaetzl, Schale,
  Scheel, Scheuer, Schmidt, Schnabel, Schofield, Schönbeck, Schreiber,
  Schulte, Schutz, Schwarm, Schwartz, Scott, Scott, Seidel, Sellers, Sengupta,
  Sennett, Sentenac, Sequino, Sergeev, Setyawati, Shaddock, Shaffer, Shahriar,
  Sharma, Sharma, Shawhan, Shen, Shikauchi, Shink, Shoemaker, Shoemaker,
  Shukla, ShyamSundar, Siellez, Sieniawska, Sigg, Singer, Singh, Singh, Singha,
  Singhal, Sintes, Sipala, Skliris, Slagmolen, Slaven-Blair, Smetana, Smith,
  Smith, Somala, Son, Soni, Sorazu, Sordini, Sorrentino, Souradeep, Sowell,
  Spencer, Spera, Srivastava, Srivastava, Staats, Stachie, Standke, Steer,
  Steinhoff, Steinke, Steinlechner, Steinlechner, Steinmeyer, Stevenson,
  Stocks, Stops, Stover, Strain, Stratta, Strunk, Sturani, Stuver, Sudhagar,
  Sudhir, Summerscales, Sun, Sunil, Sur, Suresh, Sutton, Swinkels,
  Szczepańczyk, Tacca, Tait, Talbot, Tanasijczuk, Tanner, Tao, Tápai, Tapia,
  Martin, Tasson, Taylor, Tenorio, Terkowski, Thirugnanasambandam, Thomas,
  Thomas, Thompson, Thondapu, Thorne, Thrane, Tinsman, Saravanan, Tiwari,
  Tiwari, Tiwari, Toland, Tonelli, Tornasi, Torres-Forné, Torrie, Melo,
  Töyrä, Trail, Travasso, Traylor, Tringali, Tripathee, Trovato, Trudeau,
  Tsang, Tse, Tso, Tsukada, Tsuna, Tsutsui, Turconi, Ubhi, Ueno, Ugolini,
  Unnikrishnan, Urban, Usman, Utina, Vahlbruch, Vajente, Valdes, Valentini,
  Bakel, Beuzekom, Brand, Broeck, Vander-Hyde, Schaaf, Heijningen, Veggel,
  Vardaro, Varma, Vass, Vasúth, Vecchio, Vedovato, Veitch, Veitch,
  Venkateswara, Venugopalan, Verkindt, Veske, Vetrano, Viceré, Viets,
  Vinciguerra, Vine, Vinet, Vitale, Vivanco, Vo, Vocca, Vorvick, Vyatchanin,
  Wade, Wade, Wade, Walet, Walker, Wallace, Wallace, Walsh, Wang, Wang, Wang,
  Ward, Warden, Warner, Was, Watchi, Weaver, Wei, Weinert, Weinstein, Weiss,
  Wellmann, Wen, Weßels, Westhouse, Wette, Whelan, Whiting, Whittle, Wilken,
  Williams, Willis, Willke, Winkler, Wipf, Wittel, Woan, Woehler, Wofford,
  Wong, Wright, Wu, Wysocki, Xiao, Yamamoto, Yang, Yang, Yang, Yap, Yazback,
  Yeeles, Yu, Yu, Yuen, Zadro{\textbackslash}.zny, Zadro{\textbackslash}.zny,
  Zanolin, Zelenova, Zendri, Zevin, Zhang, Zhang, Zhang, Zhao, Zhao, Zhou,
  Zhou, Zhu, Zimmerman, Zucker, \& Zweizig}]{abbott_gw190814_2020}
Abbott, R., Abbott, T.~D., Abraham, S., {et~al.} 2020{\natexlab{c}}, The
  Astrophysical Journal Letters, 896, L44, \dodoi{10.3847/2041-8213/ab960f}

\bibitem[{Abbott {et~al.}(2021{\natexlab{a}})Abbott, Abbott, Abraham, Acernese,
  Ackley, Adams, Adams, Adhikari, Adya, Affeldt, Agathos, Agatsuma, Aggarwal,
  Aguiar, Aiello, Ain, Ajith, Akcay, Allen, Allocca, Altin, Amato, Anand,
  Ananyeva, Anderson, Anderson, Angelova, Ansoldi, Antelis, Antier, Appert,
  Arai, Araya, Areeda, Arène, Arnaud, Aronson, Arun, Asali, Ascenzi, Ashton,
  Aston, Astone, Aubin, Aufmuth, AultONeal, Austin, Avendano, Babak, Badaracco,
  Bader, Bae, Baer, Bagnasco, Baird, Ball, Ballardin, Ballmer, Bals, Balsamo,
  Baltus, Banagiri, Bankar, Bankar, Barayoga, Barbieri, Barish, Barker, Barneo,
  Barnum, Barone, Barr, Barsotti, Barsuglia, Barta, Bartlett, Bartos, Bassiri,
  Basti, Bawaj, Bayley, Bazzan, Becher, Bécsy, Bedakihale, Bejger, Belahcene,
  Beniwal, Benjamin, Bennett, Bentley, Bergamin, Berger, Bergmann, Bernuzzi,
  Berry, Bersanetti, Bertolini, Betzwieser, Bhandare, Bhandari, Bhattacharjee,
  Bidler, Bilenko, Billingsley, Birney, Birnholtz, Biscans, Bischi, Biscoveanu,
  Bisht, Bitossi, Bizouard, Blackburn, Blackman, Blair, Blair, Blair, Blanch,
  Bobba, Bode, Boer, Boetzel, Bogaert, Boldrini, Bondu, Bonilla, Bonnand,
  Booker, Boom, Bork, Boschi, Bose, Bossilkov, Boudart, Bouffanais, Bozzi,
  Bradaschia, Brady, Bramley, Branchesi, Brau, Breschi, Briant, Briggs,
  Brighenti, Brillet, Brinkmann, Brockill, Brooks, Brooks, Brown, Brunett,
  Bruno, Bruntz, Buikema, Bulik, Bulten, Buonanno, Buscicchio, Buskulic, Byer,
  Cabero, Cadonati, Caesar, Cagnoli, Cahillane, Calderón~Bustillo, Callaghan,
  Callister, Calloni, Camp, Canepa, Cannon, Cao, Cao, Carapella, Carbognani,
  Carney, Carpinelli, Carullo, Carver, Casanueva~Diaz, Casentini, Caudill,
  Cavaglià, Cavalier, Cavalieri, Cella, Cerdá-Durán, Cesarini, Chaibi,
  Chakravarti, Chan, Chan, Chandra, Chanial, Chao, Charlton, Chase,
  Chassande-Mottin, Chatterjee, Chattopadhyay, Chaturvedi, Chatziioannou, Chen,
  Chen, Chen, Chen, Cheng, Cheong, Chia, Chiadini, Chierici, Chincarini,
  Chiummo, Cho, Cho, Cho, Choate, Christensen, Chu, Chua, Chung, Chung, Ciani,
  Ciecielag, Cieślar, Cifaldi, Ciobanu, Ciolfi, Cipriano, Cirone, Clara,
  Clark, Clark, Clarke, Clearwater, Clesse, Cleva, Coccia, Cohadon, Cohen,
  Colleoni, Collette, Collins, Colpi, Constancio, Conti, Cooper, Corban,
  Corbitt, Cordero-Carrión, Corezzi, Corley, Cornish, Corre, Corsi, Cortese,
  Costa, Cotesta, Coughlin, Coughlin, Coulon, Countryman, Cousins, Couvares,
  Covas, Coward, Cowart, Coyne, Coyne, Creighton, Creighton, Croquette,
  Crowder, Cudell, Cullen, Cumming, Cummings, Cunningham, Cuoco, Curyło,
  Canton, Dálya, Dana, DaneshgaranBajastani, D’Angelo, Danila, Danilishin,
  D’Antonio, Danzmann, Darsow-Fromm, Dasgupta, Datrier, Dattilo, Dave,
  Davier, Davies, Davis, Daw, Dean, DeBra, Deenadayalan, Degallaix,
  De~Laurentis, Deléglise, Del~Favero, De~Lillo, De~Lillo, Del~Pozzo,
  DeMarchi, De~Matteis, D’Emilio, Demos, Denker, Dent, Depasse, De~Pietri,
  De~Rosa, De~Rossi, DeSalvo, de~Varona, Dhurandhar, Díaz, Diaz-Ortiz, Didio,
  Dietrich, Di~Fiore, DiFronzo, Di~Giorgio, Di~Giovanni, Di~Giovanni,
  Di~Girolamo, Di~Lieto, Ding, Di~Pace, Di~Palma, Di~Renzo, Divakarla,
  Dmitriev, Doctor, D’Onofrio, Donovan, Dooley, Doravari, Dorrington, Downes,
  Drago, Driggers, Du, Ducoin, Dupej, Durante, D’Urso, Duverne, Dwyer,
  Easter, Eddolls, Edelman, Edo, Edy, Effler, Eichholz, Eikenberry, Eisenmann,
  Eisenstein, Ejlli, Errico, Essick, Estellés, Estevez, Etienne, Etzel, Evans,
  Evans, Ewing, Fafone, Fair, Fairhurst, Fan, Farah, Farinon, Farr, Farr,
  Fauchon-Jones, Favata, Fays, Fazio, Feicht, Fejer, Feng, Fenyvesi, Ferguson,
  Fernandez-Galiana, Ferrante, Ferreira, Fidecaro, Figura, Fiori, Fiorucci,
  Fishbach, Fisher, Fishner, Fittipaldi, Fitz-Axen, Fiumara, Flaminio, Floden,
  Flynn, Fong, Font, Forsyth, Fournier, Frasca, Frasconi, Frei, Freise, Frey,
  Frey, Fritschel, Frolov, Fronzé, Fulda, Fyffe, Gabbard, Gadre, Gaebel, Gair,
  Gais, Galaudage, Gamba, Ganapathy, Ganguly, Gaonkar, Garaventa,
  García-Quirós, Garufi, Gateley, Gaudio, Gayathri, Gemme, Gennai, George,
  George, George, Gergely, Ghonge, Ghosh, Ghosh, Ghosh, Giacomazzo, Giacoppo,
  Giaime, Giardina, Gibson, Gier, Gill, Giri, Glanzer, Gleckl, Godwin, Goetz,
  Goetz, Gohlke, Goncharov, González, Gopakumar, Gossan, Gosselin, Gouaty,
  Grace, Grado, Granata, Granata, Grant, Gras, Grassia, Gray, Gray, Greco,
  Green, Green, Gretarsson, Griggs, Grignani, Grimaldi, Grimes, Grimm, Grote,
  Grunewald, Gruning, Guerrero, Guidi, Guimaraes, Guixé, Gulati, Guo, Gupta,
  Gupta, Gupta, Gustafson, Gustafson, Guzman, Haegel, Halim, Hall, Hamilton,
  Hammond, Haney, Hanke, Hanks, Hanna, Hannam, Hannuksela, Hannuksela, Hansen,
  Hansen, Hanson, Harder, Hardwick, Haris, Harms, Harry, Harry, Hartwig,
  Hasskew, Haster, Haughian, Hayes, Healy, Heidmann, Heintze, Heinze, Heinzel,
  Heitmann, Hellman, Hello, Helmling-Cornell, Hemming, Hendry, Heng, Hennes,
  Hennig, Hennig, Hernandez~Vivanco, Heurs, Hild, Hill, Hines, Hochheim,
  Hofgard, Hofman, Hohmann, Holgado, Holland, Hollows, Holmes, Holt, Holz,
  Hopkins, Horst, Hough, Howell, Hoy, Hoyland, Huang, Hübner, Huddart, Huerta,
  Hughey, Hui, Husa, Huttner, Hutzler, Huxford, Huynh-Dinh, Idzkowski, Iess,
  Imperato, Inchauspe, Ingram, Intini, Isi, Iyer, JaberianHamedan, Jacqmin,
  Jadhav, Jadhav, James, Jani, Janssens, Janthalur, Jaranowski, Jariwala,
  Jaume, Jenkins, Jeunon, Jiang, Johns, Johnson-McDaniel, Jones, Jones, Jones,
  Jones, Jones, Jonker, Ju, Junker, Kalaghatgi, Kalogera, Kamai, Kandhasamy,
  Kang, Kanner, Kapadia, Kapasi, Karathanasis, Karki, Kashyap, Kasprzack,
  Kastaun, Katsanevas, Katsavounidis, Katzman, Kawabe, Kéfélian, Keitel, Key,
  Khadka, Khalili, Khan, Khan, Khazanov, Khetan, Khursheed, Kijbunchoo, Kim,
  Kim, Kim, Kim, Kim, Kim, Kimball, King, Kinley-Hanlon, Kirchhoff, Kissel,
  Kleybolte, Klimenko, Knowles, Knyazev, Koch, Koehlenbeck, Koekoek, Koley,
  Kolstein, Komori, Kondrashov, Kontos, Koper, Korobko, Korth, Kovalam, Kozak,
  Krämer, Kringel, Krishnendu, Królak, Kuehn, Kumar, Kumar, Kumar, Kumar,
  Kuns, Kwang, Lackey, Laghi, Lalande, Lam, Lamberts, Landry, Lane, Lang,
  Lange, Lantz, Lanza, La~Rosa, Lartaux-Vollard, Lasky, Laxen, Lazzarini,
  Lazzaro, Leaci, Leavey, Lecoeuche, Lee, Lee, Lee, Lee, Lehmann, Leon, Leroy,
  Letendre, Levin, Li, Li, Li, Li, Li, Linde, Linker, Linley, Littenberg, Liu,
  Liu, Llorens-Monteagudo, Lo, Lockwood, London, Longo, Lorenzini, Loriette,
  Lormand, Losurdo, Lough, Lousto, Lovelace, Lück, Lumaca, Lundgren, Ma,
  Macas, MacInnis, Macleod, MacMillan, Macquet, Magaña~Hernandez,
  Magaña-Sandoval, Magazzù, Magee, Majorana, Maksimovic, Maliakal, Malik,
  Man, Mandic, Mangano, Mansell, Manske, Mantovani, Mapelli, Marchesoni,
  Marion, Márka, Márka, Markakis, Markosyan, Markowitz, Maros, Marquina,
  Marsat, Martelli, Martin, Martin, Martinez, Martinez, Martynov, Masalehdan,
  Mason, Massera, Masserot, Massinger, Masso-Reid, Mastrogiovanni, Matas,
  Mateu-Lucena, Matichard, Matiushechkina, Mavalvala, Maynard, McCann,
  McCarthy, McClelland, McCormick, McCuller, McGuire, McIsaac, McIver, McManus,
  McRae, McWilliams, Meacher, Meadors, Mehmet, Mehta, Melatos, Melchor,
  Mendell, Menendez-Vazquez, Mercer, Mereni, Merfeld, Merilh, Merritt,
  Merzougui, Meshkov, Messenger, Messick, Metzdorff, Meyers, Meylahn, Mhaske,
  Miani, Miao, Michaloliakos, Michel, Middleton, Milano, Miller, Millhouse,
  Mills, Milotti, Milovich-Goff, Minazzoli, Minenkov, Mir, Mishkin, Mishra,
  Mistry, Mitra, Mitrofanov, Mitselmakher, Mittleman, Mo, Mogushi, Mohapatra,
  Mohite, Molina, Molina-Ruiz, Mondin, Montani, Moore, Moraru, Morawski,
  Moreno, Morisaki, Mours, Mow-Lowry, Mozzon, Muciaccia, Mukherjee, Mukherjee,
  Mukherjee, Mukherjee, Mukund, Mullavey, Munch, Muñiz, Murray, Nadji, Nagar,
  Nardecchia, Naticchioni, Nayak, Neil, Neilson, Nelemans, Nelson, Nery,
  Neunzert, Nitz, Ng, Ng, Nguyen, Nguyen, Nguyen, Nichols, Nissanke, Nocera,
  Noh, North, Nothard, Nuttall, Oberling, O’Brien, O’Dell, Oganesyan, Ogin,
  Oh, Oh, Ohme, Ohta, Okada, Olivetto, Oppermann, Oram, O’Reilly, Ormiston,
  Ortega, O’Shaughnessy, Ossokine, Osthelder, Ottaway, Overmier, Owen, Pace,
  Pagano, Page, Pagliaroli, Pai, Pai, Palamos, Palashov, Palomba, Pan, Panda,
  Pang, Pankow, Pannarale, Pant, Paoletti, Paoli, Paolone, Parker, Pascucci,
  Pasqualetti, Passaquieti, Passuello, Patel, Patricelli, Payne, Pechsiri,
  Pedraza, Pegoraro, Pele, Penn, Perego, Perez, Périgois, Perreca, Perriès,
  Petermann, Petterson, Pfeiffer, Pham, Phukon, Piccinni, Pichot, Piendibene,
  Piergiovanni, Pierini, Pierro, Pillant, Pilo, Pinard, Pinto, Piotrzkowski,
  Pirello, Pitkin, Placidi, Plastino, Pluchar, Poggiani, Polini, Pong,
  Ponrathnam, Popolizio, Porter, Poverman, Powell, Pracchia, Prajapati, Prasai,
  Prasanna, Pratten, Prestegard, Principe, Prodi, Prokhorov, Prosposito,
  Prudenzi, Puecher, Punturo, Puosi, Puppo, Pürrer, Qi, Quetschke, Quinonez,
  Quitzow-James, Raab, Raaijmakers, Radkins, Radulesco, Raffai, Rafferty, Rail,
  Raja, Rajan, Rajbhandari, Rakhmanov, Ramirez, Ramirez, Ramos-Buades, Rana,
  Rao, Rapagnani, Rapol, Ratto, Raymond, Razzano, Read, Regimbau, Rei, Reid,
  Reitze, Rettegno, Ricci, Richardson, Richardson, Richardson, Ricker,
  Riemenschneider, Riles, Rizzo, Robertson, Robinet, Rocchi, Rocha, Rodriguez,
  Rodriguez-Soto, Rolland, Rollins, Roma, Romanelli, Romano, Romel, Romero,
  Romero-Shaw, Romie, Ronchini, Rose, Rose, Rose, Rosell, Rosińska, Rosofsky,
  Ross, Rowan, Rowlinson, Roy, Roy, Ruggi, Ryan, Sachdev, Sadecki, Sadiq,
  Sakellariadou, Salafia, Salconi, Saleem, Samajdar, Sanchez, Sanchez, Sanchez,
  Sanchis-Gual, Sanders, Sandles, Santiago, Santos, Saravanan, Sarin, Sassolas,
  Sathyaprakash, Sauter, Savage, Savant, Sawant, Sayah, Schaetzl, Schale,
  Scheel, Scheuer, Schindler-Tyka, Schmidt, Schnabel, Schofield, Schönbeck,
  Schreiber, Schulte, Schutz, Schwarm, Schwartz, Scott, Scott, Seglar-Arroyo,
  Seidel, Sellers, Sengupta, Sennett, Sentenac, Sequino, Sergeev, Setyawati,
  Shaffer, Shahriar, Sharifi, Sharma, Sharma, Shawhan, Shen, Shikauchi, Shink,
  Shoemaker, Shoemaker, Shukla, ShyamSundar, Sieniawska, Sigg, Singer, Singh,
  Singh, Singha, Singhal, Sintes, Sipala, Skliris, Slagmolen, Slaven-Blair,
  Smetana, Smith, Smith, Somala, Son, Soni, Soni, Sorazu, Sordini, Sorrentino,
  Sorrentino, Soulard, Souradeep, Sowell, Spencer, Spera, Srivastava,
  Srivastava, Staats, Stachie, Steer, Steinhoff, Steinke, Steinlechner,
  Steinlechner, Steinmeyer, Stevenson, Stolle-McAllister, Stops, Stover,
  Strain, Stratta, Strunk, Sturani, Stuver, Südbeck, Sudhagar, Sudhir, Suh,
  Summerscales, Sun, Sun, Sunil, Sur, Suresh, Sutton, Swinkels, Szczepańczyk,
  Tacca, Tait, Talbot, Tanasijczuk, Tanner, Tao, Tapia, Tapia San~Martin,
  Tasson, Taylor, Tenorio, Terkowski, Thirugnanasambandam, Thomas, Thomas,
  Thomas, Thompson, Thondapu, Thorne, Thrane, Tiwari, Tiwari, Tiwari, Toland,
  Tolley, Tonelli, Tornasi, Torres-Forné, Torrie, e~Melo, Töyrä, Tran,
  Trapananti, Travasso, Traylor, Tringali, Tripathee, Trovato, Trudeau, Tsai,
  Tsang, Tse, Tso, Tsukada, Tsuna, Tsutsui, Turconi, Ubhi, Udall, Ueno,
  Ugolini, Unnikrishnan, Urban, Usman, Utina, Vahlbruch, Vajente, Vajpeyi,
  Valdes, Valentini, Valsan, van Bakel, van Beuzekom, van~den Brand, Van
  Den~Broeck, Vander-Hyde, van~der Schaaf, van Heijningen, Vardaro, Vargas,
  Varma, Vass, Vasúth, Vecchio, Vedovato, Veitch, Veitch, Venkateswara,
  Venneberg, Venugopalan, Verkindt, Verma, Veske, Vetrano, Viceré, Viets,
  Vijaykumar, Villa-Ortega, Vinet, Vitale, Vo, Vocca, Vorvick, Vyatchanin,
  Wade, Wade, Wade, Walet, Walker, Wallace, Wallace, Walsh, Wang, Wang, Wang,
  Wang, Ward, Warner, Was, Washington, Watchi, Weaver, Wei, Weinert, Weinstein,
  Weiss, Wellmann, Wen, Weßels, Westhouse, Wette, Whelan, White, White,
  Whiting, Whittle, Wilken, Williams, Williams, Williamson, Willis, Willke,
  Wilson, Wimmer, Winkler, Wipf, Woan, Woehler, Wofford, Wong, Wrangel, Wright,
  Wu, Wysocki, Xiao, Yamamoto, Yang, Yang, Yang, Yap, Yeeles, Yoon, Yu, Yu,
  Yuen, Zadrożny, Zanolin, Zelenova, Zendri, Zevin, Zhang, Zhang, Zhang,
  Zhang, Zhao, Zhao, Zheng, Zhou, Zhou, Zhu, Zimmerman, Zlochower, Zucker, \&
  Zweizig}]{abbott_gwtc-2_2021}
Abbott, R., Abbott, T., Abraham, S., {et~al.} 2021{\natexlab{a}}, Physical
  Review X, 11, 021053, \dodoi{10.1103/PhysRevX.11.021053}

\bibitem[{Abbott {et~al.}(2021{\natexlab{b}})Abbott, Abbott, Acernese, Ackley,
  Adams, Adhikari, Adhikari, Adya, Affeldt, Agarwal, Agathos, Agatsuma,
  Aggarwal, Aguiar, Aiello, Ain, Ajith, Akcay, Akutsu, Albanesi, Allocca,
  Altin, Amato, Anand, Anand, Ananyeva, Anderson, Anderson, Ando, Andrade,
  Andres, Andrić, Angelova, Ansoldi, Antelis, Antier, Appert, Arai, Arai,
  Arai, Araki, Araya, Araya, Areeda, Arène, Aritomi, Arnaud, Arogeti, Aronson,
  Arun, Asada, Asali, Ashton, Aso, Assiduo, Aston, Astone, Aubin, Austin,
  Babak, Badaracco, Bader, Badger, Bae, Bae, Baer, Bagnasco, Bai, Baiotti,
  Baird, Bajpai, Ball, Ballardin, Ballmer, Balsamo, Baltus, Banagiri, Bankar,
  Barayoga, Barbieri, Barish, Barker, Barneo, Barone, Barr, Barsotti,
  Barsuglia, Barta, Bartlett, Barton, Bartos, Bassiri, Basti, Bawaj, Bayley,
  Baylor, Bazzan, Bécsy, Bedakihale, Bejger, Belahcene, Benedetto, Beniwal,
  Bennett, Bentley, BenYaala, Bergamin, Berger, Bernuzzi, Berry, Bersanetti,
  Bertolini, Betzwieser, Beveridge, Bhandare, Bhardwaj, Bhattacharjee, Bhaumik,
  Bilenko, Billingsley, Bini, Birney, Birnholtz, Biscans, Bischi, Biscoveanu,
  Bisht, Biswas, Bitossi, Bizouard, Blackburn, Blair, Blair, Blair, Bobba,
  Bode, Boer, Bogaert, Boldrini, Bonavena, Bondu, Bonilla, Bonnand, Booker,
  Boom, Bork, Boschi, Bose, Bose, Bossilkov, Boudart, Bouffanais, Bozzi,
  Bradaschia, Brady, Bramley, Branch, Branchesi, Brandt, Brau, Breschi, Briant,
  Briggs, Brillet, Brinkmann, Brockill, Brooks, Brooks, Brown, Brunett, Bruno,
  Bruntz, Bryant, Bulik, Bulten, Buonanno, Buscicchio, Buskulic, Buy, Byer,
  Davies, Cadonati, Cagnoli, Cahillane, Bustillo, Callaghan, Callister,
  Calloni, Cameron, Camp, Canepa, Canevarolo, Cannavacciuolo, Cannon, Cao, Cao,
  Capocasa, Capote, Carapella, Carbognani, Carlin, Carney, Carpinelli,
  Carrillo, Carullo, Carver, Diaz, Casentini, Castaldi, Caudill, Cavaglià,
  Cavalier, Cavalieri, Ceasar, Cella, Cerdá-Durán, Cesarini, Chaibi,
  Chakravarti, Subrahmanya, Champion, Chan, Chan, Chan, Chan, Chan, Chandra,
  Chanial, Chao, Chapman-Bird, Charlton, Chase, Chassande-Mottin, Chatterjee,
  Chatterjee, Chatterjee, Chaturvedi, Chaty, Chatziioannou, Chen, Chen, Chen,
  Chen, Chen, Chen, Chen, Chen, Cheng, Cheong, Cheung, Chia, Chiadini, Chiang,
  Chiarini, Chierici, Chincarini, Chiofalo, Chiummo, Cho, Cho, Choudhary,
  Choudhary, Christensen, Chu, Chu, Chu, Chua, Chung, Ciani, Ciecielag,
  Cieślar, Cifaldi, Ciobanu, Ciolfi, Cipriano, Cirone, Clara, Clark, Clark,
  Clarke, Clearwater, Clesse, Cleva, Coccia, Codazzo, Cohadon, Cohen, Cohen,
  Colleoni, Collette, Colombo, Colpi, Compton, Constancio~Jr., Conti, Cooper,
  Corban, Corbitt, Cordero-Carrión, Corezzi, Corley, Cornish, Corre, Corsi,
  Cortese, Costa, Cotesta, Coughlin, Coulon, Countryman, Cousins, Couvares,
  Coward, Cowart, Coyne, Coyne, Creighton, Creighton, Criswell, Croquette,
  Crowder, Cudell, Cullen, Cumming, Cummings, Cunningham, Cuoco, Curyło,
  Dabadie, Canton, Dall'Osso, Dálya, Dana, DaneshgaranBajastani, D'Angelo,
  Danila, Danilishin, D'Antonio, Danzmann, Darsow-Fromm, Dasgupta, Datrier,
  Datta, Dattilo, Dave, Davier, Davis, Davis, Daw, de~Alarcón, Dean, DeBra,
  Deenadayalan, Degallaix, De~Laurentis, Deléglise, Del~Favero, De~Lillo,
  De~Lillo, Del~Pozzo, DeMarchi, De~Matteis, D'Emilio, Demos, Dent, Depasse,
  De~Pietri, De~Rosa, De~Rossi, DeSalvo, De~Simone, Dhurandhar, Díaz,
  Diaz-Ortiz~Jr., Didio, Dietrich, Di~Fiore, Di~Fronzo, Di~Giorgio,
  Di~Giovanni, Di~Giovanni, Di~Girolamo, Di~Lieto, Ding, Di~Pace, Di~Palma,
  Di~Renzo, Divakarla, Dmitriev, Doctor, D'Onofrio, Donovan, Dooley, Doravari,
  Dorrington, Drago, Driggers, Drori, Ducoin, Dupej, Durante, D'Urso, Duverne,
  Dwyer, Eassa, Easter, Ebersold, Eckhardt, Eddolls, Edelman, Edo, Edy, Effler,
  Eguchi, Eichholz, Eikenberry, Eisenmann, Eisenstein, Ejlli, Engelby, Enomoto,
  Errico, Essick, Estellés, Estevez, Etienne, Etzel, Evans, Evans, Ewing,
  Fafone, Fair, Fairhurst, Farah, Farinon, Farr, Farr, Farrow, Fauchon-Jones,
  Favaro, Favata, Fays, Fazio, Feicht, Fejer, Fenyvesi, Ferguson,
  Fernandez-Galiana, Ferrante, Ferreira, Fidecaro, Figura, Fiori, Fishbach,
  Fisher, Fittipaldi, Fiumara, Flaminio, Floden, Fong, Font, Fornal, Forsyth,
  Franke, Frasca, Frasconi, Frederick, Freed, Frei, Freise, Frey, Fritschel,
  Frolov, Fronzé, Fujii, Fujikawa, Fukunaga, Fukushima, Fulda, Fyffe, Gabbard,
  Gabella, Gadre, Gair, Gais, Galaudage, Gamba, Ganapathy, Ganguly, Gao,
  Gaonkar, Garaventa, García, García-Núñez, García-Quirós, Garufi,
  Gateley, Gaudio, Gayathri, Ge, Gemme, Gennai, George, George, Gerberding,
  Gergely, Gewecke, Ghonge, Ghosh, Ghosh, Ghosh, Ghosh, Giacomazzo, Giacoppo,
  Giaime, Giardina, Gibson, Gier, Giesler, Giri, Gissi, Glanzer, Gleckl,
  Godwin, Goetz, Goetz, Gohlke, Golomb, Goncharov, González, Gopakumar,
  Gosselin, Gouaty, Gould, Grace, Grado, Granata, Granata, Grant, Gras,
  Grassia, Gray, Gray, Greco, Green, Green, Gretarsson, Gretarsson, Griffith,
  Griffiths, Griggs, Grignani, Grimaldi, Grimm, Grote, Grunewald, Gruning,
  Guerra, Guidi, Guimaraes, Guixé, Gulati, Guo, Guo, Gupta, Gupta, Gupta,
  Gustafson, Gustafson, Guzman, Ha, Haegel, Hagiwara, Haino, Halim, Hall,
  Hamilton, Hammond, Han, Haney, Hanks, Hanna, Hannam, Hannuksela, Hansen,
  Hansen, Hanson, Harder, Hardwick, Haris, Harms, Harry, Harry, Hartwig,
  Hasegawa, Haskell, Hasskew, Haster, Hattori, Haughian, Hayakawa, Hayama,
  Hayes, Healy, Heidmann, Heidt, Heintze, Heinze, Heinzel, Heitmann, Hellman,
  Hello, Helmling-Cornell, Hemming, Hendry, Heng, Hennes, Hennig, Hennig,
  Hernandez, Vivanco, Heurs, Hild, Hill, Himemoto, Hines, Hiranuma, Hirata,
  Hirose, Hochheim, Hofman, Hohmann, Holcomb, Holland, Holley-Bockelmann,
  Hollows, Holmes, Holt, Holz, Hong, Hopkins, Hough, Hourihane, Howell, Hoy,
  Hoyland, Hreibi, Hsieh, Hsu, Huang, Huang, Huang, Huang, Huang, Huang,
  Hübner, Huddart, Hughey, Hui, Hui, Husa, Huttner, Huxford, Huynh-Dinh, Ide,
  Idzkowski, Iess, Ikenoue, Imam, Inayoshi, Ingram, Inoue, Ioka, Isi, Isleif,
  Ito, Itoh, Iyer, Izumi, JaberianHamedan, Jacqmin, Jadhav, Jadhav, James, Jan,
  Jani, Janquart, Janssens, Janthalur, Jaranowski, Jariwala, Jaume, Jenkins,
  Jenner, Jeon, Jeunon, Jia, Jin, Johns, Johnson-McDaniel, Jones, Jones, Jones,
  Jones, Jones, Jonker, Ju, Jung, Jung, Junker, Juste, Kaihotsu, Kajita,
  Kakizaki, Kalaghatgi, Kalogera, Kamai, Kamiizumi, Kanda, Kandhasamy, Kang,
  Kanner, Kao, Kapadia, Kapasi, Karat, Karathanasis, Karki, Kashyap, Kasprzack,
  Kastaun, Katsanevas, Katsavounidis, Katzman, Kaur, Kawabe, Kawaguchi, Kawai,
  Kawasaki, Kéfélian, Keitel, Key, Khadka, Khalili, Khan, Khazanov, Khetan,
  Khursheed, Kijbunchoo, Kim, Kim, Kim, Kim, Kim, Kim, Kimball, Kimura,
  Kinley-Hanlon, Kirchhoff, Kissel, Kita, Kitazawa, Kleybolte, Klimenko, Knee,
  Knowles, Knyazev, Koch, Koekoek, Kojima, Kokeyama, Koley, Kolitsidou,
  Kolstein, Komori, Kondrashov, Kong, Kontos, Koper, Korobko, Kotake, Kovalam,
  Kozak, Kozakai, Kozu, Kringel, Krishnendu, Królak, Kuehn, Kuei, Kuijer,
  Kulkarni, Kumar, Kumar, Kumar, Kumar, Kume, Kuns, Kuo, Kuo, Kuromiya,
  Kuroyanagi, Kusayanagi, Kuwahara, Kwak, Lagabbe, Laghi, Lalande, Lam,
  Lamberts, Landry, Lane, Lang, Lange, Lantz, La~Rosa, Lartaux-Vollard, Lasky,
  Laxen, Lazzarini, Lazzaro, Leaci, Leavey, Lecoeuche, Lee, Lee, Lee, Lee, Lee,
  Lee, Lehmann, Lemaître, Leonardi, Leroy, Letendre, Levesque, Levin, Leviton,
  Leyde, Li, Li, Li, Li, Li, Li, Lin, Lin, Lin, Lin, Lin, Linde, Linker,
  Linley, Littenberg, Liu, Liu, Liu, Liu, Llamas, Llorens-Monteagudo, Lo,
  Lockwood, Loh, London, Longo, Lopez, Portilla, Lorenzini, Loriette, Lormand,
  Losurdo, Lott, Lough, Lousto, Lovelace, Lucaccioni, Lück, Lumaca, Lundgren,
  Luo, Lynam, Macas, MacInnis, Macleod, MacMillan, Macquet, Hernandez,
  Magazzù, Magee, Maggiore, Magnozzi, Mahesh, Majorana, Makarem, Maksimovic,
  Maliakal, Malik, Man, Mandic, Mangano, Mango, Mansell, Manske, Mantovani,
  Mapelli, Marchesoni, Marchio, Marion, Mark, Márka, Márka, Markakis,
  Markosyan, Markowitz, Maros, Marquina, Marsat, Martelli, Martin, Martin,
  Martinez, Martinez, Martinez, Martinovic, Martynov, Marx, Masalehdan, Mason,
  Massera, Masserot, Massinger, Masso-Reid, Mastrogiovanni, Matas,
  Mateu-Lucena, Matichard, Matiushechkina, Mavalvala, McCann, McCarthy,
  McClelland, McClincy, McCormick, McCuller, McGhee, McGuire, McIsaac, McIver,
  McRae, McWilliams, Meacher, Mehmet, Mehta, Meijer, Melatos, Melchor, Mendell,
  Menendez-Vazquez, Menoni, Mercer, Mereni, Merfeld, Merilh, Merritt,
  Merzougui, Meshkov, Messenger, Messick, Meyers, Meylahn, Mhaske, Miani, Miao,
  Michaloliakos, Michel, Michimura, Middleton, Milano, Miller, Miller, Miller,
  Millhouse, Mills, Milotti, Minazzoli, Minenkov, Mio, Mir, Miravet-Tenés,
  Mishra, Mishra, Mistry, Mitra, Mitrofanov, Mitselmakher, Mittleman, Miyakawa,
  Miyamoto, Miyazaki, Miyo, Miyoki, Mo, Modafferi, Moguel, Mogushi, Mohapatra,
  Mohite, Molina, Molina-Ruiz, Mondin, Montani, Moore, Moraru, Morawski, More,
  Moreno, Moreno, Mori, Morisaki, Moriwaki, Morrás, Mours, Mow-Lowry, Mozzon,
  Muciaccia, Mukherjee, Mukherjee, Mukherjee, Mukherjee, Mukherjee, Mukund,
  Mullavey, Munch, Muñiz, Murray, Musenich, Muusse, Nadji, Nagano, Nagano,
  Nagar, Nakamura, Nakano, Nakano, Nakashima, Nakayama, Napolano, Nardecchia,
  Narikawa, Naticchioni, Nayak, Nayak, Negishi, Neil, Neilson, Nelemans,
  Nelson, Nery, Neubauer, Neunzert, Ng, Ng, Nguyen, Nguyen, Nguyen, Quynh, Ni,
  Nichols, Nishizawa, Nissanke, Nitoglia, Nocera, Norman, North, Nozaki, Siles,
  Nuttall, Oberling, O'Brien, Obuchi, O'Dell, Oelker, Ogaki, Oganesyan, Oh, Oh,
  Oh, Ohashi, Ohishi, Ohkawa, Ohme, Ohta, Okada, Okutani, Okutomi, Olivetto,
  Oohara, Ooi, Oram, O'Reilly, Ormiston, Ormsby, Ortega, O'Shaughnessy, O'Shea,
  Oshino, Ossokine, Osthelder, Otabe, Ottaway, Overmier, Pace, Pagano, Page,
  Pagliaroli, Pai, Pai, Palamos, Palashov, Palomba, Pan, Pan, Panda, Pang,
  Pang, Pankow, Pannarale, Pant, Panther, Paoletti, Paoli, Paolone, Parisi,
  Park, Park, Parker, Pascucci, Pasqualetti, Passaquieti, Passuello, Patel,
  Pathak, Patricelli, Patron, Paul, Payne, Pedraza, Pegoraro, Pele, Arellano,
  Penn, Perego, Pereira, Pereira, Perez, Périgois, Perkins, Perreca, Perriès,
  Petermann, Petterson, Pfeiffer, Pham, Phukon, Piccinni, Pichot, Piendibene,
  Piergiovanni, Pierini, Pierro, Pillant, Pillas, Pilo, Pinard, Pinto, Pinto,
  Piotrzkowski, Piotrzkowski, Pirello, Pitkin, Placidi, Planas, Plastino,
  Pluchar, Poggiani, Polini, Pong, Ponrathnam, Popolizio, Porter, Poulton,
  Powell, Pracchia, Pradier, Prajapati, Prasai, Prasanna, Pratten, Principe,
  Prodi, Prokhorov, Prosposito, Prudenzi, Puecher, Punturo, Puosi, Puppo,
  Pürrer, Qi, Quetschke, Quitzow-James, Qutob, Raab, Raaijmakers, Radkins,
  Radulesco, Raffai, Rail, Raja, Rajan, Ramirez, Ramirez, Ramos-Buades, Rana,
  Rapagnani, Rapol, Ray, Raymond, Raza, Razzano, Read, Rees, Regimbau, Rei,
  Reid, Reid, Reitze, Relton, Renzini, Rettegno, Reza, Rezac, Ricci, Richards,
  Richardson, Richardson, Riemenschneider, Riles, Rinaldi, Rink, Rizzo,
  Robertson, Robie, Robinet, Rocchi, Rodriguez, Rolland, Rollins, Romanelli,
  Romano, Romel, Romero-Rodríguez, Romero-Shaw, Romie, Ronchini, Rosa, Rose,
  Rosińska, Ross, Rowan, Rowlinson, Roy, Roy, Roy, Rozza, Ruggi, Ruiz-Rocha,
  Ryan, Sachdev, Sadecki, Sadiq, Sago, Saito, Saito, Sakai, Sakai,
  Sakellariadou, Sakuno, Salafia, Salconi, Saleem, Salemi, Samajdar, Sanchez,
  Sanchez, Sanchez, Sanchis-Gual, Sanders, Sanuy, Saravanan, Sarin, Sassolas,
  Satari, Sathyaprakash, Sato, Sato, Sauter, Savage, Sawada, Sawant, Sawant,
  Sayah, Schaetzl, Scheel, Scheuer, Schiworski, Schmidt, Schmidt, Schnabel,
  Schneewind, Schofield, Schönbeck, Schulte, Schutz, Schwartz, Scott, Scott,
  Seglar-Arroyo, Sekiguchi, Sekiguchi, Sellers, Sengupta, Sentenac, Seo,
  Sequino, Sergeev, Setyawati, Shaffer, Shahriar, Shams, Shao, Sharma, Sharma,
  Shawhan, Shcheblanov, Shibagaki, Shikauchi, Shimizu, Shimoda, Shimode,
  Shinkai, Shishido, Shoda, Shoemaker, Shoemaker, ShyamSundar, Sieniawska,
  Sigg, Singer, Singh, Singh, Singha, Sintes, Sipala, Skliris, Slagmolen,
  Slaven-Blair, Smetana, Smith, Smith, Soldateschi, Somala, Somiya, Son, Soni,
  Soni, Sordini, Sorrentino, Sorrentino, Sotani, Soulard, Souradeep, Sowell,
  Spagnuolo, Spencer, Spera, Srinivasan, Srivastava, Srivastava, Staats,
  Stachie, Steer, Steinhoff, Steinlechner, Steinlechner, Stevenson, Stops,
  Stover, Strain, Strang, Stratta, Strunk, Sturani, Stuver, Sudhagar, Sudhir,
  Sugimoto, Suh, Sullivan, Sullivan, Summerscales, Sun, Sun, Sunil, Sur,
  Suresh, Sutton, Suzuki, Suzuki, Swinkels, Szczepańczyk, Szewczyk, Tacca,
  Tagoshi, Tait, Takahashi, Takahashi, Takamori, Takano, Takeda, Takeda,
  Talbot, Talbot, Tanaka, Tanaka, Tanaka, Tanaka, Tanaka, Tanasijczuk, Tanioka,
  Tanner, Tao, Tao, Martín, Taranto, Tasson, Telada, Tenorio, Terhune,
  Terkowski, Thirugnanasambandam, Thomas, Thomas, Thomas, Thompson, Thondapu,
  Thorne, Thrane, Tiwari, Tiwari, Tiwari, Toivonen, Toland, Tolley, Tomaru,
  Tomigami, Tomura, Tonelli, Torres-Forné, Torrie, Melo, Töyrä, Trapananti,
  Travasso, Traylor, Trevor, Tringali, Tripathee, Troiano, Trovato, Trozzo,
  Trudeau, Tsai, Tsai, Tsang, Tsang, Tsao, Tse, Tso, Tsubono, Tsuchida,
  Tsukada, Tsuna, Tsutsui, Tsuzuki, Turbang, Turconi, Tuyenbayev, Ubhi,
  Uchikata, Uchiyama, Udall, Ueda, Uehara, Ueno, Ueshima, Unnikrishnan,
  Uraguchi, Urban, Ushiba, Utina, Vahlbruch, Vajente, Vajpeyi, Valdes,
  Valentini, Valsan, van Bakel, van Beuzekom, Brand, Broeck, Vander-Hyde,
  van~der Schaaf, van Heijningen, Vanosky, van Putten, van Remortel, Vardaro,
  Vargas, Varma, Vasúth, Vecchio, Vedovato, Veitch, Veitch, Venneberg,
  Venugopalan, Verkindt, Verma, Verma, Veske, Vetrano, Viceré, Vidyant, Viets,
  Vijaykumar, Villa-Ortega, Vinet, Virtuoso, Vitale, Vo, Vocca, von Reis, von
  Wrangel, Vorvick, Vyatchanin, Wade, Wade, Wagner, Walet, Walker, Wallace,
  Wallace, Walsh, Wang, Wang, Wang, Ward, Warner, Was, Washimi, Washington,
  Watchi, Weaver, Webster, Weinert, Weinstein, Weiss, Weller, Weller, Wellmann,
  Wen, Weßels, Wette, Whelan, White, Whiting, Whittle, Wilken, Williams,
  Williams, Williams, Williamson, Willis, Willke, Wilson, Winkler, Wipf,
  Wlodarczyk, Woan, Woehler, Wofford, Wong, Wu, Wu, Wu, Wu, Wysocki, Xiao, Xu,
  Yamada, Yamamoto, Yamamoto, Yamamoto, Yamamoto, Yamashita, Yamazaki, Yang,
  Yang, Yang, Yang, Yang, Yap, Yeeles, Yelikar, Ying, Yokogawa, Yokoyama,
  Yokozawa, Yoo, Yoshioka, Yu, Yu, Yuzurihara, Zadrożny, Zanolin, Zeidler,
  Zelenova, Zendri, Zevin, Zhan, Zhang, Zhang, Zhang, Zhang, Zhang, Zhao, Zhao,
  Zhao, Zhao, Zheng, Zhou, Zhou, Zhu, Zhu, Zimmerman, Zlochower, Zucker, \&
  Zweizig}]{abbott_gwtc-3_2021}
Abbott, R., Abbott, T.~D., Acernese, F., {et~al.} 2021{\natexlab{b}},
  arXiv:2111.03606 [astro-ph, physics:gr-qc].
\newblock \url{http://arxiv.org/abs/2111.03606}

\bibitem[{Abbott {et~al.}(2021{\natexlab{c}})Abbott, Abbott, Abraham, Acernese,
  Ackley, Adams, Adams, Adhikari, Adya, Affeldt, Agarwal, Agathos, Agatsuma,
  Aggarwal, Aguiar, Aiello, Ain, Ajith, Akutsu, Aleman, Allen, Allocca, Altin,
  Amato, Anand, Ananyeva, Anderson, Anderson, Ando, Angelova, Ansoldi, Antelis,
  Antier, Appert, Arai, Arai, Arai, Araki, Araya, Araya, Areeda, Arène,
  Aritomi, Arnaud, Aronson, Arun, Asada, Asali, Ashton, Aso, Aston, Astone,
  Aubin, Aufmuth, AultONeal, Austin, Babak, Badaracco, Bader, Bae, Bae, Baer,
  Bagnasco, Bai, Baiotti, Baird, Bajpai, Ball, Ballardin, Ballmer, Bals,
  Balsamo, Baltus, Banagiri, Bankar, Bankar, Barayoga, Barbieri, Barish,
  Barker, Barneo, Barone, Barr, Barsotti, Barsuglia, Barta, Bartlett, Barton,
  Bartos, Bassiri, Basti, Bawaj, Bayley, Baylor, Bazzan, Bécsy, Bedakihale,
  Bejger, Belahcene, Benedetto, Beniwal, Benjamin, Benkel, Bennett, Bentley,
  BenYaala, Bergamin, Berger, Bernuzzi, Berry, Bersanetti, Bertolini,
  Betzwieser, Bhandare, Bhandari, Bhattacharjee, Bhaumik, Bidler, Bilenko,
  Billingsley, Birney, Birnholtz, Biscans, Bischi, Biscoveanu, Bisht, Biswas,
  Bitossi, Bizouard, Blackburn, Blackman, Blair, Blair, Blair, Bobba, Bode,
  Boer, Bogaert, Boldrini, Bondu, Bonilla, Bonnand, Booker, Boom, Bork, Boschi,
  Bose, Bose, Bossilkov, Boudart, Bouffanais, Bozzi, Bradaschia, Brady,
  Bramley, Branch, Branchesi, Brau, Breschi, Briant, Briggs, Brillet,
  Brinkmann, Brockill, Brooks, Brooks, Brown, Brunett, Bruno, Bruntz, Bryant,
  Buikema, Bulik, Bulten, Buonanno, Buscicchio, Buskulic, Byer, Cadonati,
  Caesar, Cagnoli, Cahillane, Cain~III, Bustillo, Callaghan, Callister,
  Calloni, Camp, Canepa, Cannavacciuolo, Cannon, Cao, Cao, Cao, Capocasa,
  Capote, Carapella, Carbognani, Carlin, Carney, Carpinelli, Carullo, Carver,
  Diaz, Casentini, Castaldi, Caudill, Cavaglià, Cavalier, Cavalieri, Cella,
  Cerdá-Durán, Cesarini, Chaibi, Chakravarti, Champion, Chan, Chan, Chan,
  Chan, Chandra, Chanial, Chao, Charlton, Chase, Chassande-Mottin, Chatterjee,
  Chaturvedi, Chatziioannou, Chen, Chen, Chen, Chen, Chen, Chen, Chen, Chen,
  Chen, Cheng, Cheong, Cheung, Chia, Chiadini, Chiang, Chierici, Chincarini,
  Chiofalo, Chiummo, Cho, Cho, Choate, Choudhary, Choudhary, Christensen, Chu,
  Chu, Chu, Chua, Chung, Ciani, Ciecielag, Cieślar, Cifaldi, Ciobanu, Ciolfi,
  Cipriano, Cirone, Clara, Clark, Clark, Clarke, Clearwater, Clesse, Cleva,
  Coccia, Cohadon, Cohen, Cohen, Colleoni, Collette, Colpi, Compton,
  Constancio~Jr., Conti, Cooper, Corban, Corbitt, Cordero-Carrión, Corezzi,
  Corley, Cornish, Corre, Corsi, Cortese, Costa, Cotesta, Coughlin, Coughlin,
  Coulon, Countryman, Cousins, Couvares, Covas, Coward, Cowart, Coyne, Coyne,
  Creighton, Creighton, Criswell, Croquette, Crowder, Cudell, Cullen, Cumming,
  Cummings, Cuoco, Curyło, Canton, Dálya, Dana, DaneshgaranBajastani,
  D'Angelo, Danilishin, D'Antonio, Danzmann, Darsow-Fromm, Dasgupta, Datrier,
  Dattilo, Dave, Davier, Davies, Davis, Daw, Dean, DeBra, Deenadayalan,
  Degallaix, De~Laurentis, Deléglise, Del~Favero, De~Lillo, De~Lillo,
  Del~Pozzo, DeMarchi, De~Matteis, D'Emilio, Demos, Dent, Depasse, De~Pietri,
  De~Rosa, De~Rossi, DeSalvo, De~Simone, Dhurandhar, Díaz, Diaz-Ortiz~Jr.,
  Didio, Dietrich, Di~Fiore, Di~Fronzo, Di~Giorgio, Di~Giovanni, Di~Girolamo,
  Di~Lieto, Ding, Di~Pace, Di~Palma, Di~Renzo, Divakarla, Dmitriev, Doctor,
  D'Onofrio, Donovan, Dooley, Doravari, Dorrington, Drago, Driggers, Drori, Du,
  Ducoin, Dupej, Durante, D'Urso, Duverne, Dwyer, Easter, Ebersold, Eddolls,
  Edelman, Edo, Edy, Effler, Eguchi, Eichholz, Eikenberry, Eisenmann,
  Eisenstein, Ejlli, Enomoto, Errico, Essick, Estellés, Estevez, Etienne,
  Etzel, Evans, Evans, Ewing, Fafone, Fair, Fairhurst, Fan, Farah, Farinon,
  Farr, Farr, Farrow, Fauchon-Jones, Favata, Fays, Fazio, Feicht, Fejer, Feng,
  Fenyvesi, Ferguson, Fernandez-Galiana, Ferrante, Ferreira, Fidecaro, Figura,
  Fiori, Fishbach, Fisher, Fittipaldi, Fiumara, Flaminio, Floden, Flynn, Fong,
  Font, Fornal, Forsyth, Franke, Frasca, Frasconi, Frederick, Frei, Freise,
  Frey, Fritschel, Frolov, Fronzé, Fujii, Fujikawa, Fukunaga, Fukushima,
  Fulda, Fyffe, Gabbard, Gadre, Gaebel, Gair, Gais, Galaudage, Gamba,
  Ganapathy, Ganguly, Gao, Gaonkar, Garaventa, García-Núñez,
  García-Quirós, Garufi, Gateley, Gaudio, Gayathri, Ge, Gemme, Gennai,
  George, Gergely, Gewecke, Ghonge, Ghosh, Ghosh, Ghosh, Ghosh, Ghosh,
  Giacomazzo, Giacoppo, Giaime, Giardina, Gibson, Gier, Giesler, Giri, Gissi,
  Glanzer, Gleckl, Godwin, Goetz, Goetz, Gohlke, Goncharov, González,
  Gopakumar, Gosselin, Gouaty, Grace, Grado, Granata, Granata, Grant, Gras,
  Grassia, Gray, Gray, Greco, Green, Green, Gretarsson, Gretarsson, Griffith,
  Griffiths, Griggs, Grignani, Grimaldi, Grimes, Grimm, Grote, Grunewald,
  Gruning, Guerrero, Guidi, Guimaraes, Guixé, Gulati, Guo, Guo, Gupta, Gupta,
  Gupta, Gustafson, Gustafson, Guzman, Ha, Haegel, Hagiwara, Haino, Halim,
  Hall, Hamilton, Hammond, Han, Haney, Hanks, Hanna, Hannam, Hannuksela,
  Hansen, Hansen, Hanson, Harder, Hardwick, Haris, Harms, Harry, Harry,
  Hartwig, Hasegawa, Haskell, Hasskew, Haster, Hattori, Haughian, Hayakawa,
  Hayama, Hayes, Healy, Heidmann, Heintze, Heinze, Heinzel, Heitmann, Hellman,
  Hello, Helmling-Cornell, Hemming, Hendry, Heng, Hennes, Hennig, Hennig,
  Vivanco, Heurs, Hild, Hill, Himemoto, Hinderer, Hines, Hiranuma, Hirata,
  Hirose, Ho, Hochheim, Hofman, Hohmann, Holgado, Holland, Hollows, Holmes,
  Holt, Holz, Hong, Hopkins, Hough, Howell, Hoy, Hoyland, Hreibi, Hsieh, Hsu,
  Huang, Huang, Huang, Huang, Huang, Huang, Hübner, Huddart, Huerta, Hughey,
  Hui, Hui, Husa, Huttner, Huxford, Huynh-Dinh, Ide, Idzkowski, Iess, Ikenoue,
  Imam, Inayoshi, Inchauspe, Ingram, Inoue, Intini, Ioka, Isi, Isleif, Ito,
  Itoh, Iyer, Izumi, JaberianHamedan, Jacqmin, Jadhav, Jadhav, James, Jan,
  Jani, Janssens, Janthalur, Jaranowski, Jariwala, Jaume, Jenkins, Jeon,
  Jeunon, Jia, Jiang, Jin, Johns, Jones, Jones, Jones, Jones, Jones, Jonker,
  Ju, Jung, Jung, Junker, Kaihotsu, Kajita, Kakizaki, Kalaghatgi, Kalogera,
  Kamai, Kamiizumi, Kanda, Kandhasamy, Kang, Kanner, Kao, Kapadia, Kapasi,
  Karat, Karathanasis, Karki, Kashyap, Kasprzack, Kastaun, Katsanevas,
  Katsavounidis, Katzman, Kaur, Kawabe, Kawaguchi, Kawai, Kawasaki, Kéfélian,
  Keitel, Key, Khadka, Khalili, Khan, Khan, Khazanov, Khetan, Khursheed,
  Kijbunchoo, Kim, Kim, Kim, Kim, Kim, Kim, Kimball, Kimura, King,
  Kinley-Hanlon, Kirchhoff, Kissel, Kita, Kitazawa, Kleybolte, Klimenko, Knee,
  Knowles, Knyazev, Koch, Koekoek, Kojima, Kokeyama, Koley, Kolitsidou,
  Kolstein, Komori, Kondrashov, Kong, Kontos, Koper, Korobko, Kotake, Kovalam,
  Kozak, Kozakai, Kozu, Kringel, Krishnendu, Królak, Kuehn, Kuei, Kumar,
  Kumar, Kumar, Kumar, Kume, Kuns, Kuo, Kuo, Kuromiya, Kuroyanagi, Kusayanagi,
  Kwak, Kwang, Laghi, Lalande, Lam, Lamberts, Landry, Landry, Lane, Lang,
  Lange, Lantz, La~Rosa, Lartaux-Vollard, Lasky, Laxen, Lazzarini, Lazzaro,
  Leaci, Leavey, Lecoeuche, Lee, Lee, Lee, Lee, Lee, Lee, Lehmann, Lemaître,
  Leon, Leonardi, Leroy, Letendre, Levin, Leviton, Li, Li, Li, Li, Li, Li, Lin,
  Lin, Lin, Lin, Lin, Linde, Linker, Linley, Littenberg, Liu, Liu, Liu, Liu,
  Llorens-Monteagudo, Lo, Lockwood, Lollie, London, Longo, Lopez, Lorenzini,
  Loriette, Lormand, Losurdo, Lough, Lousto, Lovelace, Lück, Lumaca, Lundgren,
  Luo, Macas, MacInnis, Macleod, MacMillan, Macquet, Hernandez,
  Magaña-Sandoval, Magazzù, Magee, Maggiore, Majorana, Makarem, Maksimovic,
  Maliakal, Malik, Man, Mandic, Mangano, Mango, Mansell, Manske, Mantovani,
  Mapelli, Marchesoni, Marchio, Marion, Mark, Márka, Márka, Markakis,
  Markosyan, Markowitz, Maros, Marquina, Marsat, Martelli, Martin, Martin,
  Martinez, Martinez, Martinovic, Martynov, Marx, Masalehdan, Mason, Massera,
  Masserot, Massinger, Masso-Reid, Mastrogiovanni, Matas, Mateu-Lucena,
  Matichard, Matiushechkina, Mavalvala, McCann, McCarthy, McClelland, McClincy,
  McCormick, McCuller, McGhee, McGuire, McIsaac, McIver, McManus, McRae,
  McWilliams, Meacher, Mehmet, Mehta, Melatos, Melchor, Mendell,
  Menendez-Vazquez, Menoni, Mercer, Mereni, Merfeld, Merilh, Merritt,
  Merzougui, Meshkov, Messenger, Messick, Meyers, Meylahn, Mhaske, Miani, Miao,
  Michaloliakos, Michel, Michimura, Middleton, Milano, Miller, Millhouse,
  Mills, Milotti, Milovich-Goff, Minazzoli, Minenkov, Mio, Mir, Mishkin,
  Mishra, Mishra, Mistry, Mitra, Mitrofanov, Mitselmakher, Mittleman, Miyakawa,
  Miyamoto, Miyazaki, Miyo, Miyoki, Mo, Mogushi, Mohapatra, Mohite, Molina,
  Molina-Ruiz, Mondin, Montani, Moore, Moraru, Morawski, More, Moreno, Moreno,
  Mori, Morisaki, Moriwaki, Mours, Mow-Lowry, Mozzon, Muciaccia, Mukherjee,
  Mukherjee, Mukherjee, Mukherjee, Mukund, Mullavey, Munch, Muñiz, Murray,
  Musenich, Nadji, Nagano, Nagano, Nagar, Nakamura, Nakano, Nakano, Nakashima,
  Nakayama, Nardecchia, Narikawa, Naticchioni, Nayak, Nayak, Negishi, Neil,
  Neilson, Nelemans, Nelson, Nery, Neunzert, Ng, Ng, Nguyen, Nguyen, Nguyen,
  Quynh, Ni, Nichols, Nishizawa, Nissanke, Nocera, Noh, Norman, North, Nozaki,
  Nuttall, Oberling, O'Brien, Obuchi, O'Dell, Ogaki, Oganesyan, Oh, Oh, Oh,
  Ohashi, Ohishi, Ohkawa, Ohme, Ohta, Okada, Okutani, Okutomi, Olivetto,
  Oohara, Ooi, Oram, O'Reilly, Ormiston, Ormsby, Ortega, O'Shaughnessy, O'Shea,
  Oshino, Ossokine, Osthelder, Otabe, Ottaway, Overmier, Pace, Pagano, Page,
  Pagliaroli, Pai, Pai, Palamos, Palashov, Palomba, Pan, Panda, Pang, Pang,
  Pankow, Pannarale, Pant, Paoletti, Paoli, Paolone, Parisi, Park, Parker,
  Pascucci, Pasqualetti, Passaquieti, Passuello, Patel, Patricelli, Payne,
  Pechsiri, Pedraza, Pegoraro, Pele, Arellano, Penn, Perego, Pereira, Pereira,
  Perez, Périgois, Perreca, Perriès, Petermann, Petterson, Pfeiffer, Pham,
  Phukon, Piccinni, Pichot, Piendibene, Piergiovanni, Pierini, Pierro, Pillant,
  Pilo, Pinard, Pinto, Piotrzkowski, Piotrzkowski, Pirello, Pitkin, Placidi,
  Plastino, Pluchar, Poggiani, Polini, Pong, Ponrathnam, Popolizio, Porter,
  Powell, Pracchia, Pradier, Prajapati, Prasai, Prasanna, Pratten, Prestegard,
  Principe, Prodi, Prokhorov, Prosposito, Prudenzi, Puecher, Punturo, Puosi,
  Puppo, Pürrer, Qi, Quetschke, Quinonez, Quitzow-James, Raab, Raaijmakers,
  Radkins, Radulesco, Raffai, Rail, Raja, Rajan, Ramirez, Ramirez,
  Ramos-Buades, Rana, Rapagnani, Rapol, Ratto, Ray, Raymond, Raza, Razzano,
  Read, Rees, Regimbau, Rei, Reid, Reitze, Relton, Rettegno, Ricci, Richardson,
  Richardson, Richardson, Ricker, Riemenschneider, Riles, Rizzo, Robertson,
  Robie, Robinet, Rocchi, Rocha, Rodriguez, Rodriguez-Soto, Rolland, Rollins,
  Roma, Romanelli, Romano, Romel, Romero, Romero-Shaw, Romie, Rose, Rosińska,
  Rosofsky, Ross, Rowan, Rowlinson, Roy, Roy, Rozza, Ruggi, Ryan, Sachdev,
  Sadecki, Sadiq, Sago, Saito, Saito, Sakai, Sakai, Sakellariadou, Sakuno,
  Salafia, Salconi, Saleem, Salemi, Samajdar, Sanchez, Sanchez, Sanchez,
  Sanchis-Gual, Sanders, Sanuy, Saravanan, Sarin, Sassolas, Satari,
  Sathyaprakash, Sato, Sato, Sauter, Savage, Savant, Sawada, Sawant, Sawant,
  Sayah, Schaetzl, Scheel, Scheuer, Schindler-Tyka, Schmidt, Schnabel,
  Schneewind, Schofield, Schönbeck, Schulte, Schutz, Schwartz, Scott, Scott,
  Seglar-Arroyo, Seidel, Sekiguchi, Sekiguchi, Sellers, Sengupta, Sennett,
  Sentenac, Seo, Sequino, Sergeev, Setyawati, Shaffer, Shahriar, Shams, Shao,
  Sharifi, Sharma, Sharma, Shawhan, Shcheblanov, Shen, Shibagaki, Shikauchi,
  Shimizu, Shimoda, Shimode, Shink, Shinkai, Shishido, Shoda, Shoemaker,
  Shoemaker, Shukla, ShyamSundar, Sieniawska, Sigg, Singer, Singh, Singh,
  Singha, Sintes, Sipala, Skliris, Slagmolen, Slaven-Blair, Smetana, Smith,
  Smith, Somala, Somiya, Son, Soni, Soni, Sorazu, Sordini, Sorrentino,
  Sorrentino, Sotani, Soulard, Souradeep, Sowell, Spagnuolo, Spencer, Spera,
  Srivastava, Srivastava, Staats, Stachie, Steer, Steinlechner, Steinlechner,
  Stops, Stevenson, Stover, Strain, Strang, Stratta, Strunk, Sturani, Stuver,
  Südbeck, Sudhagar, Sudhir, Sugimoto, Suh, Summerscales, Sun, Sun, Sunil,
  Sur, Suresh, Sutton, Suzuki, Suzuki, Swinkels, Szczepańczyk, Szewczyk,
  Tacca, Tagoshi, Tait, Takahashi, Takahashi, Takamori, Takano, Takeda, Takeda,
  Talbot, Tanaka, Tanaka, Tanaka, Tanaka, Tanaka, Tanasijczuk, Tanioka, Tanner,
  Tao, Tapia, Martín, Tasson, Telada, Tenorio, Terkowski, Test,
  Thirugnanasambandam, Thomas, Thomas, Thompson, Thondapu, Thorne, Thrane,
  Tiwari, Tiwari, Tiwari, Toland, Tolley, Tomaru, Tomigami, Tomura, Tonelli,
  Torres-Forné, Torrie, Melo, Töyrä, Trapananti, Travasso, Traylor,
  Tringali, Tripathee, Troiano, Trovato, Trozzo, Trudeau, Tsai, Tsai, Tsang,
  Tsang, Tsao, Tse, Tso, Tsubono, Tsuchida, Tsukada, Tsuna, Tsutsui, Tsuzuki,
  Turconi, Tuyenbayev, Ubhi, Uchikata, Uchiyama, Udall, Ueda, Uehara, Ueno,
  Ueshima, Ugolini, Unnikrishnan, Uraguchi, Urban, Ushiba, Usman, Utina,
  Vahlbruch, Vajente, Vajpeyi, Valdes, Valentini, Valsan, van Bakel, van
  Beuzekom, Brand, Broeck, Vander-Hyde, van~der Schaaf, van Heijningen,
  Vanosky, van Putten, Vardaro, Vargas, Varma, Vasúth, Vecchio, Vedovato,
  Veitch, Veitch, Venkateswara, Venneberg, Venugopalan, Verkindt, Verma, Veske,
  Vetrano, Viceré, Viets, Villa-Ortega, Vinet, Vitale, Vo, Vocca, von Reis,
  von Wrangel, Vorvick, Vyatchanin, Wade, Wade, Wagner, Walet, Walker, Wallace,
  Wallace, Walsh, Wang, Wang, Wang, Ward, Warner, Was, Washimi, Washington,
  Watchi, Weaver, Wei, Weinert, Weinstein, Weiss, Weller, Wellmann, Wen,
  Weßels, Westhouse, Wette, Whelan, White, Whiting, Whittle, Wilken, Williams,
  Williams, Williamson, Willis, Willke, Wilson, Winkler, Wipf, Wlodarczyk,
  Woan, Woehler, Wofford, Wong, Wu, Wu, Wu, Wu, Wysocki, Xiao, Xu, Yamada,
  Yamamoto, Yamamoto, Yamamoto, Yamamoto, Yamashita, Yamazaki, Yang, Yang,
  Yang, Yang, Yang, Yap, Yeeles, Yelikar, Ying, Yokogawa, Yokoyama, Yokozawa,
  Yoon, Yoshioka, Yu, Yu, Yuzurihara, Zadrożny, Zanolin, Zappa, Zeidler,
  Zelenova, Zendri, Zevin, Zhan, Zhang, Zhang, Zhang, Zhang, Zhang, Zhao, Zhao,
  Zhao, Zhao, Zhou, Zhu, Zhu, Zimmerman, Zlochower, Zucker, \&
  Zweizig}]{abbott_observation_2021}
Abbott, R., Abbott, T.~D., Abraham, S., {et~al.} 2021{\natexlab{c}}, The
  Astrophysical Journal Letters, 915, L5, \dodoi{10.3847/2041-8213/ac082e}

\bibitem[{Abbott {et~al.}(2022{\natexlab{a}})Abbott, Abbott, Acernese, Ackley,
  Adams, Adhikari, Adhikari, Adya, Affeldt, Agarwal, Agathos, Agatsuma,
  Aggarwal, Aguiar, Aiello, Ain, Ajith, Albanesi, Allocca, Altin, Amato, Anand,
  Anand, Ananyeva, Anderson, Anderson, Andrade, Andres, Andrić, Angelova,
  Ansoldi, Antelis, Antier, Appert, Arai, Araya, Areeda, Arène, Arnaud,
  Aronson, Arun, Asali, Ashton, Assiduo, Aston, Astone, Aubin, Austin, Babak,
  Badaracco, Bader, Badger, Bae, Baer, Bagnasco, Bai, Baird, Ball, Ballardin,
  Ballmer, Balsamo, Baltus, Banagiri, Bankar, Barayoga, Barbieri, Barish,
  Barker, Barneo, Barone, Barr, Barsotti, Barsuglia, Barta, Bartlett, Barton,
  Bartos, Bassiri, Basti, Bawaj, Bayley, Baylor, Bazzan, Bécsy, Bedakihale,
  Bejger, Belahcene, Benedetto, Beniwal, Bennett, Bentley, BenYaala, Bergamin,
  Berger, Bernuzzi, Berry, Bersanetti, Bertolini, Betzwieser, Beveridge,
  Bhandare, Bhardwaj, Bhattacharjee, Bhaumik, Bilenko, Billingsley, Bini,
  Birney, Birnholtz, Biscans, Bischi, Biscoveanu, Bisht, Biswas, Bitossi,
  Bizouard, Blackburn, Blair, Blair, Blair, Bobba, Bode, Boer, Bogaert,
  Boldrini, Bonavena, Bondu, Bonilla, Bonnand, Booker, Boom, Bork, Boschi,
  Bose, Bose, Bossilkov, Boudart, Bouffanais, Bozzi, Bradaschia, Brady,
  Bramley, Branch, Branchesi, Brau, Breschi, Briant, Briggs, Brillet,
  Brinkmann, Brockill, Brooks, Brooks, Brown, Brunett, Bruno, Bruntz, Bryant,
  Bulik, Bulten, Buonanno, Buscicchio, Buskulic, Buy, Byer, Cadonati, Cagnoli,
  Cahillane, Bustillo, Callaghan, Callister, Calloni, Cameron, Camp, Canepa,
  Canevarolo, Cannavacciuolo, Cannon, Cao, Capote, Carapella, Carbognani,
  Carlin, Carney, Carpinelli, Carrillo, Carullo, Carver, Diaz, Casentini,
  Castaldi, Caudill, Cavaglià, Cavalier, Cavalieri, Ceasar, Cella,
  Cerdá-Durán, Cesarini, Chaibi, Chakravarti, Subrahmanya, Champion, Chan,
  Chan, Chan, Chan, Chandra, Chanial, Chao, Charlton, Chase, Chassande-Mottin,
  Chatterjee, Chatterjee, Chatterjee, Chattopadhyay, Chaturvedi, Chaty,
  Chatziioannou, Chen, Chen, Chen, Chen, Chen, Cheng, Cheong, Cheung, Chia,
  Chiadini, Chiarini, Chierici, Chincarini, Chiofalo, Chiummo, Cho, Cho,
  Choudhary, Choudhary, Christensen, Chu, Chua, Chung, Ciani, Ciecielag,
  Cieślar, Cifaldi, Ciobanu, Ciolfi, Cipriano, Cirone, Clara, Clark, Clark,
  Clarke, Clearwater, Clesse, Cleva, Coccia, Codazzo, Cohadon, Cohen, Cohen,
  Colleoni, Collette, Colombo, Colpi, Compton, Constancio~Jr., Conti, Cooper,
  Corban, Corbitt, Cordero-Carrión, Corezzi, Corley, Cornish, Corre, Corsi,
  Cortese, Costa, Cotesta, Coughlin, Coulon, Countryman, Cousins, Couvares,
  Coward, Cowart, Coyne, Coyne, Creighton, Creighton, Criswell, Croquette,
  Crowder, Cudell, Cullen, Cumming, Cummings, Cunningham, Cuoco, Curyło,
  Dabadie, Canton, Dall'Osso, Dálya, Dana, DaneshgaranBajastani, D'Angelo,
  Danila, Danilishin, D'Antonio, Danzmann, Darsow-Fromm, Dasgupta, Datrier,
  Datta, Dattilo, Dave, Davier, Davies, Davis, Davis, Daw, Dean, DeBra,
  Deenadayalan, Degallaix, De~Laurentis, Deléglise, Del~Favero, De~Lillo,
  De~Lillo, Del~Pozzo, DeMarchi, De~Matteis, D'Emilio, Demos, Dent, Depasse,
  De~Pietri, De~Rosa, De~Rossi, DeSalvo, De~Simone, Dhurandhar, Díaz,
  Diaz-Ortiz~Jr., Didio, Dietrich, Di~Fiore, Di~Fronzo, Di~Giorgio,
  Di~Giovanni, Di~Giovanni, Di~Girolamo, Di~Lieto, Ding, Di~Pace, Di~Palma,
  Di~Renzo, Divakarla, Divyajyoti, Dmitriev, Doctor, D'Onofrio, Donovan,
  Dooley, Doravari, Dorrington, Drago, Driggers, Drori, Ducoin, Dupej, Durante,
  D'Urso, Duverne, Dwyer, Eassa, Easter, Ebersold, Eckhardt, Eddolls, Edelman,
  Edo, Edy, Effler, Eichholz, Eikenberry, Eisenmann, Eisenstein, Ejlli,
  Engelby, Errico, Essick, Estellés, Estevez, Etienne, Etzel, Etzel, Evans,
  Evans, Ewing, Fafone, Fair, Fairhurst, Fanning, Farah, Farinon, Farr, Farr,
  Farrow, Fauchon-Jones, Favaro, Favata, Fays, Fazio, Feicht, Fejer, Fenyvesi,
  Ferguson, Fernandez-Galiana, Ferrante, Ferreira, Fidecaro, Figura, Fiori,
  Fishbach, Fisher, Fittipaldi, Fiumara, Flaminio, Floden, Fong, Font, Fornal,
  Forsyth, Franke, Frasca, Frasconi, Frederick, Freed, Frei, Freise, Frey,
  Fritschel, Frolov, Fronzé, Fulda, Fyffe, Gabbard, Gabella, Gadre, Gair,
  Gais, Galaudage, Gamba, Ganapathy, Ganguly, Gaonkar, Garaventa, García,
  García-Núñez, García-Quirós, Garufi, Gateley, Gaudio, Gayathri, Gemme,
  Gennai, George, George, Gerberding, Gergely, Gewecke, Ghonge, Ghosh, Ghosh,
  Ghosh, Ghosh, Giacomazzo, Giacoppo, Giaime, Giardina, Gibson, Gier, Giesler,
  Giri, Gissi, Glanzer, Gleckl, Godwin, Goetz, Goetz, Gohlke, Goncharov,
  González, Gopakumar, Gosselin, Gouaty, Gould, Grace, Grado, Granata,
  Granata, Grant, Gras, Grassia, Gray, Gray, Greco, Green, Green, Gretarsson,
  Gretarsson, Griffith, Griffiths, Griggs, Grignani, Grimaldi, Grimm, Grote,
  Grunewald, Gruning, Guerra, Guidi, Guimaraes, Guixé, Gulati, Guo, Guo,
  Gupta, Gupta, Gupta, Gustafson, Gustafson, Guzman, Haegel, Halim, Hall,
  Hamilton, Hammond, Haney, Hanks, Hanna, Hannam, Hannuksela, Hansen, Hansen,
  Hanson, Harder, Hardwick, Haris, Harms, Harry, Harry, Hartwig, Haskell,
  Hasskew, Haster, Haughian, Hayes, Healy, Heidmann, Heidt, Heintze, Heinze,
  Heinzel, Heitmann, Hellman, Hello, Helmling-Cornell, Hemming, Hendry, Heng,
  Hennes, Hennig, Hennig, Hernandez, Vivanco, Heurs, Hild, Hill, Hines,
  Hochheim, Hofman, Hohmann, Holcomb, Holland, Holley-Bockelmann, Hollows,
  Holmes, Holt, Holz, Hopkins, Hough, Hourihane, Howell, Hoy, Hoyland, Hreibi,
  Hsu, Huang, Hübner, Huddart, Hughey, Hui, Husa, Huttner, Huxford,
  Huynh-Dinh, Idzkowski, Iess, Ingram, Isi, Isleif, Iyer, JaberianHamedan,
  Jacqmin, Jadhav, Jadhav, James, Jan, Jani, Janquart, Janssens, Janthalur,
  Jaranowski, Jariwala, Jaume, Jenkins, Jenner, Jeunon, Jia, Johns,
  Johnson-McDaniel, Jones, Jones, Jones, Jones, Jones, Jonker, Ju, Junker,
  Juste, Kalaghatgi, Kalogera, Kamai, Kandhasamy, Kang, Kanner, Kao, Kapadia,
  Kapasi, Karat, Karathanasis, Karki, Kashyap, Kasprzack, Kastaun, Katsanevas,
  Katsavounidis, Katzman, Kaur, Kawabe, Kéfélian, Keitel, Key, Khadka,
  Khalili, Khan, Khazanov, Khetan, Khursheed, Kijbunchoo, Kim, Kim, Kim, Kim,
  Kim, Kimball, Kinley-Hanlon, Kirchhoff, Kissel, Kleybolte, Klimenko, Knee,
  Knowles, Knyazev, Koch, Koekoek, Koley, Kolitsidou, Kolstein, Komori,
  Kondrashov, Kontos, Koper, Korobko, Kovalam, Kozak, Kringel, Krishnendu,
  Królak, Kuehn, Kuei, Kuijer, Kumar, Kumar, Kumar, Kumar, Kuns, Kuwahara,
  Lagabbe, Laghi, Lalande, Lam, Lamberts, Landry, Lane, Lang, Lange, Lantz,
  La~Rosa, Lartaux-Vollard, Lasky, Laxen, Lazzarini, Lazzaro, Leaci, Leavey,
  Lecoeuche, Lee, Lee, Lee, Lee, Lehmann, Lemaître, Leroy, Letendre, Levesque,
  Levin, Leviton, Leyde, Li, Li, Li, Li, Li, Linde, Linker, Linley, Littenberg,
  Liu, Liu, Liu, Llamas, Llorens-Monteagudo, Lo, Lockwood, London, Longo,
  Lopez, Portilla, Lorenzini, Loriette, Lormand, Losurdo, Lott, Lough, Lousto,
  Lovelace, Lucaccioni, Lück, Lumaca, Lundgren, Lynam, Macas, MacInnis,
  Macleod, MacMillan, Macquet, Hernandez, Magazzù, Magee, Maggiore, Magnozzi,
  Mahesh, Majorana, Makarem, Maksimovic, Maliakal, Malik, Man, Mandic, Mangano,
  Mango, Mansell, Manske, Mantovani, Mapelli, Marchesoni, Marion, Mark, Márka,
  Márka, Markakis, Markosyan, Markowitz, Maros, Marquina, Marsat, Martelli,
  Martin, Martin, Martinez, Martinez, Martinez, Martinovic, Martynov, Marx,
  Masalehdan, Mason, Massera, Masserot, Massinger, Masso-Reid, Mastrogiovanni,
  Matas, Mateu-Lucena, Matichard, Matiushechkina, Mavalvala, McCann, McCarthy,
  McClelland, McClincy, McCormick, McCuller, McGhee, McGuire, McIsaac, McIver,
  McRae, McWilliams, Meacher, Mehmet, Mehta, Meijer, Melatos, Melchor, Mendell,
  Menendez-Vazquez, Menoni, Mercer, Mereni, Merfeld, Merilh, Merritt,
  Merzougui, Meshkov, Messenger, Messick, Meyers, Meylahn, Mhaske, Miani, Miao,
  Michaloliakos, Michel, Middleton, Milano, Miller, Miller, Miller, Millhouse,
  Mills, Milotti, Minazzoli, Minenkov, Mir, Miravet-Tenés, Mishra, Mishra,
  Mistry, Mitra, Mitrofanov, Mitselmakher, Mittleman, Mo, Moguel, Mogushi,
  Mohapatra, Mohite, Molina, Molina-Ruiz, Mondin, Montani, Moore, Moraru,
  Morawski, More, Moreno, Moreno, Morisaki, Mours, Mow-Lowry, Mozzon,
  Muciaccia, Mukherjee, Mukherjee, Mukherjee, Mukherjee, Mukherjee, Mukund,
  Mullavey, Munch, Muñiz, Murray, Musenich, Muusse, Nadji, Nagar, Napolano,
  Nardecchia, Naticchioni, Nayak, Nayak, Neil, Neilson, Nelemans, Nelson, Nery,
  Neubauer, Neunzert, Ng, Ng, Nguyen, Nguyen, Nguyen, Nichols, Nissanke,
  Nitoglia, Nocera, Norman, North, Nuttall, Oberling, O'Brien, O'Dell, Oelker,
  Oganesyan, Oh, Oh, Ohme, Ohta, Okada, Olivetto, Oram, O'Reilly, Ormiston,
  Ormsby, Ortega, O'Shaughnessy, O'Shea, Ossokine, Osthelder, Ottaway,
  Overmier, Pace, Pagano, Page, Pagliaroli, Pai, Pai, Palamos, Palashov,
  Palomba, Pan, Panda, Pang, Pankow, Pannarale, Pant, Panther, Paoletti, Paoli,
  Paolone, Park, Parker, Pascucci, Pasqualetti, Passaquieti, Passuello, Patel,
  Pathak, Patricelli, Patron, Patrone, Paul, Payne, Pedraza, Pegoraro, Pele,
  Penn, Perego, Pereira, Pereira, Perez, Périgois, Perkins, Perreca, Perriès,
  Petermann, Petterson, Pfeiffer, Pham, Phukon, Piccinni, Pichot, Piendibene,
  Piergiovanni, Pierini, Pierro, Pillant, Pillas, Pilo, Pinard, Pinto, Pinto,
  Piotrzkowski, Pirello, Pitkin, Placidi, Planas, Plastino, Pluchar, Poggiani,
  Polini, Pong, Ponrathnam, Popolizio, Porter, Poulton, Powell, Pracchia,
  Pradier, Prajapati, Prasai, Prasanna, Pratten, Principe, Prodi, Prokhorov,
  Prosposito, Prudenzi, Puecher, Punturo, Puosi, Puppo, Pürrer, Qi, Quetschke,
  Quitzow-James, Raab, Raaijmakers, Radkins, Radulesco, Raffai, Rail, Raja,
  Rajan, Ramirez, Ramirez, Ramos-Buades, Rana, Rapagnani, Rapol, Ray, Raymond,
  Raza, Razzano, Read, Rees, Regimbau, Rei, Reid, Reid, Reitze, Relton,
  Renzini, Rettegno, Reza, Rezac, Ricci, Richards, Richardson, Richardson,
  Riemenschneider, Riles, Rinaldi, Rink, Rizzo, Robertson, Robie, Robinet,
  Rocchi, Rodriguez, Rolland, Rollins, Romanelli, Romano, Romel,
  Romero-Rodríguez, Romero-Shaw, Romie, Ronchini, Rosa, Rose, Rosell,
  Rosińska, Ross, Rowan, Rowlinson, Roy, Roy, Roy, Rozza, Ruggi, Ruiz-Rocha,
  Ryan, Sachdev, Sadecki, Sadiq, Sakellariadou, Salafia, Salconi, Saleem,
  Salemi, Samajdar, Sanchez, Sanchez, Sanchez, Sanchis-Gual, Sanders, Sanuy,
  Saravanan, Sarin, Sassolas, Satari, Sauter, Savage, Sawant, Sawant, Sayah,
  Schaetzl, Scheel, Scheuer, Schiworski, Schmidt, Schmidt, Schnabel,
  Schneewind, Schofield, Schönbeck, Schulte, Schutz, Schwartz, Scott, Scott,
  Seglar-Arroyo, Sellers, Sengupta, Sentenac, Seo, Sequino, Sergeev, Setyawati,
  Shaffer, Shahriar, Shams, Sharma, Sharma, Shawhan, Shcheblanov, Shikauchi,
  Shoemaker, Shoemaker, ShyamSundar, Sieniawska, Sigg, Singer, Singh, Singh,
  Singha, Sintes, Sipala, Skliris, Slagmolen, Slaven-Blair, Smetana, Smith,
  Smith, Soldateschi, Somala, Son, Soni, Soni, Sordini, Sorrentino, Sorrentino,
  Soulard, Souradeep, Sowell, Spagnuolo, Spencer, Spera, Srinivasan,
  Srivastava, Srivastava, Staats, Stachie, Steer, Steinhoff, Steinlechner,
  Steinlechner, Stevenson, Stops, Stover, Strain, Strang, Stratta, Strunk,
  Sturani, Stuver, Sudhagar, Sudhir, Suh, Summerscales, Sun, Sun, Sunil, Sur,
  Suresh, Sutton, Swinkels, Szczepańczyk, Szewczyk, Tacca, Tait, Talbot,
  Talbot, Tanasijczuk, Tanner, Tao, Tao, Martín, Taranto, Tasson, Tenorio,
  Terhune, Terkowski, Thirugnanasambandam, Thomas, Thomas, Thomas, Thompson,
  Thondapu, Thorne, Thrane, Tiwari, Tiwari, Tiwari, Toivonen, Toland, Tolley,
  Tonelli, Torres-Forné, Torrie, Melo, Töyrä, Trapananti, Travasso, Traylor,
  Trevor, Tringali, Tripathee, Troiano, Trovato, Trozzo, Trudeau, Tsai, Tsai,
  Tsang, Tse, Tso, Tsukada, Tsuna, Tsutsui, Turbang, Turconi, Ubhi, Udall,
  Ueno, Unnikrishnan, Urban, Utina, Vahlbruch, Vajente, Vajpeyi, Valdes,
  Valentini, Valsan, van Bakel, van Beuzekom, Brand, Broeck, Vander-Hyde,
  van~der Schaaf, van Heijningen, Vanosky, van Remortel, Vardaro, Vargas,
  Varma, Vasúth, Vecchio, Vedovato, Veitch, Veitch, Venneberg, Venugopalan,
  Verkindt, Verma, Verma, Veske, Vetrano, Viceré, Vidyant, Viets, Vijaykumar,
  Villa-Ortega, Vinet, Virtuoso, Vitale, Vo, Vocca, von Reis, von Wrangel,
  Vorvick, Vyatchanin, Wade, Wade, Wagner, Walet, Walker, Wallace, Wallace,
  Walsh, Wang, Wang, Ward, Warner, Was, Washington, Watchi, Weaver, Webster,
  Weinert, Weinstein, Weiss, Weller, Weller, Wellmann, Wen, Weßels, Wette,
  Whelan, White, Whiting, Whittle, Wilken, Williams, Williams, Williamson,
  Willis, Willke, Wilson, Winkler, Wipf, Wlodarczyk, Woan, Woehler, Wofford,
  Wong, Wu, Wysocki, Xiao, Yamamoto, Yang, Yang, Yang, Yang, Yap, Yeeles,
  Yelikar, Ying, Yoo, Yu, Yu, Zadrożny, Zanolin, Zelenova, Zendri, Zevin,
  Zhang, Zhang, Zhang, Zhang, Zhao, Zhao, Zhao, Zhou, Zhou, Zhu, Zimmerman,
  Zlochower, Zucker, \& Zweizig}]{abbott_gwtc-21_2022}
Abbott, R., Abbott, T.~D., Acernese, F., {et~al.} 2022{\natexlab{a}},
  {GWTC}-2.1: {Deep} {Extended} {Catalog} of {Compact} {Binary} {Coalescences}
  {Observed} by {LIGO} and {Virgo} {During} the {First} {Half} of the {Third}
  {Observing} {Run},  arXiv, \dodoi{10.48550/arXiv.2108.01045}

\bibitem[{Abbott {et~al.}(2022{\natexlab{b}})Abbott, Abbott, Acernese, Ackley,
  Adams, Adhikari, Adhikari, Adya, Affeldt, Agarwal, Agathos, Agatsuma,
  Aggarwal, Aguiar, Aiello, Ain, Ajith, Akutsu, Albanesi, Allocca, Altin,
  Amato, Anand, Anand, Ananyeva, Anderson, Anderson, Ando, Andrade, Andres,
  Andrić, Angelova, Ansoldi, Antelis, Antier, Antonini, Appert, Arai, Arai,
  Arai, Araki, Araya, Araya, Areeda, Arène, Aritomi, Arnaud, Aronson, Arun,
  Asada, Asali, Ashton, Aso, Assiduo, Aston, Astone, Aubin, Austin, Babak,
  Badaracco, Bader, Badger, Bae, Bae, Baer, Bagnasco, Bai, Baiotti, Baird,
  Bajpai, Ball, Ballardin, Ballmer, Balsamo, Baltus, Banagiri, Bankar,
  Barayoga, Barbieri, Barish, Barker, Barneo, Barone, Barr, Barsotti,
  Barsuglia, Barta, Bartlett, Barton, Bartos, Bassiri, Basti, Bawaj, Bayley,
  Baylor, Bazzan, Bécsy, Bedakihale, Bejger, Belahcene, Benedetto, Beniwal,
  Bennett, Bentley, BenYaala, Bergamin, Berger, Bernuzzi, Berry, Bersanetti,
  Bertolini, Betzwieser, Beveridge, Bhandare, Bhardwaj, Bhattacharjee, Bhaumik,
  Bilenko, Billingsley, Bini, Birney, Birnholtz, Biscans, Bischi, Biscoveanu,
  Bisht, Biswas, Bitossi, Bizouard, Blackburn, Blair, Blair, Blair, Bobba,
  Bode, Boer, Bogaert, Boldrini, Bonavena, Bondu, Bonilla, Bonnand, Booker,
  Boom, Bork, Boschi, Bose, Bose, Bossilkov, Boudart, Bouffanais, Bozzi,
  Bradaschia, Brady, Bramley, Branch, Branchesi, Brau, Breschi, Briant, Briggs,
  Brillet, Brinkmann, Brockill, Brooks, Brooks, Brown, Brunett, Bruno, Bruntz,
  Bryant, Bulik, Bulten, Buonanno, Buscicchio, Buskulic, Buy, Byer, Cadonati,
  Cagnoli, Cahillane, Bustillo, Callaghan, Callister, Calloni, Cameron, Camp,
  Canepa, Canevarolo, Cannavacciuolo, Cannon, Cao, Cao, Capocasa, Capote,
  Carapella, Carbognani, Carlin, Carney, Carpinelli, Carrillo, Carullo, Carver,
  Diaz, Casentini, Castaldi, Caudill, Cavaglià, Cavalier, Cavalieri, Ceasar,
  Cella, Cerdá-Durán, Cesarini, Chaibi, Chakravarti, Subrahmanya, Champion,
  Chan, Chan, Chan, Chan, Chan, Chandra, Chanial, Chao, Charlton, Chase,
  Chassande-Mottin, Chatterjee, Chatterjee, Chatterjee, Chaturvedi, Chaty,
  Chatziioannou, Chen, Chen, Chen, Chen, Chen, Chen, Chen, Chen, Cheng, Cheong,
  Cheung, Chia, Chiadini, Chiang, Chiarini, Chierici, Chincarini, Chiofalo,
  Chiummo, Cho, Cho, Choudhary, Choudhary, Christensen, Chu, Chu, Chu, Chua,
  Chung, Ciani, Ciecielag, Cieślar, Cifaldi, Ciobanu, Ciolfi, Cipriano,
  Cirone, Clara, Clark, Clark, Clarke, Clearwater, Clesse, Cleva, Coccia,
  Codazzo, Cohadon, Cohen, Cohen, Colleoni, Collette, Colombo, Colpi, Compton,
  Constancio~Jr., Conti, Cooper, Corban, Corbitt, Cordero-Carrión, Corezzi,
  Corley, Cornish, Corre, Corsi, Cortese, Costa, Cotesta, Coughlin, Coulon,
  Countryman, Cousins, Couvares, Coward, Cowart, Coyne, Coyne, Creighton,
  Creighton, Criswell, Croquette, Crowder, Cudell, Cullen, Cumming, Cummings,
  Cunningham, Cuoco, Curyło, Dabadie, Canton, Dall'Osso, Dálya, Dana,
  DaneshgaranBajastani, D'Angelo, Danilishin, D'Antonio, Danzmann,
  Darsow-Fromm, Dasgupta, Datrier, Datta, Dattilo, Dave, Davier, Davies, Davis,
  Davis, Daw, Dean, DeBra, Deenadayalan, Degallaix, De~Laurentis, Deléglise,
  Del~Favero, De~Lillo, De~Lillo, Del~Pozzo, DeMarchi, De~Matteis, D'Emilio,
  Demos, Dent, Depasse, De~Pietri, De~Rosa, De~Rossi, DeSalvo, De~Simone,
  Dhurandhar, Díaz, Diaz-Ortiz~Jr., Didio, Dietrich, Di~Fiore, Di~Fronzo,
  Di~Giorgio, Di~Giovanni, Di~Giovanni, Di~Girolamo, Di~Lieto, Ding, Di~Pace,
  Di~Palma, Di~Renzo, Divakarla, Dmitriev, Doctor, D'Onofrio, Donovan, Dooley,
  Doravari, Dorrington, Drago, Driggers, Drori, Ducoin, Dupej, Durante, D'Urso,
  Duverne, Dwyer, Eassa, Easter, Ebersold, Eckhardt, Eddolls, Edelman, Edo,
  Edy, Effler, Eguchi, Eichholz, Eikenberry, Eisenmann, Eisenstein, Ejlli,
  Engelby, Enomoto, Errico, Essick, Estellés, Estevez, Etienne, Etzel, Evans,
  Evans, Ewing, Fafone, Fair, Fairhurst, Farah, Farinon, Farr, Farr, Farrow,
  Fauchon-Jones, Favaro, Favata, Fays, Fazio, Feicht, Fejer, Fenyvesi,
  Ferguson, Fernandez-Galiana, Ferrante, Ferreira, Fidecaro, Figura, Fiori,
  Fishbach, Fisher, Fittipaldi, Fiumara, Flaminio, Floden, Fong, Font, Fornal,
  Forsyth, Franke, Frasca, Frasconi, Frederick, Freed, Frei, Freise, Frey,
  Fritschel, Frolov, Fronzé, Fujii, Fujikawa, Fukunaga, Fukushima, Fulda,
  Fyffe, Gabbard, Gadre, Gair, Gais, Galaudage, Gamba, Ganapathy, Ganguly, Gao,
  Gaonkar, Garaventa, García-Núñez, García-Quirós, Garufi, Gateley,
  Gaudio, Gayathri, Ge, Gemme, Gennai, George, Gerberding, Gergely, Gewecke,
  Ghonge, Ghosh, Ghosh, Ghosh, Ghosh, Giacomazzo, Giacoppo, Giaime, Giardina,
  Gibson, Gier, Giesler, Giri, Gissi, Glanzer, Gleckl, Godwin, Goetz, Goetz,
  Gohlke, Golomb, Goncharov, González, Gopakumar, Gosselin, Gouaty, Gould,
  Grace, Grado, Granata, Granata, Grant, Gras, Grassia, Gray, Gray, Greco,
  Green, Green, Gretarsson, Gretarsson, Griffith, Griffiths, Griggs, Grignani,
  Grimaldi, Grimm, Grote, Grunewald, Gruning, Guerra, Guidi, Guimaraes, Guixé,
  Gulati, Guo, Guo, Gupta, Gupta, Gupta, Gustafson, Gustafson, Guzman, Ha,
  Haegel, Hagiwara, Haino, Halim, Hall, Hamilton, Hammond, Han, Haney, Hanks,
  Hanna, Hannam, Hannuksela, Hansen, Hansen, Hanson, Harder, Hardwick, Haris,
  Harms, Harry, Harry, Hartwig, Hasegawa, Haskell, Hasskew, Haster, Hattori,
  Haughian, Hayakawa, Hayama, Hayes, Healy, Heidmann, Heidt, Heintze, Heinze,
  Heinzel, Heitmann, Hellman, Hello, Helmling-Cornell, Hemming, Hendry, Heng,
  Hennes, Hennig, Hennig, Hernandez, Vivanco, Heurs, Hild, Hill, Himemoto,
  Hines, Hiranuma, Hirata, Hirose, Hochheim, Hofman, Hohmann, Holcomb, Holland,
  Hollows, Holmes, Holt, Holz, Hong, Hopkins, Hough, Hourihane, Howell, Hoy,
  Hoyland, Hreibi, Hsieh, Hsu, Huang, Huang, Huang, Huang, Huang, Huang,
  Hübner, Huddart, Hughey, Hui, Hui, Husa, Huttner, Huxford, Huynh-Dinh, Ide,
  Idzkowski, Iess, Ikenoue, Imam, Inayoshi, Ingram, Inoue, Ioka, Isi, Isleif,
  Ito, Itoh, Iyer, Izumi, JaberianHamedan, Jacqmin, Jadhav, Jadhav, James, Jan,
  Jani, Janquart, Janssens, Janthalur, Jaranowski, Jariwala, Jaume, Jenkins,
  Jenner, Jeon, Jeunon, Jia, Jin, Johns, Jones, Jones, Jones, Jones, Jones,
  Jonker, Ju, Jung, Jung, Junker, Juste, Kaihotsu, Kajita, Kakizaki,
  Kalaghatgi, Kalogera, Kamai, Kamiizumi, Kanda, Kandhasamy, Kang, Kanner, Kao,
  Kapadia, Kapasi, Karat, Karathanasis, Karki, Kashyap, Kasprzack, Kastaun,
  Katsanevas, Katsavounidis, Katzman, Kaur, Kawabe, Kawaguchi, Kawai, Kawasaki,
  Kéfélian, Keitel, Key, Khadka, Khalili, Khan, Khazanov, Khetan, Khursheed,
  Kijbunchoo, Kim, Kim, Kim, Kim, Kim, Kim, Kimball, Kimura, Kinley-Hanlon,
  Kirchhoff, Kissel, Kita, Kitazawa, Kleybolte, Klimenko, Knee, Knowles,
  Knyazev, Koch, Koekoek, Kojima, Kokeyama, Koley, Kolitsidou, Kolstein,
  Komori, Kondrashov, Kong, Kontos, Koper, Korobko, Kotake, Kovalam, Kozak,
  Kozakai, Kozu, Kringel, Krishnendu, Królak, Kuehn, Kuei, Kuijer, Kumar,
  Kumar, Kumar, Kumar, Kume, Kuns, Kuo, Kuo, Kuromiya, Kuroyanagi, Kusayanagi,
  Kuwahara, Kwak, Lagabbe, Laghi, Lalande, Lam, Lamberts, Landry, Landry, Lane,
  Lang, Lange, Lantz, La~Rosa, Lartaux-Vollard, Lasky, Laxen, Lazzarini,
  Lazzaro, Leaci, Leavey, Lecoeuche, Lee, Lee, Lee, Lee, Lee, Lee, Lehmann,
  Lemaître, Leonardi, Leroy, Letendre, Levesque, Levin, Leviton, Leyde, Li,
  Li, Li, Li, Li, Li, Lin, Lin, Lin, Lin, Lin, Linde, Linker, Linley,
  Littenberg, Liu, Liu, Liu, Liu, Llamas, Llorens-Monteagudo, Lo, Lockwood,
  London, Longo, Lopez, Portilla, Lorenzini, Loriette, Lormand, Losurdo, Lott,
  Lough, Lousto, Lovelace, Lucaccioni, Lück, Lumaca, Lundgren, Luo, Lynam,
  Macas, MacInnis, Macleod, MacMillan, Macquet, Hernandez, Magazzù, Magee,
  Maggiore, Magnozzi, Mahesh, Majorana, Makarem, Maksimovic, Maliakal, Malik,
  Man, Mandic, Mangano, Mango, Mansell, Manske, Mantovani, Mapelli, Marchesoni,
  Marchio, Marion, Mark, Márka, Márka, Markakis, Markosyan, Markowitz, Maros,
  Marquina, Marsat, Martelli, Martin, Martin, Martinez, Martinez, Martinez,
  Martinovic, Martynov, Marx, Masalehdan, Mason, Massera, Masserot, Massinger,
  Masso-Reid, Mastrogiovanni, Matas, Mateu-Lucena, Matichard, Matiushechkina,
  Mavalvala, McCann, McCarthy, McClelland, McClincy, McCormick, McCuller,
  McGhee, McGuire, McIsaac, McIver, McRae, McWilliams, Meacher, Mehmet, Mehta,
  Meijer, Melatos, Melchor, Mendell, Menendez-Vazquez, Menoni, Mercer, Mereni,
  Merfeld, Merilh, Merritt, Merzougui, Meshkov, Messenger, Messick, Meyers,
  Meylahn, Mhaske, Miani, Miao, Michaloliakos, Michel, Michimura, Middleton,
  Milano, Miller, Miller, Miller, Miller, Millhouse, Mills, Milotti, Minazzoli,
  Minenkov, Mio, Mir, Miravet-Tenés, Mishra, Mishra, Mistry, Mitra,
  Mitrofanov, Mitselmakher, Mittleman, Miyakawa, Miyamoto, Miyazaki, Miyo,
  Miyoki, Mo, Moguel, Mogushi, Mohapatra, Mohite, Molina, Molina-Ruiz, Mondin,
  Montani, Moore, Moraru, Morawski, More, Moreno, Moreno, Mori, Morisaki,
  Moriwaki, Mours, Mow-Lowry, Mozzon, Muciaccia, Mukherjee, Mukherjee,
  Mukherjee, Mukherjee, Mukherjee, Mukund, Mullavey, Munch, Muñiz, Murray,
  Musenich, Muusse, Nadji, Nagano, Nagano, Nagar, Nakamura, Nakano, Nakano,
  Nakashima, Nakayama, Napolano, Nardecchia, Narikawa, Naticchioni, Nayak,
  Nayak, Negishi, Neil, Neilson, Nelemans, Nelson, Nery, Neubauer, Neunzert,
  Ng, Ng, Nguyen, Nguyen, Nguyen, Quynh, Ni, Nichols, Nishizawa, Nissanke,
  Nitoglia, Nocera, Norman, North, Nozaki, Nuttall, Oberling, O'Brien, Obuchi,
  O'Dell, Oelker, Ogaki, Oganesyan, Oh, Oh, Oh, Ohashi, Ohishi, Ohkawa, Ohme,
  Ohta, Okada, Okutani, Okutomi, Olivetto, Oohara, Ooi, Oram, O'Reilly,
  Ormiston, Ormsby, Ortega, O'Shaughnessy, O'Shea, Oshino, Ossokine, Osthelder,
  Otabe, Ottaway, Overmier, Pace, Pagano, Page, Pagliaroli, Pai, Pai, Palamos,
  Palashov, Palomba, Pan, Pan, Panda, Pang, Pang, Pankow, Pannarale, Pant,
  Panther, Paoletti, Paoli, Paolone, Parisi, Park, Park, Parker, Pascucci,
  Pasqualetti, Passaquieti, Passuello, Patel, Pathak, Patricelli, Patron, Paul,
  Payne, Pedraza, Pegoraro, Pele, Arellano, Penn, Perego, Pereira, Pereira,
  Perez, Périgois, Perkins, Perreca, Perriès, Petermann, Petterson, Pfeiffer,
  Pham, Phukon, Piccinni, Pichot, Piendibene, Piergiovanni, Pierini, Pierro,
  Pillant, Pillas, Pilo, Pinard, Pinto, Pinto, Piotrzkowski, Pirello, Pitkin,
  Placidi, Planas, Plastino, Pluchar, Poggiani, Polini, Pong, Ponrathnam,
  Popolizio, Porter, Poulton, Powell, Pracchia, Pradier, Prajapati, Prasai,
  Prasanna, Pratten, Principe, Prodi, Prokhorov, Prosposito, Prudenzi, Puecher,
  Punturo, Puosi, Puppo, Pürrer, Qi, Quetschke, Quitzow-James, Raab,
  Raaijmakers, Radkins, Radulesco, Raffai, Rail, Raja, Rajan, Ramirez, Ramirez,
  Ramos-Buades, Rana, Rapagnani, Rapol, Ray, Raymond, Raza, Razzano, Read,
  Rees, Regimbau, Rei, Reid, Reid, Reitze, Relton, Renzini, Rettegno, Rezac,
  Ricci, Richards, Richardson, Richardson, Riemenschneider, Riles, Rinaldi,
  Rink, Rizzo, Robertson, Robie, Robinet, Rocchi, Rodriguez, Rolland, Rollins,
  Romanelli, Romano, Romel, Romero-Rodríguez, Romero-Shaw, Romie, Ronchini,
  Rosa, Rose, Rosińska, Ross, Rowan, Rowlinson, Roy, Roy, Roy, Rozza, Ruggi,
  Ryan, Sachdev, Sadecki, Sadiq, Sago, Saito, Saito, Sakai, Sakai,
  Sakellariadou, Sakuno, Salafia, Salconi, Saleem, Salemi, Samajdar, Sanchez,
  Sanchez, Sanchez, Sanchis-Gual, Sanders, Sanuy, Saravanan, Sarin, Sassolas,
  Satari, Sathyaprakash, Sato, Sato, Sauter, Savage, Sawada, Sawant, Sawant,
  Sayah, Schaetzl, Scheel, Scheuer, Schiworski, Schmidt, Schmidt, Schnabel,
  Schneewind, Schofield, Schönbeck, Schulte, Schutz, Schwartz, Scott, Scott,
  Seglar-Arroyo, Sekiguchi, Sekiguchi, Sellers, Sengupta, Sentenac, Seo,
  Sequino, Sergeev, Setyawati, Shaffer, Shahriar, Shams, Shao, Sharma, Sharma,
  Shawhan, Shcheblanov, Shibagaki, Shikauchi, Shimizu, Shimoda, Shimode,
  Shinkai, Shishido, Shoda, Shoemaker, Shoemaker, ShyamSundar, Sieniawska,
  Sigg, Singer, Singh, Singh, Singha, Sintes, Sipala, Skliris, Slagmolen,
  Slaven-Blair, Smetana, Smith, Smith, Soldateschi, Somala, Somiya, Son, Soni,
  Soni, Sordini, Sorrentino, Sorrentino, Sotani, Soulard, Souradeep, Sowell,
  Spagnuolo, Spencer, Spera, Srinivasan, Srivastava, Srivastava, Staats,
  Stachie, Steer, Steinlechner, Steinlechner, Stops, Stover, Strain, Strang,
  Stratta, Strunk, Sturani, Stuver, Sudhagar, Sudhir, Sugimoto, Suh,
  Summerscales, Sun, Sun, Sunil, Sur, Suresh, Sutton, Suzuki, Suzuki, Swinkels,
  Szczepańczyk, Szewczyk, Tacca, Tagoshi, Tait, Takahashi, Takahashi,
  Takamori, Takano, Takeda, Takeda, Talbot, Talbot, Tanaka, Tanaka, Tanaka,
  Tanaka, Tanaka, Tanasijczuk, Tanioka, Tanner, Tao, Tao, Martín, Taranto,
  Tasson, Telada, Tenorio, Terhune, Terkowski, Thirugnanasambandam, Thomas,
  Thomas, Thompson, Thondapu, Thorne, Thrane, Tiwari, Tiwari, Tiwari, Toivonen,
  Toland, Tolley, Tomaru, Tomigami, Tomura, Tonelli, Torres-Forné, Torrie,
  Melo, Töyrä, Trapananti, Travasso, Traylor, Trevor, Tringali, Tripathee,
  Troiano, Trovato, Trozzo, Trudeau, Tsai, Tsai, Tsang, Tsang, Tsao, Tse, Tso,
  Tsubono, Tsuchida, Tsukada, Tsuna, Tsutsui, Tsuzuki, Turbang, Turconi,
  Tuyenbayev, Ubhi, Uchikata, Uchiyama, Udall, Ueda, Uehara, Ueno, Ueshima,
  Unnikrishnan, Uraguchi, Urban, Ushiba, Utina, Vahlbruch, Vajente, Vajpeyi,
  Valdes, Valentini, Valsan, van Bakel, van Beuzekom, Brand, Broeck,
  Vander-Hyde, van~der Schaaf, van Heijningen, Vanosky, van Putten, van
  Remortel, Vardaro, Vargas, Varma, Vasúth, Vecchio, Vedovato, Veitch, Veitch,
  Venneberg, Venugopalan, Verkindt, Verma, Verma, Veske, Vetrano, Viceré,
  Vidyant, Viets, Vijaykumar, Villa-Ortega, Vinet, Virtuoso, Vitale, Vo, Vocca,
  von Reis, von Wrangel, Vorvick, Vyatchanin, Wade, Wade, Wagner, Walet,
  Walker, Wallace, Wallace, Walsh, Wang, Wang, Wang, Ward, Warner, Was,
  Washimi, Washington, Watchi, Weaver, Webster, Weinert, Weinstein, Weiss,
  Weller, Wellmann, Wen, Weßels, Wette, Whelan, White, Whiting, Whittle,
  Wilken, Williams, Williams, Williamson, Willis, Willke, Wilson, Winkler,
  Wipf, Wlodarczyk, Woan, Woehler, Wofford, Wong, Wu, Wu, Wu, Wu, Wysocki,
  Xiao, Xu, Yamada, Yamamoto, Yamamoto, Yamamoto, Yamamoto, Yamashita,
  Yamazaki, Yang, Yang, Yang, Yang, Yang, Yap, Yeeles, Yelikar, Ying, Yokogawa,
  Yokoyama, Yokozawa, Yoo, Yoshioka, Yu, Yu, Yuzurihara, Zadrożny, Zanolin,
  Zeidler, Zelenova, Zendri, Zevin, Zhan, Zhang, Zhang, Zhang, Zhang, Zhang,
  Zhao, Zhao, Zhao, Zhao, Zhou, Zhou, Zhu, Zhu, Zimmerman, Zlochower, Zucker,
  \& Zweizig}]{abbott_population_2022}
---. 2022{\natexlab{b}}, The population of merging compact binaries inferred
  using gravitational waves through {GWTC}-3,  arXiv,
  \dodoi{10.48550/arXiv.2111.03634}

\bibitem[{Acernese {et~al.}(2015)Acernese, Agathos, Agatsuma, Aisa, Allemandou,
  Allocca, Amarni, Astone, Balestri, Ballardin, Barone, Baronick, Barsuglia,
  Basti, Basti, Bauer, Bavigadda, Bejger, Beker, Belczynski, Bersanetti,
  Bertolini, Bitossi, Bizouard, Bloemen, Blom, Boer, Bogaert, Bondi, Bondu,
  Bonelli, Bonnand, Boschi, Bosi, Bouedo, Bradaschia, Branchesi, Briant,
  Brillet, Brisson, Bulik, Bulten, Buskulic, Buy, Cagnoli, Calloni, Campeggi,
  Canuel, Carbognani, Cavalier, Cavalieri, Cella, Cesarini, Chassande-Mottin,
  Chincarini, Chiummo, Chua, Cleva, Coccia, Cohadon, Colla, Colombini, Conte,
  Coulon, Cuoco, Dalmaz, D'Antonio, Dattilo, Davier, Day, Debreczeni,
  Degallaix, Deléglise, Del~Pozzo, Dereli, De~Rosa, Di~Fiore, Di~Lieto,
  Di~Virgilio, Doets, Dolique, Drago, Ducrot, Endrőczi, Fafone, Farinon,
  Ferrante, Ferrini, Fidecaro, Fiori, Flaminio, Fournier, Franco, Frasca,
  Frasconi, Gammaitoni, Garufi, Gaspard, Gatto, Gemme, Gendre, Genin, Gennai,
  Ghosh, Giacobone, Giazotto, Gouaty, Granata, Greco, Groot, Guidi, Harms,
  Heidmann, Heitmann, Hello, Hemming, Hennes, Hofman, Jaranowski, Jonker,
  Kasprzack, Kéfélian, Kowalska, Kraan, Królak, Kutynia, Lazzaro, Leonardi,
  Leroy, Letendre, Li, Lieunard, Lorenzini, Loriette, Losurdo, Magazzù,
  Majorana, Maksimovic, Malvezzi, Man, Mangano, Mantovani, Marchesoni, Marion,
  Marque, Martelli, Martellini, Masserot, Meacher, Meidam, Mezzani, Michel,
  Milano, Minenkov, Moggi, Mohan, Montani, Morgado, Mours, Mul, Nagy,
  Nardecchia, Naticchioni, Nelemans, Neri, Neri, Nocera, Pacaud, Palomba,
  Paoletti, Paoli, Pasqualetti, Passaquieti, Passuello, Perciballi, Petit,
  Pichot, Piergiovanni, Pillant, Piluso, Pinard, Poggiani, Prijatelj, Prodi,
  Punturo, Puppo, Rabeling, Rácz, Rapagnani, Razzano, Re, Regimbau, Ricci,
  Robinet, Rocchi, Rolland, Romano, Rosińska, Ruggi, Saracco, Sassolas,
  Schimmel, Sentenac, Sequino, Shah, Siellez, Straniero, Swinkels, Tacca,
  Tonelli, Travasso, Turconi, Vajente, van Bakel, van Beuzekom, Brand, Broeck,
  van~der Sluys, van Heijningen, Vasúth, Vedovato, Veitch, Verkindt, Vetrano,
  Viceré, Vinet, Visser, Vocca, Ward, Was, Wei, Yvert, Zadrożny, \&
  Zendri}]{acernese_advanced_2015}
Acernese, F., Agathos, M., Agatsuma, K., {et~al.} 2015, Classical and Quantum
  Gravity, 32, 024001, \dodoi{10.1088/0264-9381/32/2/024001}

\bibitem[{Aghanim {et~al.}(2020)Aghanim, Akrami, Ashdown, Aumont, Baccigalupi,
  Ballardini, Banday, Barreiro, Bartolo, Basak, Battye, Benabed, Bernard,
  Bersanelli, Bielewicz, Bock, Bond, Borrill, Bouchet, Boulanger, Bucher,
  Burigana, Butler, Calabrese, Cardoso, Carron, Challinor, Chiang, Chluba,
  Colombo, Combet, Contreras, Crill, Cuttaia, de~Bernardis, de~Zotti,
  Delabrouille, Delouis, Di~Valentino, Diego, Doré, Douspis, Ducout, Dupac,
  Dusini, Efstathiou, Elsner, Enßlin, Eriksen, Fantaye, Farhang, Fergusson,
  Fernandez-Cobos, Finelli, Forastieri, Frailis, Fraisse, Franceschi, Frolov,
  Galeotta, Galli, Ganga, Génova-Santos, Gerbino, Ghosh, González-Nuevo,
  Górski, Gratton, Gruppuso, Gudmundsson, Hamann, Handley, Hansen, Herranz,
  Hildebrandt, Hivon, Huang, Jaffe, Jones, Karakci, Keihänen, Keskitalo,
  Kiiveri, Kim, Kisner, Knox, Krachmalnicoff, Kunz, Kurki-Suonio, Lagache,
  Lamarre, Lasenby, Lattanzi, Lawrence, Le~Jeune, Lemos, Lesgourgues, Levrier,
  Lewis, Liguori, Lilje, Lilley, Lindholm, López-Caniego, Lubin, Ma,
  Macías-Pérez, Maggio, Maino, Mandolesi, Mangilli, Marcos-Caballero, Maris,
  Martin, Martinelli, Martínez-González, Matarrese, Mauri, McEwen, Meinhold,
  Melchiorri, Mennella, Migliaccio, Millea, Mitra, Miville-Deschênes,
  Molinari, Montier, Morgante, Moss, Natoli, Nørgaard-Nielsen, Pagano,
  Paoletti, Partridge, Patanchon, Peiris, Perrotta, Pettorino, Piacentini,
  Polastri, Polenta, Puget, Rachen, Reinecke, Remazeilles, Renzi, Rocha,
  Rosset, Roudier, Rubiño-Martín, Ruiz-Granados, Salvati, Sandri, Savelainen,
  Scott, Shellard, Sirignano, Sirri, Spencer, Sunyaev, Suur-Uski, Tauber,
  Tavagnacco, Tenti, Toffolatti, Tomasi, Trombetti, Valenziano, Valiviita,
  Van~Tent, Vibert, Vielva, Villa, Vittorio, Wandelt, Wehus, White, White,
  Zacchei, \& Zonca}]{aghanim_planck_2020}
Aghanim, N., Akrami, Y., Ashdown, M., {et~al.} 2020, Astronomy and
  Astrophysics, 641, A6, \dodoi{10.1051/0004-6361/201833910}

\bibitem[{Arnason {et~al.}(2021)Arnason, Papei, Barmby, Bahramian, \&
  Gorski}]{arnason_distances_2021}
Arnason, R.~M., Papei, H., Barmby, P., Bahramian, A., \& Gorski, M.~D. 2021,
  Monthly Notices of the Royal Astronomical Society, 502, 5455,
  \dodoi{10.1093/mnras/stab345}

\bibitem[{Bardeen(1970)}]{bardeen_kerr_1970}
Bardeen, J.~M. 1970, Nature, 226, 64, \dodoi{10.1038/226064a0}

\bibitem[{Barrett {et~al.}(2018)Barrett, Gaebel, Neijssel, Vigna-Gómez,
  Stevenson, Berry, Farr, \& Mandel}]{barrett_accuracy_2018}
Barrett, J.~W., Gaebel, S.~M., Neijssel, C.~J., {et~al.} 2018, Monthly Notices
  of the Royal Astronomical Society, 477, 4685, \dodoi{10.1093/mnras/sty908}

\bibitem[{Basu-Zych {et~al.}(2020)Basu-Zych, Hornschemeier, Haberl, Vulic,
  Wilms, Zezas, Kovlakas, Ptak, \& Dauser}]{basu-zych_next_2020}
Basu-Zych, A.~R., Hornschemeier, A.~E., Haberl, F., {et~al.} 2020, Monthly
  Notices of the Royal Astronomical Society, 498, 1651,
  \dodoi{10.1093/mnras/staa2343}

\bibitem[{Belczynski {et~al.}(2012)Belczynski, Bulik, \&
  Fryer}]{belczynski_high_2012}
Belczynski, K., Bulik, T., \& Fryer, C.~L. 2012, arXiv:1208.2422 [astro-ph].
\newblock \url{http://arxiv.org/abs/1208.2422}

\bibitem[{Belczynski {et~al.}(2013)Belczynski, Bulik, Mandel, Sathyaprakash,
  Zdziarski, \& Mikolajewska}]{belczynski_cyg_2013}
Belczynski, K., Bulik, T., Mandel, I., {et~al.} 2013, The Astrophysical
  Journal, 764, 96, \dodoi{10.1088/0004-637X/764/1/96}

\bibitem[{Belczynski {et~al.}(2016)Belczynski, Holz, Bulik, \&
  O'Shaughnessy}]{belczynski_first_2016}
Belczynski, K., Holz, D.~E., Bulik, T., \& O'Shaughnessy, R. 2016, Nature, 534,
  512, \dodoi{10.1038/nature18322}

\bibitem[{Belczynski {et~al.}(2008)Belczynski, Kalogera, Rasio, Taam, Zezas,
  Bulik, Maccarone, \& Ivanova}]{belczynski_compact_2008}
Belczynski, K., Kalogera, V., Rasio, F.~A., {et~al.} 2008, The Astrophysical
  Journal Supplement Series, 174, 223, \dodoi{10.1086/521026}

\bibitem[{Belczynski {et~al.}(2022)Belczynski, Romagnolo, Olejak, Klencki,
  Chattopadhyay, Stevenson, Miller, Lasota, \&
  Crowther}]{belczynski_uncertain_2022}
Belczynski, K., Romagnolo, A., Olejak, A., {et~al.} 2022, The Astrophysical
  Journal, 925, 69, \dodoi{10.3847/1538-4357/ac375a}

\bibitem[{Blondin \& Owen(1997)}]{blondin_wind_1997}
Blondin, J.~M., \& Owen, M.~P. 1997, 121, 361.
\newblock \url{https://ui.adsabs.harvard.edu/abs/1997ASPC..121..361B}

\bibitem[{Bolton(1972)}]{bolton_identification_1972}
Bolton, C.~T. 1972, Nature, 235, 271, \dodoi{10.1038/235271b0}

\bibitem[{Bondi \& Hoyle(1944)}]{1944MNRAS.104..273B}
Bondi, H., \& Hoyle, F. 1944, Monthly Notices of the Royal Astronomical
  Society, 104, 273, \dodoi{10.1093/mnras/104.5.273}

\bibitem[{Breivik {et~al.}(2020)Breivik, Coughlin, Zevin, Rodriguez, Kremer,
  Ye, Andrews, Kurkowski, Digman, Larson, \& Rasio}]{breivik_cosmic_2020}
Breivik, K., Coughlin, S., Zevin, M., {et~al.} 2020, The Astrophysical Journal,
  898, 71, \dodoi{10.3847/1538-4357/ab9d85}

\bibitem[{Broekgaarden {et~al.}(2022)Broekgaarden, Berger, Stevenson, Justham,
  Mandel, Chruślińska, van Son, Wagg, Vigna-Gómez, de~Mink, Chattopadhyay,
  \& Neijssel}]{broekgaarden_impact_2022}
Broekgaarden, F.~S., Berger, E., Stevenson, S., {et~al.} 2022, Monthly Notices
  of the Royal Astronomical Society, \dodoi{10.1093/mnras/stac1677}

\bibitem[{{Chandra X-ray Center} {et~al.}(2021){Chandra X-ray Center}, {Chandra
  Project Science, MSFC}, \& {Chandra IPI
  Teams}}]{chandra_x-ray_center_chandra_2021}
{Chandra X-ray Center}, {Chandra Project Science, MSFC}, \& {Chandra IPI
  Teams}. 2021, Chandra {Cycle} 24 {Proposers}' {Observatory} {Guide}.
\newblock \url{https://cxc.harvard.edu/proposer/POG/}

\bibitem[{Chen {et~al.}(2021)Chen, Holz, Miller, Evans, Vitale, \&
  Creighton}]{chen_distance_2021}
Chen, H.-Y., Holz, D.~E., Miller, J., {et~al.} 2021, Classical and Quantum
  Gravity, 38, 055010, \dodoi{10.1088/1361-6382/abd594}

\bibitem[{Chruślińska(2022)}]{chruslinska_chemical_2022}
Chruślińska, M. 2022, Chemical evolution of the {Universe} and its
  consequences for gravitational-wave astrophysics,  arXiv,
  \dodoi{10.48550/arXiv.2206.10622}

\bibitem[{Claeys {et~al.}(2014)Claeys, Pols, Izzard, Vink, \&
  Verbunt}]{claeys_theoretical_2014}
Claeys, J. S.~W., Pols, O.~R., Izzard, R.~G., Vink, J., \& Verbunt, F. W.~M.
  2014, Astronomy and Astrophysics, 563, A83,
  \dodoi{10.1051/0004-6361/201322714}

\bibitem[{Clavel {et~al.}(2019)Clavel, Tomsick, Hare, Krivonos, Mori, \&
  Stern}]{clavel_nustar_2019}
Clavel, M., Tomsick, J.~A., Hare, J., {et~al.} 2019, The Astrophysical Journal,
  887, 32, \dodoi{10.3847/1538-4357/ab4b55}

\bibitem[{Dominik {et~al.}(2015)Dominik, Berti, O'Shaughnessy, Mandel,
  Belczynski, Fryer, Holz, Bulik, \& Pannarale}]{dominik_double_2015}
Dominik, M., Berti, E., O'Shaughnessy, R., {et~al.} 2015, The Astrophysical
  Journal, 806, 263, \dodoi{10.1088/0004-637X/806/2/263}

\bibitem[{Edelman {et~al.}(2022)Edelman, Doctor, Godfrey, \&
  Farr}]{edelman_aint_2022}
Edelman, B., Doctor, Z., Godfrey, J., \& Farr, B. 2022, The Astrophysical
  Journal, 924, 101, \dodoi{10.3847/1538-4357/ac3667}

\bibitem[{Evans {et~al.}(2010)Evans, Primini, Glotfelty, Anderson, Bonaventura,
  Chen, Davis, Doe, Evans, Fabbiano, Galle, Gibbs, Grier, Hain, Hall, Harbo,
  He, Houck, Karovska, Kashyap, Lauer, McCollough, McDowell, Miller, Mitschang,
  Morgan, Mossman, Nichols, Nowak, Plummer, Refsdal, Rots, Siemiginowska,
  Sundheim, Tibbetts, Van~Stone, Winkelman, \& Zografou}]{evans_chandra_2010}
Evans, I.~N., Primini, F.~A., Glotfelty, K.~J., {et~al.} 2010, The
  Astrophysical Journal Supplement Series, 189, 37,
  \dodoi{10.1088/0067-0049/189/1/37}

\bibitem[{Fabbiano(2006)}]{fabbiano_populations_2006}
Fabbiano, G. 2006, Annual Review of Astronomy and Astrophysics, 44, 323,
  \dodoi{10.1146/annurev.astro.44.051905.092519}

\bibitem[{Finn \& Chernoff(1993)}]{finn_observing_1993}
Finn, L.~S., \& Chernoff, D.~F. 1993, Physical Review D, 47, 2198,
  \dodoi{10.1103/PhysRevD.47.2198}

\bibitem[{Fishbach {et~al.}(2020)Fishbach, Farr, \& Holz}]{fishbach_most_2020}
Fishbach, M., Farr, W.~M., \& Holz, D.~E. 2020, The Astrophysical Journal, 891,
  L31, \dodoi{10.3847/2041-8213/ab77c9}

\bibitem[{Fishbach {et~al.}(2018)Fishbach, Holz, \& Farr}]{fishbach_does_2018}
Fishbach, M., Holz, D.~E., \& Farr, W.~M. 2018, The Astrophysical Journal, 863,
  L41, \dodoi{10.3847/2041-8213/aad800}

\bibitem[{Fishbach \& Kalogera(2022)}]{fishbach_apples_2022}
Fishbach, M., \& Kalogera, V. 2022, The Astrophysical Journal Letters, 929,
  L26, \dodoi{10.3847/2041-8213/ac64a5}

\bibitem[{Fragos {et~al.}(2022)Fragos, Andrews, Bavera, Berry, Coughlin,
  Dotter, Giri, Kalogera, Katsaggelos, Kovlakas, Lalvani, Misra, Srivastava,
  Qin, Rocha, Roman-Garza, Serra, Stahle, Sun, Teng, Trajcevski, Tran, Xing,
  Zapartas, \& Zevin}]{fragos_posydon_2022}
Fragos, T., Andrews, J.~J., Bavera, S.~S., {et~al.} 2022, {POSYDON}: {A}
  {General}-{Purpose} {Population} {Synthesis} {Code} with {Detailed}
  {Binary}-{Evolution} {Simulations},  arXiv, \dodoi{10.48550/arXiv.2202.05892}

\bibitem[{Fryer {et~al.}(2012)Fryer, Belczynski, Wiktorowicz, Dominik,
  Kalogera, \& Holz}]{fryer_compact_2012}
Fryer, C.~L., Belczynski, K., Wiktorowicz, G., {et~al.} 2012, The Astrophysical
  Journal, 749, 91, \dodoi{10.1088/0004-637X/749/1/91}

\bibitem[{Gallegos-Garcia {et~al.}(2021)Gallegos-Garcia, Berry, Marchant, \&
  Kalogera}]{gallegos-garcia_binary_2021}
Gallegos-Garcia, M., Berry, C. P.~L., Marchant, P., \& Kalogera, V. 2021, The
  Astrophysical Journal, 922, 110, \dodoi{10.3847/1538-4357/ac2610}

\bibitem[{Gallegos-Garcia {et~al.}(2022)Gallegos-Garcia, Fishbach, Kalogera,
  Berry, \& Doctor}]{gallegos-garcia_high-spin_2022}
Gallegos-Garcia, M., Fishbach, M., Kalogera, V., Berry, C. P.~L., \& Doctor, Z.
  2022, Do high-spin high mass {X}-ray binaries contribute to the population of
  merging binary black holes?,  arXiv, \dodoi{10.48550/arXiv.2207.14290}

\bibitem[{Grevesse \& Sauval(1998)}]{grevesse_standard_1998}
Grevesse, N., \& Sauval, A.~J. 1998, Space Science Reviews, 85, 161,
  \dodoi{10.1023/A:1005161325181}

\bibitem[{Haberl \& Sturm(2016)}]{haberl_high-mass_2016}
Haberl, F., \& Sturm, R. 2016, Astronomy \& Astrophysics, 586, A81,
  \dodoi{10.1051/0004-6361/201527326}

\bibitem[{Heuvel(2018)}]{heuvel_high-mass_2018}
Heuvel, E. P. J. v.~d. 2018, Proceedings of the International Astronomical
  Union, 14, 1, \dodoi{10.1017/S1743921319001315}

\bibitem[{Hirai \& Mandel(2021)}]{hirai_conditions_2021}
Hirai, R., \& Mandel, I. 2021, Publications of the Astronomical Society of
  Australia, 38, e056, \dodoi{10.1017/pasa.2021.53}

\bibitem[{Hunter(2007)}]{hunter_matplotlib_2007}
Hunter, J.~D. 2007, Computing in Science \& Engineering, 9, 90,
  \dodoi{10.1109/MCSE.2007.55}

\bibitem[{Hurley {et~al.}(2000)Hurley, Pols, \&
  Tout}]{hurley_comprehensive_2000}
Hurley, J.~R., Pols, O.~R., \& Tout, C.~A. 2000, Monthly Notices of the Royal
  Astronomical Society, 315, 543, \dodoi{10.1046/j.1365-8711.2000.03426.x}

\bibitem[{Hurley {et~al.}(2002)Hurley, Tout, \& Pols}]{hurley_evolution_2002}
Hurley, J.~R., Tout, C.~A., \& Pols, O.~R. 2002, Monthly Notices of the Royal
  Astronomical Society, 329, 897, \dodoi{10.1046/j.1365-8711.2002.05038.x}

\bibitem[{Ivanova {et~al.}(2013)Ivanova, Justham, Chen, De~Marco, Fryer,
  Gaburov, Ge, Glebbeek, Han, Li, Lu, Marsh, Podsiadlowski, Potter, Soker,
  Taam, Tauris, van~den Heuvel, \& Webbink}]{ivanova_common_2013}
Ivanova, N., Justham, S., Chen, X., {et~al.} 2013, Astronomy and Astrophysics
  Review, 21, 59, \dodoi{10.1007/s00159-013-0059-2}

\bibitem[{Kalogera \& Baym(1996)}]{kalogera_maximum_1996}
Kalogera, V., \& Baym, G. 1996, The Astrophysical Journal Letters, 470, L61,
  \dodoi{10.1086/310296}

\bibitem[{Kratter(2011)}]{kratter_formation_2011}
Kratter, K.~M. 2011, arXiv:1109.3740 [astro-ph].
\newblock \url{http://arxiv.org/abs/1109.3740}

\bibitem[{Kretschmar {et~al.}(2019)Kretschmar, Fürst, Sidoli, Bozzo,
  Alfonso-Garzón, Bodaghee, Chaty, Chernyakova, Ferrigno, Manousakis,
  Negueruela, Postnov, Paizis, Reig, Rodes-Roca, Tsygankov, Bird, Kühnel,
  Blay, Caballero, Coe, Domingo, Doroshenko, Ducci, Falanga, Grebenev,
  Grinberg, Hemphill, Kreykenbohm, Fritz, Li, Lutovinov, Martínez-Núñez,
  Mas-Hesse, Masetti, McBride, Neronov, Pottschmidt, Rodriguez, Romano,
  Rothschild, Santangelo, Sguera, Staubert, Tomsick, Torrejón, Torres, Walter,
  Wilms, Wilson-Hodge, \& Zhang}]{kretschmar_advances_2019}
Kretschmar, P., Fürst, F., Sidoli, L., {et~al.} 2019, New Astronomy Reviews,
  86, 101546, \dodoi{10.1016/j.newar.2020.101546}

\bibitem[{Krivonos {et~al.}(2012)Krivonos, Tsygankov, Lutovinov, Revnivtsev,
  Churazov, \& Sunyaev}]{krivonos_integralibis_2012}
Krivonos, R., Tsygankov, S., Lutovinov, A., {et~al.} 2012, Astronomy and
  Astrophysics, 545, A27, \dodoi{10.1051/0004-6361/201219617}

\bibitem[{Krivonos {et~al.}(2021)Krivonos, Bird, Churazov, Tomsick, Bazzano,
  Beckmann, Belanger, Bodaghee, Chaty, Kuulkers, Lutovinov, Malizia, Masetti,
  Mereminskiy, Sunyaev, Tsygankov, Ubertini, \& Winkler}]{krivonos_15_2021}
Krivonos, R.~A., Bird, A.~J., Churazov, E.~M., {et~al.} 2021, New Astronomy
  Reviews, 92, 101612, \dodoi{10.1016/j.newar.2021.101612}

\bibitem[{Kroupa(2001)}]{kroupa_variation_2001}
Kroupa, P. 2001, Monthly Notices of the Royal Astronomical Society, 322, 231,
  \dodoi{10.1046/j.1365-8711.2001.04022.x}

\bibitem[{Lattimer(2021)}]{lattimer_neutron_2021}
Lattimer, J. 2021, Annual Review of Nuclear and Particle Science, 71, 433,
  \dodoi{10.1146/annurev-nucl-102419-124827}

\bibitem[{Lazzarini {et~al.}(2018)Lazzarini, Hornschemeier, Williams, Wik,
  Vulic, Yukita, Zezas, Lewis, Durbin, Ptak, Bodaghee, Lehmer, Antoniou, \&
  Maccarone}]{lazzarini_young_2018}
Lazzarini, M., Hornschemeier, A.~E., Williams, B.~F., {et~al.} 2018, The
  Astrophysical Journal, 862, 28, \dodoi{10.3847/1538-4357/aacb2a}

\bibitem[{Livio \& Soker(1988)}]{livio_common_1988}
Livio, M., \& Soker, N. 1988, The Astrophysical Journal, 329, 764,
  \dodoi{10.1086/166419}

\bibitem[{Lutovinov {et~al.}(2013)Lutovinov, Revnivtsev, Tsygankov, \&
  Krivonos}]{lutovinov_population_2013}
Lutovinov, A.~A., Revnivtsev, M.~G., Tsygankov, S.~S., \& Krivonos, R.~A. 2013,
  Monthly Notices of the Royal Astronomical Society, 431, 327,
  \dodoi{10.1093/mnras/stt168}

\bibitem[{Mandel \& Farmer(2018)}]{mandel_merging_2018}
Mandel, I., \& Farmer, A. 2018, arXiv:1806.05820 [astro-ph, physics:gr-qc].
\newblock \url{http://arxiv.org/abs/1806.05820}

\bibitem[{Marchant {et~al.}(2021)Marchant, Pappas, Gallegos-Garcia, Berry,
  Taam, Kalogera, \& Podsiadlowski}]{marchant_role_2021}
Marchant, P., Pappas, K. M.~W., Gallegos-Garcia, M., {et~al.} 2021, Astronomy
  \& Astrophysics, 650, A107, \dodoi{10.1051/0004-6361/202039992}

\bibitem[{Marchant {et~al.}(2019)Marchant, Renzo, Farmer, Pappas, Taam,
  de~Mink, \& Kalogera}]{marchant_pulsational_2019}
Marchant, P., Renzo, M., Farmer, R., {et~al.} 2019, The Astrophysical Journal,
  882, 36, \dodoi{10.3847/1538-4357/ab3426}

\bibitem[{Mark {et~al.}(1969)Mark, Price, Rodrigues, Seward, \&
  Swift}]{mark_detection_1969}
Mark, H., Price, R., Rodrigues, R., Seward, F.~D., \& Swift, C.~D. 1969, The
  Astrophysical Journal, 155, L143, \dodoi{10.1086/180322}

\bibitem[{McKinney(2010)}]{mckinney_data_2010}
McKinney, W. 2010, Proceedings of the 9th Python in Science Conference, 56,
  \dodoi{10.25080/Majora-92bf1922-00a}

\bibitem[{Miller \& Miller(2015)}]{miller_masses_2015}
Miller, M.~C., \& Miller, J.~M. 2015, Physics Reports, 548, 1,
  \dodoi{10.1016/j.physrep.2014.09.003}

\bibitem[{Miller-Jones {et~al.}(2021)Miller-Jones, Bahramian, Orosz, Mandel,
  Gou, Maccarone, Neijssel, Zhao, Ziółkowski, Reid, Uttley, Zheng, Byun,
  Dodson, Grinberg, Jung, Kim, Marcote, Markoff, Rioja, Rushton, Russell,
  Sivakoff, Tetarenko, Tudose, \& Wilms}]{miller-jones_cygnus_2021}
Miller-Jones, J. C.~A., Bahramian, A., Orosz, J.~A., {et~al.} 2021, Science,
  371, 1046, \dodoi{10.1126/science.abb3363}

\bibitem[{Motta {et~al.}(2021)Motta, Rodriguez, Jourdain, Del~Santo, Belanger,
  Cangemi, Grinberg, Kajava, Kuulkers, Malzac, Pottschmidt, Roques,
  Sánchez-Fernández, \& Wilms}]{motta_integral_2021}
Motta, S.~E., Rodriguez, J., Jourdain, E., {et~al.} 2021, New Astronomy
  Reviews, 93, 101618, \dodoi{10.1016/j.newar.2021.101618}

\bibitem[{Neijssel {et~al.}(2021)Neijssel, Vinciguerra, Vigna-Gomez, Hirai,
  Miller-Jones, Bahramian, Maccarone, \& Mandel}]{neijssel_wind_2021}
Neijssel, C.~J., Vinciguerra, S., Vigna-Gomez, A., {et~al.} 2021, The
  Astrophysical Journal, 908, 118, \dodoi{10.3847/1538-4357/abde4a}

\bibitem[{Neijssel {et~al.}(2019)Neijssel, Vigna-Gómez, Stevenson, Barrett,
  Gaebel, Broekgaarden, de~Mink, Szécsi, Vinciguerra, \&
  Mandel}]{neijssel_effect_2019}
Neijssel, C.~J., Vigna-Gómez, A., Stevenson, S., {et~al.} 2019, Monthly
  Notices of the Royal Astronomical Society, 490, 3740,
  \dodoi{10.1093/mnras/stz2840}

\bibitem[{Nelson {et~al.}(2019)Nelson, Springel, Pillepich, Rodriguez-Gomez,
  Torrey, Genel, Vogelsberger, Pakmor, Marinacci, Weinberger, Kelley, Lovell,
  Diemer, \& Hernquist}]{nelson_illustristng_2019}
Nelson, D., Springel, V., Pillepich, A., {et~al.} 2019, Computational
  Astrophysics and Cosmology, 6, 2, \dodoi{10.1186/s40668-019-0028-x}

\bibitem[{Nelson {et~al.}(2021)Nelson, Springel, Pillepich, Rodriguez-Gomez,
  Torrey, Genel, Vogelsberger, Pakmor, Marinacci, Weinberger, Kelley, Lovell,
  Diemer, \& Hernquist}]{nelson_illustristng_2021}
---. 2021, The {IllustrisTNG} {Simulations}: {Public} {Data} {Release},  arXiv,
  \dodoi{10.48550/arXiv.1812.05609}

\bibitem[{Oh {et~al.}(2018)Oh, Koss, Markwardt, Schawinski, Baumgartner,
  Barthelmy, Cenko, Gehrels, Mushotzky, Petulante, Ricci, Lien, \&
  Trakhtenbrot}]{oh_105-month_2018}
Oh, K., Koss, M., Markwardt, C.~B., {et~al.} 2018, The Astrophysical Journal
  Supplement Series, 235, 4, \dodoi{10.3847/1538-4365/aaa7fd}

\bibitem[{Orosz {et~al.}(2007)Orosz, McClintock, Narayan, Bailyn, Hartman,
  Macri, Liu, Pietsch, Remillard, Shporer, \&
  Mazeh}]{orosz_1565-solar-mass_2007}
Orosz, J.~A., McClintock, J.~E., Narayan, R., {et~al.} 2007, Nature, 449, 872,
  \dodoi{10.1038/nature06218}

\bibitem[{Orosz {et~al.}(2009)Orosz, Steeghs, McClintock, Torres, Bochkov, Gou,
  Narayan, Blaschak, Levine, Remillard, Bailyn, Dwyer, \&
  Buxton}]{orosz_new_2009}
Orosz, J.~A., Steeghs, D., McClintock, J.~E., {et~al.} 2009, The Astrophysical
  Journal, 697, 573, \dodoi{10.1088/0004-637X/697/1/573}

\bibitem[{Pavlinsky {et~al.}(2021)Pavlinsky, Tkachenko, Levin, Alexandrovich,
  Arefiev, Babyshkin, Batanov, Bodnar, Bogomolov, Bubnov, Buntov, Burenin,
  Chelovekov, Chen, Drozdova, Ehlert, Filippova, Frolov, Gamkov, Garanin,
  Garin, Glushenko, Gorelov, Grebenev, Grigorovich, Gureev, Gurova, Ilkaev,
  Katasonov, Krivchenko, Krivonos, Korotkov, Kudelin, Kuznetsova, Lazarchuk,
  Lomakin, Lapshov, Lipilin, Lutovinov, Mereminskiy, Molkov, Nazarov,
  Oleinikov, Pikalov, Ramsey, Roiz, Rotin, Ryadov, Sankin, Sazonov, Sedov,
  Semena, Semena, Serbinov, Shirshakov, Shtykovsky, Shvetsov, Sunyaev, Swartz,
  Tambov, Voron, \& Yaskovich}]{pavlinsky_art-xc_2021}
Pavlinsky, M., Tkachenko, A., Levin, V., {et~al.} 2021, Astronomy \&
  Astrophysics, 650, A42, \dodoi{10.1051/0004-6361/202040265}

\bibitem[{Paxton {et~al.}(2011)Paxton, Bildsten, Dotter, Herwig, Lesaffre, \&
  Timmes}]{paxton_modules_2011}
Paxton, B., Bildsten, L., Dotter, A., {et~al.} 2011, The Astrophysical Journal
  Supplement Series, 192, 3, \dodoi{10.1088/0067-0049/192/1/3}

\bibitem[{Podsiadlowski {et~al.}(2003)Podsiadlowski, Rappaport, \&
  Han}]{podsiadlowski_formation_2003}
Podsiadlowski, P., Rappaport, S., \& Han, Z. 2003, Monthly Notices of the Royal
  Astronomical Society, 341, 385, \dodoi{10.1046/j.1365-8711.2003.06464.x}

\bibitem[{Predehl {et~al.}(2021)Predehl, Andritschke, Arefiev, Babyshkin,
  Batanov, Becker, Böhringer, Bogomolov, Boller, Borm, Bornemann, Bräuninger,
  Brüggen, Brunner, Brusa, Bulbul, Buntov, Burwitz, Burkert, Clerc, Churazov,
  Coutinho, Dauser, Dennerl, Doroshenko, Eder, Emberger, Eraerds, Finoguenov,
  Freyberg, Friedrich, Friedrich, Fürmetz, Georgakakis, Gilfanov, Granato,
  Grossberger, Gueguen, Gureev, Haberl, Hälker, Hartner, Hasinger, Huber, Ji,
  Kienlin, Kink, Korotkov, Kreykenbohm, Lamer, Lomakin, Lapshov, Liu, Maitra,
  Meidinger, Menz, Merloni, Mernik, Mican, Mohr, Müller, Nandra, Nazarov,
  Pacaud, Pavlinsky, Perinati, Pfeffermann, Pietschner, Ramos-Ceja, Rau,
  Reiffers, Reiprich, Robrade, Salvato, Sanders, Santangelo, Sasaki, Scheuerle,
  Schmid, Schmitt, Schwope, Shirshakov, Steinmetz, Stewart, Strüder, Sunyaev,
  Tenzer, Tiedemann, Trümper, Voron, Weber, Wilms, \&
  Yaroshenko}]{predehl_erosita_2021}
Predehl, P., Andritschke, R., Arefiev, V., {et~al.} 2021, Astronomy \&
  Astrophysics, 647, A1, \dodoi{10.1051/0004-6361/202039313}

\bibitem[{Ramachandran {et~al.}(2022)Ramachandran, Oskinova, Hamann, Sander,
  Todt, Pauli, Shenar, Torrejón, Postnov, Blondin, Bozzo, Hainich, \&
  Massa}]{ramachandran_phase-resolved_2022}
Ramachandran, V., Oskinova, L.~M., Hamann, W.-R., {et~al.} 2022, Astronomy \&
  Astrophysics, \dodoi{10.1051/0004-6361/202243683}

\bibitem[{Remillard \& McClintock(2006)}]{remillard_x-ray_2006}
Remillard, R.~A., \& McClintock, J.~E. 2006, Annual Review of Astronomy and
  Astrophysics, 44, 49, \dodoi{10.1146/annurev.astro.44.051905.092532}

\bibitem[{Rhoades \& Ruffini(1974)}]{rhoades_maximum_1974}
Rhoades, C.~E., \& Ruffini, R. 1974, Physical Review Letters, 32, 324,
  \dodoi{10.1103/PhysRevLett.32.324}

\bibitem[{Rice {et~al.}(2021)Rice, Rangelov, Prestwich, Chandar, Bichon, \&
  Boldt}]{rice_x-ray_2021}
Rice, J.~R., Rangelov, B., Prestwich, A., {et~al.} 2021, The Astrophysical
  Journal, 922, 178, \dodoi{10.3847/1538-4357/ac22ac}

\bibitem[{Riley {et~al.}(2022)Riley, Agrawal, Barrett, Boyett, Broekgaarden,
  Chattopadhyay, Gaebel, Gittins, Hirai, Howitt, Justham, Khandelwal, Kummer,
  Lau, Mandel, de~Mink, Neijssel, Riley, van Son, Stevenson, Vigna-Gomez,
  Vinciguerra, Wagg, \& Willcox}]{riley_rapid_2022}
Riley, J., Agrawal, P., Barrett, J.~W., {et~al.} 2022, The Astrophysical
  Journal Supplement Series, 258, 34, \dodoi{10.3847/1538-4365/ac416c}

\bibitem[{Rosen {et~al.}(2016)Rosen, Webb, Watson, Ballet, Barret, Braito,
  Carrera, Ceballos, Coriat, Della~Ceca, Denkinson, Esquej, Farrell, Freyberg,
  Grisé, Guillout, Heil, Koliopanos, Law-Green, Lamer, Lin, Martino, Michel,
  Motch, Gomez-Moran, Page, Page, Page, Pakull, Pye, Read, Rodriguez, Sakano,
  Saxton, Schwope, Scott, Sturm, Traulsen, Yershov, \&
  Zolotukhin}]{rosen_xmm-newton_2016}
Rosen, S.~R., Webb, N.~A., Watson, M.~G., {et~al.} 2016, Astronomy \&
  Astrophysics, 590, A1, \dodoi{10.1051/0004-6361/201526416}

\bibitem[{Ryden(2002)}]{ryden_introduction_2002}
Ryden, B.~S. 2002, Introduction to {Cosmology}: {Barbara} {Ryden}, 1st edn.
  (San Francisco: Addison-Wesley)

\bibitem[{Sana {et~al.}(2012)Sana, de~Mink, de~Koter, Langer, Evans, Gieles,
  Gosset, Izzard, Le~Bouquin, \& Schneider}]{sana_binary_2012}
Sana, H., de~Mink, S.~E., de~Koter, A., {et~al.} 2012, Science, 337, 444,
  \dodoi{10.1126/science.1223344}

\bibitem[{Shao \& Li(2019)}]{shao_population_2019}
Shao, Y., \& Li, X.-D. 2019, The Astrophysical Journal, 885, 151,
  \dodoi{10.3847/1538-4357/ab4816}

\bibitem[{Spera {et~al.}(2019)Spera, Mapelli, Giacobbo, Trani, Bressan, \&
  Costa}]{spera_merging_2019}
Spera, M., Mapelli, M., Giacobbo, N., {et~al.} 2019, Monthly Notices of the
  Royal Astronomical Society, 485, 889, \dodoi{10.1093/mnras/stz359}

\bibitem[{Taibi {et~al.}(2022)Taibi, Battaglia, Leaman, Brooks, Riggs, Munshi,
  Revaz, \& Jablonka}]{taibi_stellar_2022}
Taibi, S., Battaglia, G., Leaman, R., {et~al.} 2022, The stellar metallicity
  gradients of {Local} {Group} dwarf galaxies,  arXiv,
  \dodoi{10.48550/arXiv.2206.08988}

\bibitem[{Tauris \& Heuvel(2003)}]{tauris_formation_2003}
Tauris, T.~M., \& Heuvel, E. v.~d. 2003, arXiv:astro-ph/0303456.
\newblock \url{http://arxiv.org/abs/astro-ph/0303456}

\bibitem[{Thorne(1997)}]{thorne_gravitational_1997}
Thorne, K.~S. 1997, Gravitational {Radiation} -- {A} {New} {Window} {Onto} the
  {Universe},  arXiv, \dodoi{10.48550/arXiv.gr-qc/9704042}

\bibitem[{Tiwari \& Fairhurst(2021)}]{tiwari_emergence_2021}
Tiwari, V., \& Fairhurst, S. 2021, The Astrophysical Journal Letters, 913, L19,
  \dodoi{10.3847/2041-8213/abfbe7}

\bibitem[{van~der Walt {et~al.}(2011)van~der Walt, Colbert, \&
  Varoquaux}]{van_der_walt_numpy_2011}
van~der Walt, S., Colbert, S.~C., \& Varoquaux, G. 2011, Computing in Science
  \& Engineering, 13, 22, \dodoi{10.1109/MCSE.2011.37}

\bibitem[{van Son {et~al.}(2022{\natexlab{a}})van Son, de~Mink, Chruslinska,
  Conroy, Pakmor, \& Hernquist}]{van_son_locations_2022}
van Son, L. A.~C., de~Mink, S.~E., Chruslinska, M., {et~al.}
  2022{\natexlab{a}}, The locations of features in the mass distribution of
  merging binary black holes are robust against uncertainties in the
  metallicity-dependent cosmic star formation history,  arXiv.
\newblock \url{http://arxiv.org/abs/2209.03385}

\bibitem[{van Son {et~al.}(2022{\natexlab{b}})van Son, de~Mink, Callister,
  Justham, Renzo, Wagg, Broekgaarden, Kummer, Pakmor, \&
  Mandel}]{van_son_redshift_2022}
van Son, L. A.~C., de~Mink, S.~E., Callister, T., {et~al.} 2022{\natexlab{b}},
  The Astrophysical Journal, 931, 17, \dodoi{10.3847/1538-4357/ac64a3}

\bibitem[{Verbunt(1993)}]{verbunt_origin_1993}
Verbunt, F. 1993, Annual Review of Astronomy and Astrophysics, 31, 93,
  \dodoi{10.1146/annurev.aa.31.090193.000521}

\bibitem[{Vink \& de~Koter(2005)}]{vink_metallicity_2005}
Vink, J.~S., \& de~Koter, A. 2005, Astronomy and Astrophysics, 442, 587,
  \dodoi{10.1051/0004-6361:20052862}

\bibitem[{Vink {et~al.}(2001)Vink, de~Koter, \& Lamers}]{vink_mass-loss_2001}
Vink, J.~S., de~Koter, A., \& Lamers, H. J. G. L.~M. 2001, Astronomy and
  Astrophysics, 369, 574, \dodoi{10.1051/0004-6361:20010127}

\bibitem[{Waskom(2021)}]{waskom_seaborn_2021}
Waskom, M. 2021, The Journal of Open Source Software, 6, 3021,
  \dodoi{10.21105/joss.03021}

\bibitem[{Webster \& Murdin(1972)}]{webster_cygnus_1972}
Webster, B.~L., \& Murdin, P. 1972, Nature, 235, 37, \dodoi{10.1038/235037a0}

\bibitem[{Williams {et~al.}(2021)Williams, Kreckel, Belfiore, Groves,
  Sandstrom, Santoro, Blanc, Bigiel, Boquien, Chevance, Congiu, Emsellem,
  Glover, Grasha, Klessen, Koch, Kruijssen, Leroy, Liu, Meidt, Pan, Querejeta,
  Rosolowsky, Saito, Sánchez-Blázquez, Schinnerer, Schruba, \&
  Watkins}]{williams_two-dimensional_2021}
Williams, T.~G., Kreckel, K., Belfiore, F., {et~al.} 2021, Monthly Notices of
  the Royal Astronomical Society, 509, 1303, \dodoi{10.1093/mnras/stab3082}

\bibitem[{Woosley(2017)}]{woosley_pulsational_2017}
Woosley, S.~E. 2017, The Astrophysical Journal, 836, 244,
  \dodoi{10.3847/1538-4357/836/2/244}

\bibitem[{Zdziarski {et~al.}(2013)Zdziarski, Mikolajewska, \&
  Belczynski}]{zdziarski_cyg_2013}
Zdziarski, A.~A., Mikolajewska, J., \& Belczynski, K. 2013, Monthly Notices of
  the Royal Astronomical Society: Letters, 429, L104,
  \dodoi{10.1093/mnrasl/sls035}

\bibitem[{Zevin \& Bavera(2022)}]{zevin_suspicious_2022}
Zevin, M., \& Bavera, S.~S. 2022, The Astrophysical Journal, 933, 86,
  \dodoi{10.3847/1538-4357/ac6f5d}

\bibitem[{Zevin {et~al.}(2020)Zevin, Spera, Berry, \&
  Kalogera}]{zevin_exploring_2020}
Zevin, M., Spera, M., Berry, C. P.~L., \& Kalogera, V. 2020, The Astrophysical
  Journal Letters, 899, L1, \dodoi{10.3847/2041-8213/aba74e}

\bibitem[{Zevin {et~al.}(2021)Zevin, Bavera, Berry, Kalogera, Fragos, Marchant,
  Rodriguez, Antonini, Holz, \& Pankow}]{zevin_one_2021}
Zevin, M., Bavera, S.~S., Berry, C. P.~L., {et~al.} 2021, The Astrophysical
  Journal, 910, 152, \dodoi{10.3847/1538-4357/abe40e}

\end{thebibliography}
\bibliographystyle{aasjournal}


\end{document}